\documentclass[final,3p,times,twocolumn]{elsarticle}

\usepackage[T1]{fontenc}
\usepackage{graphicx}
\usepackage[intlimits]{amsmath}
\usepackage{amssymb,amsfonts}
\usepackage{bbm}
\usepackage{siunitx}
\usepackage{float}
\usepackage{footnote}
\usepackage{tikz}
\usetikzlibrary{arrows.meta,calc,patterns}
\usepackage{cuted}
\usepackage{caption}
\usepackage{cancel}
\usepackage{dblfloatfix}
\usepackage{needspace}
\usepackage{wasysym}
\usepackage[caption=false]{subfig}
\usetikzlibrary{positioning}
\usetikzlibrary{calc}
\usepackage{tabularx}
\usepackage{longtable}
\usepackage{ltxtable}
\newcolumntype{L}[1]{>{\raggedright\arraybackslash}p{#1}}
\newcolumntype{C}[1]{>{\centering\arraybackslash}p{#1}}
\newcolumntype{R}[1]{>{\raggedleft\arraybackslash}p{#1}}
\newcolumntype{J}[1]{>{\justifying\arraybackslash}p{#1}}
\usepackage[table]{xcolor}
\makesavenoteenv{table}
\makesavenoteenv{tabular}
\usepackage{pgfplots}
\usepackage{todonotes}
\usepackage{pgfplots}
\usepackage{tikz}
\usepackage{tikzscale}
\newlength\tkwi
\newlength\tkhe
\usetikzlibrary{external}
\usepackage{amsmath}
\usepackage{xurl}
\usepackage{hyperref}

\newcommand{\ov}{\overline}

\newcommand{\pfrac}[2]{\frac{\partial #1}{\partial #2}}

\newcommand{\bbm}[1]{\mathbbm{#1}}

\newcommand{\tab}[1]{Tab.~\ref{#1}}

\pgfplotsset{compat=1.18}

\begin{document}

\begin{frontmatter}



\title{Influence of Ideal, Cubic, and Multiparameter Helmholtz Equations of State on the Static and Dynamic Performance of Porous Aerostatic Journal Bearings}


\author{Laurenz Stellmach,
Steffen Nitzschke, and
Elmar Woschke} 

\affiliation{organization={Otto von Guericke University Magdeburg},
            addressline={Universitaetsplatz 2}, 
            city={Magdeburg},
            postcode={39106}, 
            state={Saxony-Anhalt},
            country={Germany}}

\begin{abstract}
Porous aerostatic gas bearings are commonly modeled using the ideal-gas
law, although non-ideal thermodynamic behavior may become relevant for
alternative working fluids. This study quantifies the influence of the
thermodynamic closure on the stationary and dynamic response of porous
aerostatic journal bearings. A finite-volume formulation couples the
compressible Reynolds equation with three-dimensional Darcy flow, retaining
transient gas storage in both domains. Density is used as the primary
thermodynamic variable, allowing the ideal-gas law, three cubic equations
of state---van der Waals, Soave--Redlich--Kwong, and Peng--Robinson---and
multiparameter Helmholtz equations of state (EOS) to be evaluated within
an otherwise unchanged bearing model. Dry air and the refrigerants R1234yf and R1234ze(E) are considered as working fluids to assess the influence of the thermodynamic closure across fluids with different degrees of non-ideal behavior. For air, EOS-induced
differences remain negligible over the investigated range. For the
refrigerants, the ideal-gas approximation produces several-percent
deviations, with supplied mass flow and dynamic force response being
particularly sensitive. The cubic equations substantially reduce these
deviations, with the employed Peng--Robinson formulation providing the
closest agreement with the Helmholtz reference for both refrigerants at a
computational cost comparable to the other cubic models. The results
identify density and differential compressibility as the main
thermodynamic quantities governing the propagation of EOS errors into mass
transport and transient gas storage.
\end{abstract}



\begin{keyword}
Air bearing \sep Porous bearing shell \sep Real gas \sep Peng-Robinson \sep Helmholtz-EOS \sep Bearing coefficients.


\end{keyword}

\end{frontmatter}



%
\section{Introduction}
Gas-lubricated bearings provide contact-free support with low friction,
negligible lubricant contamination, and the capability for high-speed and
high-precision operation. Aerostatic bearings generate the load-carrying
pressure by supplying externally pressurized gas to the lubrication gap and
are therefore able to support the rotor independently of its rotational
speed. They are used in a broad range of precision and high-speed
applications, including ultra-precision machine tools,
semiconductor manufacturing, turbomachinery, and medical systems
\cite{Gao2019,SanAndres2021}. In porous aerostatic bearings, the gas is
supplied through a permeable bearing shell that acts as a distributed
restrictor. Compared with discrete orifice or slot restrictors, the porous
material enables gas to be supplied over a large part of the bearing surface
and thereby promotes a comparatively uniform pressure distribution
\cite{Kwan1998,Gao2019,SanAndres2021,Buuren2014}.

The numerical analysis of porous gas journal bearings is commonly based on a
coupled description of the lubrication film and the porous restrictor.
Analytical descriptions of gas-lubricated porous journal bearings date back
several decades, including early investigations of their steady-state
pressure distribution and load-carrying characteristics~\cite{Sun1975}.
The pressure field in the thin gas film is generally described by the
compressible Reynolds equation, whereas Darcy-type relations are employed
for the flow through the porous material. Established analytical and
numerical models frequently assume isothermal conditions and use the
ideal-gas relation to close the governing equations. Such approaches have,
for example, been employed for the static and dynamic analysis of porous gas
journal bearings by \textsc{San Andr\'es} et al.~\cite{SanAndres2021} and
for a spatially resolved Reynolds--Darcy formulation by \textsc{van Buuren}~\cite{Buuren2014} or \textsc{Gu} et al.~\cite{Gu2022}.
The ideal-gas approximation provides a simple and
computationally efficient relation between pressure and density and is
typically well suited to conventional air-bearing applications when the
operating conditions remain sufficiently far from thermodynamic regions
with pronounced non-ideal behavior.

The ideal-gas approximation becomes less reliable for non-air working fluids and at elevated reduced pressures or near the saturation or critical region. Under these conditions, density and differential compressibility can deviate noticeably from ideal-gas behavior, affecting mass transport and, in transient models, gas storage within the porous medium. Consequently, the required thermodynamic fidelity depends on the working fluid, operating state, and whether stationary or dynamic bearing quantities are considered.

Real-gas effects have already received considerable attention in other
classes of gas-lubricated bearings, particularly in applications involving
refrigerants and supercritical fluids. \textsc{Garcia} et
al.~\cite{Garcia2013} developed a thermo-hydrodynamic model for
refrigerant-lubricated foil gas bearings in which the lubricant properties are
described using a nonlinear cubic equation of state (EOS) and the possibility of
a vapor--liquid transition is considered. \textsc{Bouchehit} et
al.~\cite{Bouchehit2016} subsequently investigated the static and dynamic
performance of refrigerant-lubricated bearings using a modified
Peng--Robinson equation of state. For gas foil bearings operating with
high-pressure or supercritical carbon dioxide,
\textsc{Conboy}~\cite{Conboy2013} accounted for real-gas properties using
the REFPROP fluid-property database, while \textsc{Kim}~\cite{Kim2015}
developed a three-dimensional thermo-hydrodynamic model of radial foil
bearings explicitly accounting for non-ideal gas behavior. Real-gas effects have also been investigated by \textsc{Guenat} et al.~\cite{Guenat2018} for plain aerodynamic and herringbone-groove journal gas bearings, where they were shown to influence both static and dynamic bearing characteristics.

More recently, \textsc{de Waart} and \textsc{Pini}~\cite{deWaart2024}
systematically investigated the influence of the working fluid and
non-ideal thermodynamic behavior on rigid and gas foil bearings. Their
analysis employed the Peng--Robinson equation of state to identify the
underlying thermodynamic trends and high-accuracy multiparameter
Helmholtz-energy equations of state to verify the resulting observations.
The existing literature therefore demonstrates that both cubic equations
of state and high-accuracy multiparameter thermodynamic models are
established approaches for describing non-ideal behavior in gas and foil
bearing applications.

For porous gas journal bearings, explicit real-gas modeling has only recently
been introduced. \textsc{Bechiri} et al.~\cite{Bechiri2026} investigated porous gas
bearings lubricated with air, R-134a, helium, and hydrogen using the
Peng--Robinson equation of state. However, several aspects limit the
conclusions that can be drawn regarding real-gas effects in porous
aerostatic bearings. Although the model is described as employing
Darcy--Forchheimer flow, the presented porous-medium relation corresponds
to the linear Darcy law without a Forchheimer inertial term. Moreover, the
quantitative validation was performed against a refrigerant-lubricated gas
foil bearing rather than a porous gas bearing and therefore does not
directly validate the porous feeding model. The reference configuration
further employed $p_s/p_\mathrm{amb}=1$ and a comparatively large clearance
ratio of $c/R_s=0.003$, substantially reducing the aerostatic contribution
of the porous feeding system. Under these conditions, varying the
permeability from $8.2\times10^{-16}$ to $10^{-13}\,\mathrm{m^2}$ produced
only minor changes in load capacity, making it difficult to assess the
representativeness of the configuration and to isolate the reported
real-gas effects. The present work addresses these limitations using a
systematically verified porous-bearing formulation and a controlled
comparison of ideal-gas, cubic, and Helmholtz-energy equations of state
under representative aerostatic operating conditions.

Recent work has also extended porous aerostatic-bearing analysis toward
application-specific transient simulations. \textsc{Wu} et al.~\cite{Wu2026}
developed an experimentally validated CFD model for porous aerostatic
bearings in Joule--Thomson linear compressors, accounting for transient fluid--structure
interaction and investigating the influence of operating and structural
parameters. Their study focuses on transient bearing dynamics and design
optimization rather than on the influence of the thermodynamic equation of
state.

However, the influence of the equation of state model itself has not yet
been systematically analysed for porous gas journal bearings. Existing
porous-bearing formulations predominantly employ either the ideal-gas
approximation or, in the recent real-gas formulation, a single cubic
equation of state. Consequently, it remains unclear how large the error
introduced by the ideal-gas approximation becomes for different porous
bearing quantities, how accurately different cubic equations of state
recover the predictions of a high-accuracy thermodynamic reference, and
whether the required level of thermodynamic modeling differs between
stationary and dynamic analyses. In particular, a controlled comparison of several cubic equations of state against multiparameter Helmholtz-energy-based reference models within the same spatially resolved porous-bearing formulation is currently lacking in the literature. This distinction is especially relevant when transient gas
storage within the porous shell is retained, since the dynamic pressure
response then also depends on the compressibility of the gas stored within
the porous structure.

The present study addresses this gap by systematically varying the
thermodynamic closure while retaining an otherwise identical bearing
geometry, porous-domain formulation, numerical discretization, and set of
operating conditions. The ideal-gas law is compared with the van der
Waals~\cite{Waals1873}, Soave--Redlich--Kwong~\cite{Soave1972}, and
Peng--Robinson~\cite{Peng1976} cubic equations of state, while
high-accuracy multiparameter Helmholtz equations of state
\cite{Lemmon2000,Lemmon2022,Thol2016} are employed as reference models.
Dry air~\cite{Lemmon2000} is considered as the conventional bearing gas,
whereas R1234yf~\cite{Lemmon2022} and R1234ze(E)~\cite{Thol2016} are selected
as low-global-warming-potential hydrofluoroolefin (HFO) refrigerants used in automotive and
stationary refrigeration applications, respectively. Both fluids exhibit
considerably more pronounced non-ideal thermodynamic behavior than air
within the investigated pressure and temperature range.
This selection enables the transition from nearly ideal
to distinctly non-ideal gas behavior to be examined without changing the
underlying bearing model.

The investigation comprises three complementary components. First, the
numerical and thermodynamic implementations are verified independently.
The influence of the thermodynamic closure on the bearing response is then
quantified under stationary and dynamic conditions in terms of load
capacity, supplied mass flow, and complex dynamic force response. Finally,
the observed bearing-level differences are interpreted in terms of the
underlying thermodynamic properties and evaluated together with the
computational cost to derive practical EOS-selection criteria.
\section{Porous gas bearing model}
The present study employs the spatially resolved Reynolds--Darcy model
introduced and verified in a preceding investigation
\cite{Stellmach2026a}. Since the focus of the present work lies on the
thermodynamic closure, the bearing formulation is summarized here briefly.

\subsection{Geometry and governing equations}\label{sec:geometry_governing_equations}
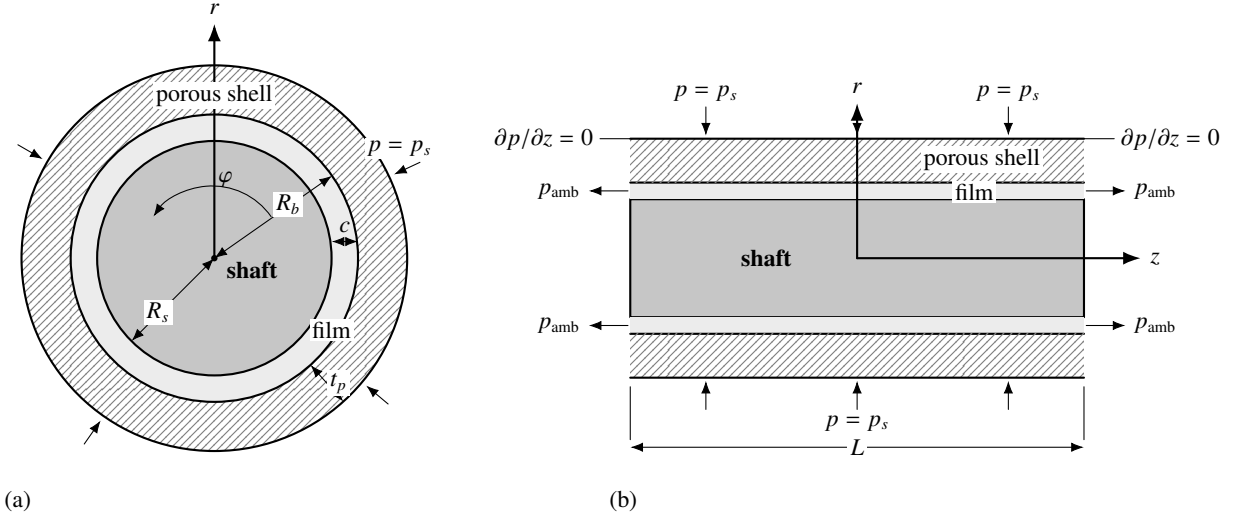
\begin{figure*}[h]
    \centering
    \begin{tikzpicture}[
        >=Latex,
        line cap=round,
        line join=round,
        every node/.style={font=\small},
        dim/.style={<->,thin},
        bc/.style={->,thin},
        labelbox/.style={fill=white,inner sep=1pt}
    ]

    \begin{scope}[shift={(0,0)}]

        \def\Rs{1.55}
        \def\Rb{1.90}
        \def\Ro{2.55}

        \path[pattern=north east lines, pattern color=gray, even odd rule]
            (0,0) circle (\Ro)
            (0,0) circle (\Rb);

        \fill[gray!15] (0,0) circle (\Rb);
        \fill[gray!45] (0,0) circle (\Rs);

        \draw[thick] (0,0) circle (\Rs);
        \draw[thick] (0,0) circle (\Rb);
        \draw[thick] (0,0) circle (\Ro);

        \draw[->,thick] (0,0) -- (0,3.1) node[above] {$r$};

        \draw[->] (0.75,0.55) arc[start angle=35,end angle=145,radius=0.95];
        \node at (0.15,1.05) {$\varphi$};

        \fill (0,0) circle (1.2pt);

        \node[labelbox] at (0,2.15) {porous shell};
        \node[labelbox] at (1.55,-0.95) {film};
        \node at (0.5,-0.15) {\textbf{shaft}};

        \draw[dim] (0,0) -- ({\Rs*cos(225)},{\Rs*sin(225)})
            node[labelbox,midway,below left] {$R_s$};

        \draw[dim] (0,0) -- ({\Rb*cos(35)},{\Rb*sin(35)})
            node[labelbox,midway,above right] {$R_b$};

        \draw[dim] (\Rs,0.22) -- (\Rb,0.22)
            node[midway,above] {$c$};

        \draw[dim] ({\Rb*cos(-48)},{\Rb*sin(-48)})
                   -- ({\Ro*cos(-48)},{\Ro*sin(-48)})
            node[labelbox,midway,right] {$t_p$};

        \draw[bc] ({3.00*cos(45)},{3.00*sin(45)}) -- ({2.62*cos(45)},{2.62*sin(45)});
        \draw[bc] ({3.00*cos(135)},{3.00*sin(135)}) -- ({2.62*cos(135)},{2.62*sin(135)});
        \draw[bc] ({3.00*cos(225)},{3.00*sin(225)}) -- ({2.62*cos(225)},{2.62*sin(225)});
        \draw[bc] ({3.00*cos(315)},{3.00*sin(315)}) -- ({2.62*cos(315)},{2.62*sin(315)});

        \node[labelbox] at (2.6,2.3) {$p=p_s$};

        \node[anchor=north west] at (-2.9,-2.95) {(a)};

    \end{scope}

    \begin{scope}[shift={(5.5,0)}]

        \def\xL{0}
        \def\xR{6.0}
        \def\ys{0.78}
        \def\yf{1.00}
        \def\yo{1.58}

        \fill[gray!45] (\xL,-\ys) rectangle (\xR,\ys);
        \draw[thick] (\xL,-\ys) rectangle (\xR,\ys);

        \fill[gray!15] (\xL,\ys) rectangle (\xR,\yf);
        \fill[gray!15] (\xL,-\yf) rectangle (\xR,-\ys);
        \draw[thick] (\xL,\yf) -- (\xR,\yf);
        \draw[thick] (\xL,-\yf) -- (\xR,-\yf);

        \path[pattern=north east lines, pattern color=gray]
            (\xL,\yf) rectangle (\xR,\yo);
        \path[pattern=north east lines, pattern color=gray]
            (\xL,-\yo) rectangle (\xR,-\yf);

        \draw[thick] (\xL,\yo) -- (\xR,\yo);
        \draw[thick] (\xL,-\yo) -- (\xR,-\yo);

        \node at (1.8,0) {\textbf{shaft}};
        \node[labelbox] at (4.55,0.88) {film};
        \node[labelbox] at (4.65,1.30) {porous shell};

        \draw[->,thick] (3.0,0) -- (6.75,0) node[right] {$z$};
        \draw[->,thick] (3.0,0) -- (3.0,2.05) node[above] {$r$};

        \draw[bc] (1.0,2.00) -- (1.0,1.62);
        \draw[bc] (2.0,2.00) -- (2.0,1.62);
        \draw[bc] (4.0,2.00) -- (4.0,1.62);
        \draw[bc] (5.0,2.00) -- (5.0,1.62);

        \draw[bc] (1.0,-2.00) -- (1.0,-1.62);
        \draw[bc] (2.0,-2.00) -- (2.0,-1.62);
        \draw[bc] (4.0,-2.00) -- (4.0,-1.62);
        \draw[bc] (5.0,-2.00) -- (5.0,-1.62);

        \node[labelbox] at (1.50,2.18) {$p=p_s$};
        \node[labelbox] at (4.50,-2.18) {$p=p_s$};
        \node[labelbox] at (1.50,-2.18) {$p=p_s$};
        \node[labelbox] at (4.50,2.18) {$p=p_s$};

        \draw[bc] (-0.03,0.89) -- (-0.55,0.89);
        \draw[bc] (6.03,0.89) -- (6.55,0.89);
        \draw[bc] (-0.03,-0.89) -- (-0.55,-0.89);
        \draw[bc] (6.03,-0.89) -- (6.55,-0.89);

        \node[labelbox,anchor=east] at (-0.62,0.89) {$p_\mathrm{amb}$};
        \node[labelbox,anchor=west] at (6.62,0.89) {$p_\mathrm{amb}$};

        \node[labelbox,anchor=east] at (-0.62,-0.89) {$p_\mathrm{amb}$};
        \node[labelbox,anchor=west] at (6.62,-0.89) {$p_\mathrm{amb}$};

        \draw[thick] (\xL,\yf) -- (\xL,\yo);
        \draw[thick] (\xR,\yf) -- (\xR,\yo);
        \draw[thick] (\xL,-\yf) -- (\xL,-\yo);
        \draw[thick] (\xR,-\yf) -- (\xR,-\yo);

        \node[labelbox,anchor=east]
            at ({\xL-0.10},{0.5*(\yf+\yo)})
            {$\partial p/\partial z=0$};

        \node[labelbox,anchor=west]
            at ({\xR+0.10},{0.5*(\yf+\yo)})
            {$\partial p/\partial z=0$};

        \draw[dim] (\xL,-2.5) -- (\xR,-2.5)
            node[midway,labelbox] {$L$};
        \draw[thin] (\xL,-1.70) -- (\xL,-2.40);
        \draw[thin] (\xR,-1.70) -- (\xR,-2.40);

        \node[anchor=north west] at (-0.4,-2.95) {(b)};

    \end{scope}

    \end{tikzpicture}

    \caption{Schematic of the porous aerostatic journal bearing and the applied
    boundary conditions: (a) radial cross-section and (b) longitudinal
    cross section. Supply pressure is prescribed at the outer radial surface of the porous shell,
    its axial ends are impermeable, and ambient pressure is imposed at the
    axial film boundaries.}
    \label{fig:bearing_geometry}
\end{figure*}

An aerostatic journal bearing consists of a rotating shaft and
a cylindrical porous bearing shell separated by a thin air film.
The inner radius of the porous shell is denoted by $R_b$, the shaft radius
by $R_s$, the axial bearing length by $L$, the radial bearing clearance by
$c$, and the thickness of the porous layer by $t_p$ (cf. Fig.~\ref{fig:bearing_geometry} and \tab{tab:bearing_data}). The pressure fields in
the lubrication film and the porous shell are described in cylindrical
coordinates $(r,\varphi,z)$, where the radial coordinate is directed
outwards from the lubrication film through the porous material. The
geometry and coordinate system correspond to the formulation presented in
\cite{Stellmach2026a}.

\begin{table}[h]
\caption{Bearing geometry, porous-material parameters, and reference operating
conditions used in the present study, adapted from \cite{SanAndres2021} and \cite{Liu2022}.}
\label{tab:bearing_data}
\centering
\renewcommand*{\arraystretch}{1.2}
\begin{tabular}{L{0.70cm}R{2.0cm}L{0.7cm}L{2.6cm}}
\hline
\textbf{Parameter} & \textbf{Value} & & \textbf{Description}\\
\hline
\hline
\(R_s\) & 38.09 & mm & shaft radius\\
\(c\) & 0.01 & mm & radial bearing clearance\\
\(t_p\) & 2.71 & mm & thickness of the porous layer\\
\(L\) & 88.8 & mm & axial bearing length\\
\(K_{r,\varphi,z}\) & \(8.2 \times 10^{-16} \) & \(\text{m}^2\) & permeabilities of the porous shell\\
\(\varepsilon_p\) & 0.08 & - & porosity of the porous layer according to~\cite{Liu2022}\\
\(n_s\) & 25000 & \(\text{min}^{-1}\) & dynamic reference shaft speed\\
\(p_{\mathrm{amb}}\) & 1 & bar & ambient pressure\\
\(p_{s}\) & 4 & bar & reference supply pressure\\
\(T_\mathrm{amb}\) & 293.15 & K & reference temperature\\
\hline
\cline{1-4}
\end{tabular}
\end{table}

Assuming a thin lubrication film, the compressible Reynolds equation is used to describe the gas flow in the lubrication film. With $\varrho$
denoting the fluid density, $p$ the pressure, $\eta$ the dynamic viscosity,
$h$ the local film thickness, and $u_m=(u_s+u_b)/2$ the mean
circumferential surface velocity, the governing equation reads
\begin{align}\label{eq:reynolds_dim}
    \begin{split}
        \frac{1}{r} \pfrac{}{\varphi} \left( \frac{\varrho h^3}{12 \eta} \frac{1}{r}\pfrac{p}{\varphi} \right)
        &+ \pfrac{}{z} \left( \frac{\varrho h^3}{12 \eta} \pfrac{p}{z} \right)\\
        &= \frac{1}{r} \pfrac{}{\varphi} \left( \varrho h u_m \right) + \varrho w_D + \pfrac{\left(\varrho h \right)}{t} \, .
    \end{split}
\end{align}
The terms on the left-hand side describe the pressure-driven Poiseuille flow
in the circumferential and axial directions. The right-hand side contains
the circumferential Couette contribution, the mass exchange with the porous
shell, and the squeeze flow within the lubrication film.

The gas flow through the porous restrictor is described by Darcy's
law~\cite{Darcy1856,Whitaker1986},
\begin{align}\label{eq:darcy_law}
    \bbm{v}_D = \begin{bmatrix} u_D\\ v_D\\ w_D \end{bmatrix} = -\frac{\bbm{K}}{\eta}\nabla p,
\end{align}
where $\bbm{v}_D$ denotes the superficial Darcy velocity and $\bbm{K}$ the permeability tensor.
The velocities components $u_D$, $v_D$, and $w_D$ denote the circumferential,
axial, and radial superficial Darcy velocities, respectively.
The corresponding mean pore velocity is obtained by dividing the
superficial Darcy velocity by the porosity $\varepsilon_p = V_\mathrm{pore} / V_\mathrm{total}$, whereas the permeability $\bbm{K}$ characterizes the hydraulic resistance of the connected pore structure~\cite{Stellmach2026a}.

For a rigid porous matrix, conservation of gas mass yields
\begin{align}
    \nabla\cdot
    \left(
        \varrho\frac{\bbm{K}}{\eta}\nabla p
    \right)
    =
    \varepsilon_p
    \frac{\partial\varrho}{\partial t}.
    \label{eq:darcy_continuity}
\end{align}
The transient term represents gas storage within the porous medium and is retained in the dynamic calculations because it depends directly on the pressure--density relation~\cite{Stellmach2026a}.
For the assumed homogeneous and isotropic porous shell the permeability tensor reduces to the scalar value $\bbm{K} = K_{r,\varphi,z} \bbm{I}$.

The lubrication film and porous shell are coupled by continuity of pressure and normal mass flux using the Morgan-type interface treatment of \cite{Stellmach2026a}. With the radial coordinate directed outwards, the usual aerostatic pressure gradient yields $w_D<0$, corresponding to flow from the porous shell into the lubrication film. Accordingly, the interface term appears as $+\varrho w_D$ in Eq.~\eqref{eq:reynolds_dim}.

Periodic boundary conditions are imposed in the circumferential direction, while the remaining boundary conditions are summarized in Fig.~\ref{fig:bearing_geometry}.

\subsection{Thermodynamic closure}
Eqs.~\eqref{eq:reynolds_dim} and \eqref{eq:darcy_continuity} contain both
pressure and density and therefore require an additional thermodynamic
relation for closure. In the present formulation, the fluid is described
by a general thermal equation of state, with pressure treated as a function of density and temperature, $p = p(\varrho,T)$.

Density is employed as the primary thermodynamic solution variable in the
finite-volume formulation. Consequently, pressure gradients occurring in
the Reynolds and Darcy equations are evaluated through
\begin{align}
    \nabla p
    =
    \left(
        \frac{\partial p}{\partial\varrho}
    \right)_T
    \nabla\varrho
    +
    \left(
        \frac{\partial p}{\partial T}
    \right)_\varrho
    \nabla T \, .
    \label{eq:pressure_gradient_general}
\end{align}
This formulation separates the discretization of the governing flow
equations from the particular thermodynamic model. Any equation of state
can therefore be introduced by providing the pressure and the required
thermodynamic derivatives at the local state.

The present investigation is restricted to isothermal conditions,
$T=T_0$. Eq.~\eqref{eq:pressure_gradient_general} consequently
reduces to
\begin{align}\label{eq:pressure_gradient_isothermal}
    \nabla p
    =
    p_\varrho\,\nabla\varrho,
    \qquad
    p_\varrho
    =
    \left(
        \frac{\partial p}{\partial\varrho}
    \right)_T \, .
\end{align}
The derivative $p_\varrho$ links density and pressure gradients, while its reciprocal characterizes the local isothermal density response to pressure changes. Thus, the EOS affects both pressure-driven transport and transient gas storage. The Reynolds--Darcy formulation itself remains unchanged; only the relations for $p$ and $p_\varrho$ differ between the thermodynamic models.

\subsection{Equations of state}\label{sec:eos}
%
The five different equations of state considered in this study are described in the following.

The parameters required by the cubic equations of state are summarized in
Tab.~\ref{tab:fluid_parameters}. For dry air, effective pseudo-pure-fluid
properties are employed, whereas R1234yf and R1234ze(E) are treated as pure
fluids. The thermodynamic data are consistent with the reference
formulations of \textsc{Lemmon} et al.~\cite{Lemmon2000},
\textsc{Lemmon} and \textsc{Akasaka}~\cite{Lemmon2022}, and
\textsc{Thol} and \textsc{Lemmon}~\cite{Thol2016}, respectively.

\begin{table}[h]
    \centering
    \caption{Fluid parameters used for the cubic equations of state.}
    \renewcommand*{\arraystretch}{1.2}
    \label{tab:fluid_parameters}
    \begin{tabular}{L{1.5cm}C{1.1cm}C{1cm}C{1cm}C{1cm}}
        \hline
        Fluid
        & $M$
        & $T_c$
        & $p_c$
        & $\omega_\mathrm{ac}$ \\
        & [kg/mol]
        & [K]
        & [MPa]
        & [-] \\
        \hline
        Air
        & 0.028965
        & 132.53
        & 3.786
        & 0.0335 \\
        R1234yf
        & 0.11404
        & 367.85
        & 3.384
        & 0.276 \\
        R1234ze(E)
        & 0.11404
        & 382.52
        & 3.635
        & 0.3131 \\
        \hline
    \end{tabular}
\end{table}

The ideal-gas equation requires only the molar mass $M$, whereas the cubic
equations additionally use the critical parameters and acentric factors
listed in Tab.~\ref{tab:fluid_parameters}, whereas the Helmholtz formulations employ fluid-specific reducing parameters defined below and \ref{app:helmholtz_parameters}.

\subsubsection{Ideal-gas equation of state}

The ideal-gas law serves as the reference for the conventional treatment of
gas-lubricated bearings. In terms of the mass density, it reads
\begin{align}
    p
    &=
    \frac{\varrho R T}{M},
    \label{eq:eos_ideal}
    \\
    p_\varrho
    &=
    \left(
        \frac{\partial p}{\partial\varrho}
    \right)_T
    =
    \frac{R T}{M},
    \label{eq:eos_ideal_dpdrho}
\end{align}
where $R$ denotes the universal gas constant and $M$ the molar mass of the
fluid. The pressure derivative is therefore independent of density.

\subsubsection{Van der Waals equation of state}

The van der Waals (VDW) equation represents the simplest real-gas model considered
in the present study and introduces corrections for intermolecular
attraction and finite molecular volume~\cite{Waals1873}. Expressed in terms
of mass density, it is written as
\begin{align}
    p
    &=
    \frac{\varrho R T}{M-b_{\mathrm{VDW}}\varrho}
    -
    a_{\mathrm{VDW}}
    \frac{\varrho^2}{M^2},
    \label{eq:eos_vdw}
\end{align}
with
\begin{align}
    a_{\mathrm{VDW}}
    &=
    \frac{27 R^2 T_c^2}{64 p_c},
    &
    b_{\mathrm{VDW}}
    &=
    \frac{R T_c}{8p_c},
    \label{eq:eos_vdw_parameters}
\end{align}
where $T_c$ and $p_c$ denote the critical temperature and pressure of the fluid,
respectively. The required derivative reads
\begin{align}
    p_\varrho
    &=
    \frac{MRT}
         {\left(M-b_{\mathrm{VDW}}\varrho\right)^2}
    -
    2a_{\mathrm{VDW}}\frac{\varrho}{M^2}.
    \label{eq:eos_vdw_dpdrho}
\end{align}
The van der Waals equation is included primarily as a low-order real-gas
model and provides a useful intermediate level between the ideal-gas
approximation and more accurate cubic equations of state.

\subsubsection{Soave--Redlich--Kwong equation of state}

The Soave--Redlich--Kwong (SRK) equation of state consists of a positive repulsive contribution, accounting for the finite molecular volume, and a negative attractive contribution representing intermolecular forces. Compared with the original Redlich--Kwong formulation~\cite{Redlich1949}, Soave introduced an acentric-factor-dependent temperature correction of the attractive contribution~\cite{Soave1972}.
In density form, the equation reads
\begin{align}
    p
    &=
    \frac{\varrho R T}{M-b_{\mathrm{SRK}}\varrho}
    -
    \frac{
        a_{\mathrm{SRK}}\alpha_{\mathrm{SRK}}(T)\varrho^2
    }{
        M\left(M+b_{\mathrm{SRK}}\varrho\right)
    },
    \label{eq:eos_srk}
\end{align}
where
%
\begin{align}\label{eq:eos_srk_ab_omega}
\begin{split}
    a_{\mathrm{SRK}} &= \frac{1}{9\left(2^{1/3}-1\right)} \frac{R^2T_c^2}{p_c} \, , \, \, \quad
     b_{\mathrm{SRK}} = \frac{2^{1/3}-1}{3} \frac{RT_c}{p_c} \, .
\end{split}
\end{align}
The temperature-dependent attraction correction is
\begin{align}\label{eq:eos_srk_m_alpha}
\begin{split}
    \alpha_{\mathrm{SRK}}(T)
    &=
    \left[
        1+
        m_{\mathrm{SRK}}
        \left(
            1-\sqrt{\frac{T}{T_c}}
        \right)
    \right]^2,
    \nonumber\\
    m_{\mathrm{SRK}}
    &=
    0.480
    +
    1.574\omega_\mathrm{ac}
    -
    0.176\omega_\mathrm{ac}^2,
\end{split}
\end{align}
with $\omega_\mathrm{ac}$ denoting the dimensionless fluid-specific acentric factor.
The corresponding isothermal density derivative reads
\begin{align} 
    p_\varrho
    &=
    \frac{MRT}
         {\left(M-b_{\mathrm{SRK}}\varrho\right)^2}
    -
    a_{\mathrm{SRK}}\alpha_{\mathrm{SRK}}
    \frac{
        \varrho\left(2M+b_{\mathrm{SRK}}\varrho\right)
    }{
        M\left(M+b_{\mathrm{SRK}}\varrho\right)^2
    }.
\end{align}

\subsubsection{Peng--Robinson equation of state}

The Peng--Robinson (PR) equation of state provides a widely used cubic
description of real-fluid behavior~\cite{Peng1976}. In the present study,
the standard Peng--Robinson coefficients from the critical-point
conditions are employed, while the
temperature-dependent attraction correction uses the Stryjek--Vera base
acentric-factor correlation as described below.

In terms of mass density, the Peng--Robinson equation can be written as
\begin{align}
    p
    &=
    \frac{\varrho RT}
         {M-b_{\mathrm{PR}}\varrho}
    -
    \frac{
        a_{\mathrm{PR}}\alpha_{\mathrm{PR}}(T)\varrho^2
    }{
        M^2
        +2b_{\mathrm{PR}}M\varrho
        -b_{\mathrm{PR}}^2\varrho^2
    },
    \label{eq:eos_pr}
\end{align}
with
\begin{align}\label{eq:eos_pr_ab}
    \begin{split}
        a_{\mathrm{PR}} &= 0.457235528921 \frac{R^2T_c^2}{p_c} \, , \\
        b_{\mathrm{PR}} &= 0.0777960739039 \frac{RT_c}{p_c} \, .
    \end{split}
\end{align}
%

The temperature dependence of the attractive contribution is described by
\begin{align}
    \alpha_{\mathrm{PR}}(T)
    &=
    \left[
        1+
        m_{\mathrm{PR}}
        \left(
            1-\sqrt{\frac{T}{T_c}}
        \right)
    \right]^2.
    \label{eq:eos_pr_alpha}
\end{align}
In the original Peng--Robinson formulation~\cite{Peng1976}, the
acentric-factor correlation is
\begin{align}
    m_{\mathrm{PR,orig}}
    &=
    0.37464
    +
    1.54226\omega_\mathrm{ac}
    -
    0.26992\omega_\mathrm{ac}^2,
    \label{eq:eos_pr_m_original}
\end{align}
where $\omega_\mathrm{ac}$ denotes the dimensionless fluid-specific acentric factor. \textsc{Stryjek} and
\textsc{Vera}~\cite{Stryjek1986} proposed the modified base correlation
\begin{align}\label{eq:eos_pr_m_sv}
    \begin{split}
    m_{\mathrm{SV}} = &0.378893
    + 1.4897153 \omega_\mathrm{ac} \\
    &- 0.17131848 \omega_\mathrm{ac}^2
    + 0.0196554 \omega_\mathrm{ac}^3 \, .
    \end{split}
\end{align}
Following Bechiri et al.~\cite{Bechiri2026}, the present model uses
$m_\mathrm{PR}=m_\mathrm{SV}$. Since the additional fluid-specific Peng--Robinson--Stryjek--Vera (PRSV)
correction parameters are omitted, the formulation corresponds to a
Peng--Robinson EOS using the Stryjek--Vera base acentric-factor correlation
rather than the full PRSV formulation. Throughout the remainder of this
work, the abbreviation PR refers to this specific parameterization.

The corresponding isothermal density derivative required by the present
Reynolds--Darcy formulation follows as
\begin{align} \label{eq:eos_pr_dpdrho}
    p_\varrho
    &=
    \frac{MRT}{\left(M-b_{\mathrm{PR}}\varrho\right)^2}
    -
    \frac{
        2a_{\mathrm{PR}}\alpha_{\mathrm{PR}}(T)
        M\varrho\left(M+b_{\mathrm{PR}}\varrho\right)
    }{
        \left(
            M^2
            +2b_{\mathrm{PR}}M\varrho
            -b_{\mathrm{PR}}^2\varrho^2
        \right)^2
    } \, .
\end{align}

\subsubsection{Multiparameter Helmholtz-energy equations of state}\label{sec:eos_helmholtz}

High-accuracy multiparameter equations of state formulated in terms of the
Helmholtz energy are employed as the thermodynamic reference. The Helmholtz
energy is a thermodynamic potential expressed as a function of density and
temperature, from which pressure and other thermodynamic properties are
obtained through analytical derivatives.
Such formulations describe the dimensionless molar Helmholtz energy as the sum
of an ideal-gas contribution $\alpha^0$ and a residual contribution $\alpha^r$,
\begin{align}
    \alpha(\delta,\tau_H)
    =
    \alpha^0(\delta,\tau_H)
    +
    \alpha^r(\delta,\tau_H),
    \label{eq:eos_helmholtz_alpha}
\end{align}
with the reduced density and inverse reduced temperature
\begin{align}
    \delta
    &=
    \frac{\varrho/M}{\varrho_{\mathrm{r,mol}}},
    &
    \tau_H
    &=
    \frac{T_r}{T}.
    \label{eq:eos_helmholtz_reduced}
\end{align}
Here, $\varrho_{\mathrm{r,mol}}$ and $T_r$ denote the fluid-specific reducing parameters
for molar density and temperature, respectively. The residual Helmholtz energy is composed of polynomial terms $\mathcal{P}$, exponentially damped terms $\mathcal{E}$, and, where required, Gaussian terms $\mathcal{G}$ (cf. \ref{app:helmholtz_parameters}).
It is given by
\begin{align}
    \alpha^r
    ={}&
    \sum_{i\in\mathcal{P}}
        n_i\delta^{d_i}\tau_H^{t_i}
    +
    \sum_{i\in\mathcal{E}}
        n_i\delta^{d_i}\tau_H^{t_i}
        \exp\left(-\delta^{l_i}\right)
    \nonumber\\
    &+
    \sum_{i\in\mathcal{G}}
        n_i\delta^{d_i}\tau_H^{t_i}
        \exp
        \left[
            -\eta_i(\delta-\epsilon_i)^2
            -\beta_i(\tau_H-\gamma_i)^2
        \right].
    \label{eq:eos_helmholtz_residual}
\end{align}
Using the decomposition of the Helmholtz energy into ideal and residual
contributions, the pressure is obtained as
\begin{align}
    p
    &=
    \varrho R_\mathrm{spc} T
    \left(
        1+
        \delta\alpha^r_\delta
    \right),
    \label{eq:eos_helmholtz_pressure}
\end{align}
where $R_\mathrm{spc}=R/M$ and
$\alpha^r_\delta=\partial\alpha^r/\partial\delta$.
Here, the unity term represents the ideal-gas contribution, whereas
$\delta\alpha^r_\delta$ accounts for the additional contribution arising
from non-ideal fluid behavior.

Accordingly, the isothermal density derivative required by the present
bearing formulation is
\begin{align}
    p_\varrho
    &=
    R_\mathrm{spc}T
    \left(
        1
        +
        2\delta\alpha^r_\delta
        +
        \delta^2\alpha^r_{\delta\delta}
    \right),
    \label{eq:eos_helmholtz_dpdrho}
\end{align}
with
$\alpha^r_{\delta\delta}
=\partial^2\alpha^r/\partial\delta^2$.
Only derivatives of the residual Helmholtz contribution are required for
the pressure and its isothermal density derivative, since the ideal-gas
contribution is represented explicitly by the unity terms. All derivatives
are evaluated analytically in the present implementation.

Fluid-specific Helmholtz formulations are employed for each investigated
working fluid. Dry air is described according to \textsc{Lemmon} et
al.~\cite{Lemmon2000}, R1234yf according to \textsc{Lemmon} and
\textsc{Akasaka}~\cite{Lemmon2022}, and R1234ze(E) according to
\textsc{Thol} and \textsc{Lemmon}~\cite{Thol2016}. The corresponding
reducing parameters and coefficient sets are taken directly from these
reference formulations. Details are given in the \ref{app:helmholtz_parameters}.

\subsubsection{Thermodynamic comparison quantities}

To characterize the degree of non-ideal behavior independently of the
bearing solution, two additional thermodynamic quantities are considered.
The compressibility factor is defined as
\begin{align}
    Z
    =
    \frac{p}{\varrho R_\mathrm{spc} T}.
    \label{eq:compressibility_factor}
\end{align}
In addition, the dimensionless isothermal compressibility measure
\begin{align}
    \chi_T
    =
    p\kappa_T
    =
    \frac{p}{\varrho p_\varrho},
    \label{eq:chi_T}
\end{align}
is evaluated, where
$\kappa_T=(1/\varrho)(\partial\varrho/\partial p)_T$ denotes the isothermal
compressibility. For an ideal gas, both $Z$ and $\chi_T$ are equal to unity. While $Z$
characterizes deviations of the pressure--density relation from ideal-gas
behavior, $\chi_T$ additionally reflects deviations in its local slope and is
therefore particularly relevant for the transient storage response of the
compressible gas. In the bearing formulation, $Z$ is used as a diagnostic
measure of thermodynamic non-ideality, whereas the influence of $\chi_T$ on
transient gas storage is derived in Eqs.~\eqref{eq:storage_compliance_chi}~and~\eqref{eq:film_shell_storage_linearized}.

\subsubsection{Thermodynamic validity and phase selection}

The present Reynolds--Darcy formulation is restricted to single-phase gas
flow and does not account for condensation or two-phase effects
\cite{Guenat2019}. Hence, refrigerant states must satisfy
$p<p_{\mathrm{sat}}(T)$. For the present parameter study, the stricter
criterion
\begin{align}
    p \leq 0.95\,p_{\mathrm{sat}}(T)
\end{align}
is applied, with $p_{\mathrm{sat}}$ obtained from fluid-specific
vapor-pressure correlations~\cite{Lemmon2022,Thol2016}. For the cubic equations of state, the low-density vapor root is selected whenever the pressure--density relation admits multiple mathematical roots. The criterion is evaluated over the entire pressure field for stationary calculations and at the maximum instantaneous film pressure for dynamic simulations.

\subsection{Transport properties}
\label{sec:transport_properties}
For each fluid, the same viscosity correlation is used for all equations of state, isolating the influence of the thermodynamic closure.

\subsubsection{Dynamic viscosity}

For dry air, the dynamic viscosity is described by Sutherland's
law~\cite{Sutherland1893},
\begin{align}
    \eta(T)
    &=
    \eta_{\mathrm{ref}}
    \left(
        \frac{T}{T_{\mathrm{ref}}}
    \right)^{3/2}
    \frac{T_{\mathrm{ref}}+S}
         {T+S},
    \label{eq:sutherland}
\end{align}
where $\eta_{\mathrm{ref}}$ denotes the viscosity at the reference
temperature $T_{\mathrm{ref}}$. The Sutherland constant is set to
$S=110.4~\mathrm{K}$~\cite{White1991}. Since the present calculations are isothermal,
the viscosity of air is spatially constant for each investigated
temperature.

For R1234yf and R1234ze(E), the wide-ranging correlations of
\textsc{Huber} and \textsc{Assael}~\cite{Huber2016} are employed.
In contrast to the Sutherland relation, these correlations account for
both temperature and density dependence. Their general structure can be
written as
\begin{align}
    \eta(\varrho_{\mathrm{mol}},T)
    =
    \eta_0(T)
    +
    \eta_1(T)\varrho_{\mathrm{mol}}
    +
    \Delta\eta(\varrho_{\mathrm{mol}},T),
    \label{eq:hfo_viscosity}
\end{align}
where $\varrho_{\mathrm{mol}}=\varrho/M$ is the molar density,
$\eta_0$ represents the dilute-gas contribution corresponding to the
zero-density limit,
$\eta_1\varrho_{\mathrm{mol}}$ the initial-density contribution, and
$\Delta\eta$ the residual dense-fluid contribution. The critical
enhancement is neglected consistently with the correlation proposed in
\cite{Huber2016}. The fluid-specific coefficients are taken directly
from the respective correlations and are not refitted in the present work.

Although the model is isothermal, the density dependence of the refrigerant viscosity is retained. Thus, EOS-dependent density differences may indirectly affect the local viscosity, while the underlying viscosity correlation remains unchanged.

All investigated operating states remain within the applicable ranges of the employed viscosity correlations~\cite{Huber2016,Sutherland1893}.

\subsection{Nondimensional formulation}
\label{sec:nondimensionalization}
The governing Reynolds and Darcy equations are nondimensionalized to
separate the relevant geometric and transport parameter ratios from the
absolute scales of the bearing and to improve the numerical conditioning
of the resulting system. The scaling follows the formulation introduced
in \cite{Stellmach2026a}, but is extended here to retain the general
thermodynamic relation introduced in Sec.~\ref{sec:eos}.

The density, temperature, viscosity, film thickness, spatial coordinates,
permeability, and time are scaled according to
\begin{align}\label{eq:nondim_scaling}
    \begin{split}
    \varrho &= \varrho_{\mathrm{amb}} P, \qquad T = T_{\mathrm{amb}} \Theta, \qquad \eta = \eta_{\mathrm{amb}} E,\\
    h &= cH, \qquad \quad \ \ r = R_b \bar r, \qquad \quad \, z = L \ov{z},\\
    K_{r,\varphi,z} &= K_0 \kappa_{r,\varphi,z}, \quad \ \, p = p_0 \Pi, \qquad \quad \, t = t_0 \tau ,\\
    K_0 &= \left(K_r+K_\varphi+K_z\right)/3,
    \end{split}
\end{align}
where
\begin{align}
    p_0
    &=
    \varrho_{\mathrm{amb}}\frac{RT_{\mathrm{amb}}}{M},
    \label{eq:p0}
    \\
    t_0
    &=
    \frac{12\eta_{\mathrm{amb}}R_b^2}
         {p_0c^2}.
    \label{eq:t0}
\end{align}
Here, $\varrho_{\mathrm{amb}}$ and $\eta_{\mathrm{amb}}$ denote the density
and dynamic viscosity at the ambient thermodynamic state. For each
equation of state, $\varrho_{\mathrm{amb}}$ is determined consistently from
the prescribed ambient pressure $p_\mathrm{amb}$ and temperature $T_\mathrm{amb}$ using the respective
thermodynamic closure. The physical boundary pressures are therefore
identical for all calculations, whereas the corresponding boundary
densities are EOS dependent.

It should be noted that the pressure scale $p_0$ of
Eq.~\eqref{eq:p0} coincides with the ambient pressure for an ideal gas.
For a real gas, however, $p_0$ is a reference scale derived from the
ambient density and does not generally equal $p_{\mathrm{amb}}$.
Accordingly, the dimensionless ambient pressure $\Pi_{\mathrm{amb}}$
is not necessarily unity.

\subsubsection{Dimensionless equation of state interface}

The nondimensionalization provides a convenient interface between the
flow equations and the thermodynamic closure.
Using Eqs.~\eqref{eq:nondim_scaling} and \eqref{eq:p0}, the relation in Eq.~\eqref{eq:pressure_gradient_isothermal} can be expressed as
\begin{align}
    \frac{\partial\Pi}{\partial P}
    &=
    G(P)
    =
    \frac{M}{RT_{\mathrm{amb}}}
    \left(
        \frac{\partial p}{\partial\varrho}
    \right)_T ,
    \label{eq:G_definition}
\end{align}
so that,
\begin{align}
    \frac{\partial\Pi}{\partial X}
    =
    G(P)
    \frac{\partial P}{\partial X},
    \label{eq:dimless_pressure_gradient}
\end{align}
holds for any dimensionless spatial coordinate $X$.
Hence, the entire direct influence of the equation of state on the
pressure-gradient terms is represented by the dimensionless response
function $G(P)$. For the ideal-gas law,
\begin{align}
    \Pi = P\Theta,
    \qquad
    G(P) = \Theta.
\end{align}
Under the isothermal conditions considered here, $\Theta=1$, and therefore $\Pi=P$, $G=1$.
For the real-gas equations of state considered in
Sec.~\ref{sec:eos}, $G$ generally depends on density. The corresponding
dimensionless pressure itself is obtained from
\begin{align}
    \Pi(P)
    =
    \frac{
        p\!\left(\varrho_{\mathrm{amb}}P,T_{\mathrm{amb}}\right)
    }{p_0}.
    \label{eq:Pi_EOS}
\end{align}
This representation enables the Reynolds and Darcy operators to retain
the same algebraic structure for all thermodynamic closures.

\subsubsection{Dimensionless Reynolds equation}

Introducing the dimensionless quantities into
Eq.~\eqref{eq:reynolds_dim} yields
\begin{align}
    &
    \frac{1}{\bar r}
    \frac{\partial}{\partial\varphi}
    \left[
        \frac{PH^3}{E\bar r}
        G(P)
        \frac{\partial P}{\partial\varphi}
    \right]
    +
    \Lambda_z
    \frac{\partial}{\partial \ov{z}}
    \left[
        \frac{PH^3}{E}
        G(P)
        \frac{\partial P}{\partial \ov{z}}
    \right]
    \nonumber\\
    &\qquad =
    \Lambda_\Omega
    \frac{1}{\bar r}
    \frac{\partial(PHU)}{\partial\varphi}
    -
    \Lambda_\kappa
    \frac{P\kappa_r}{E}
    G(P)
    \frac{\partial P}{\partial\bar r}
    +
    \frac{\partial(PH)}{\partial\tau},
    \label{eq:reynolds_dimless}
\end{align}
where $U$ denotes the dimensionless mean circumferential velocity. The
constant coefficients are
\begin{align}
    \Lambda_z
    &=
    \left(
        \frac{R_b}{L}
    \right)^2,
    &
    \Lambda_\Omega
    &=
    \frac{12u_{m,0}\eta_{\mathrm{amb}}R_b}
         {p_0c^2},
    &
    \Lambda_\kappa
    &=
    \frac{12K_0R_b}{c^3},
    \label{eq:lambda_parameters}
\end{align}
with $u_{m,0}$ denoting the characteristic circumferential velocity used
for scaling. The three parameters quantify the relative contributions of
axial pressure-driven transport, journal rotation, and gas supply through
the porous restrictor, respectively. The radial derivative in the porous
source term is evaluated at the film--shell interface.

\subsubsection{Dimensionless porous-shell equation}

Applying the same scaling to the porous-domain mass balance
Eq.~\eqref{eq:darcy_continuity} gives
\begin{align}
    &
    \mathcal{P}_{\varphi}
    \frac{1}{\bar r}
    \frac{\partial}{\partial\varphi}
    \left[
        \frac{P\kappa_\varphi}{E\bar r}
        G(P)
        \frac{\partial P}{\partial\varphi}
    \right]
    +
    \mathcal{P}_{z}
    \frac{\partial}{\partial \ov{z}}
    \left[
        \frac{P\kappa_z}{E}
        G(P)
        \frac{\partial P}{\partial \ov{z}}
    \right]
    \nonumber\\
    &\qquad +
    \mathcal{P}_{r}
    \frac{1}{\bar r}
    \frac{\partial}{\partial\bar r}
    \left[
        \bar r
        \frac{P\kappa_r}{E}
        G(P)
        \frac{\partial P}{\partial\bar r}
    \right]
    =
    \frac{\partial P}{\partial\tau},
    \label{eq:darcy_dimless}
\end{align}
with
\begin{align}
    \mathcal{P}_{\varphi}
    &=
    \mathcal{P}_{r}
    =
    \frac{12K_0}{\varepsilon_pc^2},
    &
    \mathcal{P}_{z}
    &=
    \frac{12K_0R_b^2}
         {\varepsilon_pc^2L^2}.
    \label{eq:darcy_prefactors}
\end{align}
These parameters characterize pressure-driven transport within the porous
medium relative to the characteristic response time $t_0$ of the
lubrication film.

Eqs.~\eqref{eq:reynolds_dimless} and
\eqref{eq:darcy_dimless} show that the thermodynamic model enters the
pressure-driven transport through $G(P)$, while density itself appears in
the mass-flux and transient-storage terms. For the refrigerants, the selected equation of state also affects the viscosity indirectly through the density-dependent transport correlation and may therefore contribute to differences in the bearing response.
\subsection{Numerical discretization and solution procedure}
\label{sec:numerical_solution}
The coupled Reynolds--Darcy system is discretized using a cell-centered
finite-volume method on a structured grid in the circumferential, axial,
and radial directions~\cite{Moukalled2016,Ferziger2020b}. Integration over
non-overlapping control volumes converts the divergence terms into fluxes
across the control-volume faces and ensures local conservation of mass
\cite{Moukalled2016}. The numerical formulation follows the method
introduced and verified in \cite{Stellmach2026a}; therefore, only aspects
relevant to the incorporation of different thermodynamic closures are
summarized here.

The lubrication film is discretized by a two-dimensional
$(\varphi,z)$ grid at the inner surface of the porous shell, whereas the
porous domain is represented by three-dimensional $(\varphi,z,r)$ control
volumes. The boundary conditions introduced in
Sec.~\ref{sec:geometry_governing_equations} are imposed at the respective
control-volume faces. Both domains are coupled through their common radial
interface and solved simultaneously within one global system.

Integration of Eqs.~\eqref{eq:reynolds_dimless} and
\eqref{eq:darcy_dimless} yields pressure-driven mass fluxes across the
control-volume faces. A generic flux across face $f$ connecting adjacent
control volumes $i$ and $j$ can be written as
\begin{align}
    \Phi_f
    =
    \Gamma_f\,\mathcal{M}_f
    \left(
        \Pi_j-\Pi_i
    \right),
    \label{eq:generic_face_flux}
\end{align}
where $\Gamma_f$ contains the geometric and operator-specific coefficients
and $\mathcal{M}_f$ denotes the density--viscosity mobility at the face.
Analogous expressions are obtained for the circumferential, axial, and
radial fluxes.

Since the dimensionless density $P$ is used as the primary thermodynamic
variable, the pressures at the adjacent cell centers are obtained from the
selected equation of state. The finite pressure difference entering
Eq.~\eqref{eq:generic_face_flux} is therefore evaluated directly as
\begin{align}
    \Pi_j-\Pi_i = \Pi(P_j)-\Pi(P_i),
    \label{eq:true_pressure_difference}
\end{align}
rather than approximated by
$G(P_i) (P_j-P_i)$. Although both expressions coincide in the differential
limit, the latter represents a local linearization of the
pressure--density relation. The direct evaluation in
Eq.~\eqref{eq:true_pressure_difference} therefore retains the nonlinear
EOS behavior across the finite control-volume spacing.

The resulting nonlinear algebraic equations of the lubrication film and
porous shell are assembled into a common residual system and solved
monolithically within the same nonlinear iteration. The derivative
$p_\varrho$ enters the consistent linearization of the thermodynamic
relations, whereas the residual fluxes themselves are evaluated from the
actual EOS pressures according to Eq.~\eqref{eq:true_pressure_difference}.

\subsubsection{Nonlinear solution}

The dimensionless density is the primary unknown of the coupled system.
The nonlinear finite-volume system is solved using a Newton-type iteration,
\begin{align}
    \bbm{P}^{k+1}
    =
    \bbm{P}^{k}
    +
    \Delta\bbm{P}^{k},
    \label{eq:newton_update}
\end{align}
where $k$ denotes the nonlinear iteration.
The Reynolds and Darcy equations are assembled into one monolithic sparse
system such that the film and porous-domain density fields, and consequently
the corresponding pressures and interface fluxes, are updated consistently
within the same nonlinear iteration.

For the equation of state contribution, the pressure is linearized locally
according to
\begin{align}
    \Pi
    \left(
        P^k+\Delta P
    \right)
    \approx
    \Pi(P^k)
    +
    G(P^k)\Delta P,
    \label{eq:eos_newton}
\end{align}
where
\begin{align}
    G(P)
    =
    \frac{\partial\Pi}{\partial P}
\end{align}
is given by Eq.~\eqref{eq:G_definition}. Hence, only the pressure
$\Pi(P)$ and its first density derivative $G(P)$ are required from the
thermodynamic model for the nonlinear solution. No numerical
differentiation of the equation of state is employed.

Eq.~\eqref{eq:eos_newton} is used only for the Jacobian, while the nonlinear residual is evaluated from the actual EOS pressures according to Eq.~\eqref{eq:true_pressure_difference}. Thus, the converged solution reflects the full nonlinear equation of state.

For the refrigerants, viscosity also depends on density and is updated before each system assembly. While the thermodynamic pressure dependence is included in the Newton linearization, the viscosity derivative is neglected, resulting in a lagged quasi-Newton coupling.

\subsubsection{Transient terms}

For transient calculations, the density-storage terms in both the
lubrication film and the porous shell are discretized implicitly using a
first-order backward-Euler scheme. In the porous domain, this yields
\begin{align}\label{eq:backward_euler_shell}
    \left.
    \frac{\partial P}{\partial\tau}
    \right|^{n+1}
    \approx
    \frac{
        P^{n+1}-P^n
    }{\Delta\tau},
\end{align}
where $n$ and $n+1$ denote consecutive time levels. In the
lubrication film, the transient contribution additionally contains the
prescribed variation of the film thickness. Following
\cite{Stellmach2026a}, the squeeze term is evaluated as
\begin{align}\label{eq:backward_euler_film}
    \left.
    \frac{\partial(PH)}{\partial\tau}
    \right|^{n+1}
    \approx
    H^{n+1}
    \frac{
        P^{n+1}-P^n
    }{\Delta\tau}
    +
    P^{n+1}
    \dot{H}^{n+1}.
\end{align}
The converged density field of the preceding physical time step is used
as the initial state for the nonlinear iteration at the subsequent time
step.

\subsubsection{Equation-of-state evaluation}

The multiparameter Helmholtz equations require substantially more
operations per state evaluation than the ideal-gas and cubic equations.
To avoid repeated thermodynamic evaluations during finite-volume
assembly, the complete thermodynamic state is therefore prepared once
per nonlinear iteration. For every film and porous control volume,
\begin{align}
    \Pi_i
    &=
    \Pi(P_i),
    &
    G_i
    &=
    \left.
    \frac{\partial\Pi}{\partial P}
    \right|_{P_i}
    \label{eq:eos_precompute}
\end{align}
are evaluated and stored. The subsequent
Reynolds and Darcy flux operators use only these precomputed fields, thereby
avoiding repeated EOS evaluations for neighboring face fluxes. This is
particularly beneficial for the computationally more expensive Helmholtz
equations of state.


After assembly, the sparse linear system for $\Delta P$ is solved using
BiCGStab with ILU preconditioning and a relative linear-solver tolerance
of $10^{-10}$. The nonlinear iteration is continued until
$\|\Delta P\|_2<10^{-9}$.
\subsection{Bearing quantities and dynamic force coefficients}
\label{sec:bearing_quantities}
The thermodynamic closure is assessed through both thermodynamic quantities and the resulting bearing response, represented by load capacity, supplied mass flow, and dynamic stiffness and damping coefficients.

\subsubsection{Bearing force and supplied mass flow}

The pressure field obtained from the converged density solution is
integrated over the lubrication-film surface to determine the bearing
reaction force, yielding in Cartesian coordinates for the force components
\begin{align}\label{eq:force_xy}
\begin{split}
    F_x
    &=
    R_b
    \int_0^L
    \int_0^{2\pi}
    \left[p(\varphi,z) - p_\mathrm{amb}\right]
    \cos\varphi
    \,
    \mathrm{d}\varphi
    \mathrm{d}z,
    \\
    F_y
    &=
    R_b
    \int_0^L
    \int_0^{2\pi}
    \left[p(\varphi,z) - p_\mathrm{amb}\right]
    \sin\varphi
    \,
    \mathrm{d}\varphi
    \mathrm{d}z.
\end{split}
\end{align}
The resulting load capacity is
\begin{align}
    F_B
    =
    \sqrt{
        F_x^2+F_y^2
    }.
    \label{eq:load_capacity}
\end{align}
The supplied mass flow is obtained directly from the Darcy mass flux
through the supply boundary $\Gamma_s$,
\begin{align}
    \dot m_{s}
    =
    -
    \int_{\Gamma_s}
    \varrho\,
    \bbm{v}_D\cdot\bbm{n}
    \,
    \mathrm{d}A,
    \label{eq:supply_mass_flow}
\end{align}
where $\bbm{n}$ denotes the outward normal of the porous domain.
Accordingly, $\dot m_{s}>0$ represents gas supplied to the
bearing. In the finite-volume formulation,
Eq.~\eqref{eq:supply_mass_flow} is evaluated from the sum of the
corresponding boundary-face mass fluxes.

\subsubsection{Dynamic force coefficients}

The dynamic bearing behavior is characterized by linearized stiffness and
damping coefficients relating small journal perturbations about an
operating point to the resulting dynamic bearing forces
\cite{Lund1968,Rao1977,Rao1978,Lund1987,Stellmach2026a},
\begin{align}
    \begin{bmatrix}
        F_x'\\
        F_y'
    \end{bmatrix}
    =
    -
    \begin{bmatrix}
        K_{xx} & K_{xy}\\
        K_{yx} & K_{yy}
    \end{bmatrix}
    \begin{bmatrix}
        x'\\
        y'
    \end{bmatrix}
    -
    \begin{bmatrix}
        D_{xx} & D_{xy}\\
        D_{yx} & D_{yy}
    \end{bmatrix}
    \begin{bmatrix}
        \dot{x}'\\
        \dot{y}'
    \end{bmatrix},
    \label{eq:linear_force_coefficients}
\end{align}
where the primes denote perturbations about the stationary operating
point.

In the present study, the dynamic force coefficients are evaluated at the centered bearing position to provide a symmetric reference configuration for isolating the influence of the thermodynamic closure. Owing to the rotational symmetry of the
cylindrical geometry and boundary conditions, the coefficient matrices
satisfy
\begin{align}\label{eq:dynamic_symmetry}
    \begin{split}
        K_{xx} &= K_{yy}, \qquad K_{xy} = -K_{yx},\\
        D_{xx} &= D_{yy}, \qquad D_{xy} = -D_{yx}.
    \end{split}
\end{align}
Frequency-response-based identification provides the stiffness and damping
information simultaneously from the amplitude and phase of the dynamic
force response to a prescribed harmonic perturbation
\cite{Snyder2019,Stellmach2026a}.
Consequently, an independent excitation in both lateral directions is not
required. A harmonic displacement in the $y$-direction,
\begin{align}
    y(t)
    &=
    y_0+A\cos(\omega_\mathrm{ex} t) = A\cos(\omega_\mathrm{ex} t),
    &
    x(t)
    &=
    x_0 = 0,
    \label{eq:harmonic_excitation}
\end{align}
is sufficient to determine both the direct and cross-coupled dynamic
coefficients. The excitation amplitude $A$ is chosen sufficiently small
to remain within the linear perturbation range.

After the initial disturbances have decayed, the mean bearing force
is removed,
\begin{align}
    F_j'(t)
    =
    F_j(t)-\overline{F_j},
    \qquad
    j\in\{x,y\},
    \label{eq:force_fluctuation}
\end{align}
and the complex amplitudes at the excitation frequency are determined over
an evaluation interval containing an integer number of excitation periods,
\begin{align}
    \hat q(\omega_\mathrm{ex})
    =
    \frac{2}{T_{\mathrm{eval}}}
    \int_{t_\mathrm{start}}^{t_\mathrm{start}+T_{\mathrm{eval}}}
    q(t)
    e^{-\mathrm{i}\omega_\mathrm{ex} t}
    \,
    \mathrm{d}t,
    \label{eq:complex_fourier}
\end{align}
where $q$ represents the displacement or one of the dynamic force
components.

For the prescribed $y$-excitation, the corresponding complex dynamic
stiffnesses are
\begin{align}
    H_{yy}(\omega_\mathrm{ex})
    &=
    -
    \frac{\hat F_y(\omega_\mathrm{ex})}
         {\hat y(\omega_\mathrm{ex})},
    &
    H_{xy}(\omega_\mathrm{ex})
    &=
    -
    \frac{\hat F_x(\omega_\mathrm{ex})}
         {\hat y(\omega_\mathrm{ex})}.
    \label{eq:complex_stiffness}
\end{align}
With the Fourier convention of Eq.~\eqref{eq:complex_fourier}, they are
related to the stiffness and damping coefficients by
\begin{align}
    H_{yy}
    &=
    K_{yy}
    +
    \mathrm{i}\omega_\mathrm{ex} D_{yy},
    &
    H_{xy}
    &=
    K_{xy}
    +
    \mathrm{i}\omega_\mathrm{ex} D_{xy}.
    \label{eq:H_KD}
\end{align}
The individual coefficients therefore follow as
\begin{align}\label{eq:KD_from_H}
    \begin{split}
        K_{yy} &= \Re(H_{yy}), \qquad D_{yy} = \frac{\Im(H_{yy})}{\omega_\mathrm{ex}}, \\
        K_{xy} &= \Re(H_{xy}), \qquad D_{xy} = \frac{\Im(H_{xy})}{\omega_\mathrm{ex}}.
    \end{split}
\end{align}

\subsubsection{Comparison of thermodynamic closures}

For a bearing quantity $q$, the deviation of a simplified equation of
state from the corresponding Helmholtz reference solution is expressed as
\begin{align}
    \varepsilon_q
    = 100
    \frac{
        q_{\mathrm{EOS}}-q_{\mathrm{Helm}}
    }{
        q_{\mathrm{Helm}}
    }
    \label{eq:relative_deviation}
\end{align}
in percent.
%
For the dynamic response, relative deviations of individual stiffness or
damping coefficients can become ill-conditioned when their reference values
approach or cross zero. Therefore, the direct and cross-coupled complex
responses to the prescribed $y$-excitation are combined into the relative
response error
\begin{align}
    \varepsilon_{H,y}
    =
    100
    \frac{
        \sqrt{
            \left|H_{yy}-H_{yy,\mathrm{ref}}\right|^2
            +
            \left|H_{xy}-H_{xy,\mathrm{ref}}\right|^2
        }
    }{
        \sqrt{
            \left|H_{yy,\mathrm{ref}}\right|^2
            +
            \left|H_{xy,\mathrm{ref}}\right|^2
        }
    } .
    \label{eq:H_vector_error}
\end{align}
Here, the reference response is chosen according to the respective
comparison. For the EOS assessment, the Helmholtz solution is used as
reference, whereas the finest investigated discretization serves as
reference in the grid-convergence analysis.
This measure evaluates the error in the complete linearized force response
without artificially amplifying deviations of individual stiffness or
damping components that approach or cross zero.
%
\section{Verification and model assessment}\label{kap:verification}
The verification and model assessment comprise four stages. First, the
underlying flow assumptions are assessed for the investigated operating
range. The numerical implementation is then examined with respect to mass
conservation, discretization, and perturbation amplitude. Subsequently, the
Helmholtz-energy implementation and the coupled Reynolds--Darcy formulation
are verified independently. Finally, the bearing-level reference comparison reported in Stellmach et al.~\cite{Stellmach2026a} is briefly summarized.

\subsection{Flow-regime assessment}\label{sec:flow_regime_assessment}

For the porous shell, the maximum area-averaged superficial Darcy velocity
at the outer surface remains below approximately
$1.7\times10^{-3}~\mathrm{m\,s^{-1}}$ over all investigated dynamic cases.
Using this area-averaged value as a characteristic velocity in the
permeability-based Reynolds number
$Re_K=\varrho |\bbm{v}_D|\sqrt{K}/\eta$ gives
$Re_K\lesssim10^{-4}$, indicating that inertia effects are negligible and
supporting the use of Darcy's law without a Forchheimer correction~\cite{DePaoli2023}.

The rotational film-flow regime is assessed using the Taylor number
~\cite{Bouchehit2021}
\begin{align}
    T\!a = \frac{\varrho \Omega R_s h}{\eta} \sqrt{\frac{h}{R_s}},
\end{align}
where $\Omega$ denotes the shaft angular velocity.
For the centered configuration, the critical value is approximately
$T\!a_c=41.2$~\cite{Bouchehit2021}. The maximum values obtained for air,
R1234yf, and R1234ze(E) are $4.3$, $30.5$, and $28.9$, respectively.
Thus, all investigated cases remain below the onset of Taylor-vortex
instability, so that the thin-film Reynolds formulation does not require an
additional correction for Taylor-vortex-induced transport.

\subsection{Numerical verification}\label{kap:numerical_verification}

The numerical verification is performed for R1234yf using the
multiparameter Helmholtz-energy equation of state at
$T=293.15~\mathrm{K}$ and the bearing data from
Tab.~\ref{tab:bearing_data}.
Both squeeze flow in the lubrication film and gas storage within the porous
shell are retained.
Unless varied explicitly, the journal is excited harmonically about the
centered position with $A/c=0.01$.
Three excitation angular frequencies,
$\omega_\mathrm{ex}=\{10,300,3000\}~\mathrm{rad\,s^{-1}}$, are considered
to cover low-, intermediate-, and high-frequency operating conditions.

The complex response functions $H_{yy}$ and $H_{xy}$ are used as the primary
dynamic quantities because they combine stiffness and damping while avoiding
ill-conditioning of individual coefficients near zero crossings.

\subsubsection{Mass conservation}
\label{kap:mass_conservation}

The cell-centered finite-volume discretization is conservative by
construction. For the transient porous domain, this property is checked
directly from the discrete mass balance. The gas mass stored within the porous
shell at time step $n$ is~\cite{Stellmach2026a}
\begin{align}
    M_\mathrm{p}^{n}
    =
    \sum_{P\in\Omega_\mathrm{p}}
    \varepsilon_\mathrm{p}\,
    \varrho_P^{n}\,
    V_P
    \, ,
    \label{gl:porous_mass}
\end{align}
where $\varepsilon_\mathrm{p}$ denotes the porosity and $V_P$ the control
volume~\cite{Stellmach2026a}. With $\dot m_\mathrm{s}$ denoting the mass flow entering the porous
shell through the supply boundary and $\dot m_{\mathrm{p\rightarrow f}}$ the
mass flow leaving the shell toward the lubrication film, the backward-Euler
balance reads~\cite{Stellmach2026a}
\begin{align}
    R_\mathrm{m}^{n}
    =
    \frac{M_\mathrm{p}^{n}-M_\mathrm{p}^{n-1}}{\Delta t}
    -
    \dot m_\mathrm{s}^{n}
    +
    \dot m_{\mathrm{p\rightarrow f}}^{n}
    =
    0
    \, .
    \label{gl:porous_mass_balance}
\end{align}
To obtain a dimensionless measure that is comparable between operating
points, the mass-balance residual is normalized by a representative scale
of the underlying balance, following common practice for finite-volume
residuals~\cite{Moukalled2016}. Here, the largest instantaneous contribution
from storage, supply, or interface mass flow is used as normalization scale,
\begin{align}
    \varepsilon_\mathrm{m}^{n}
    = 100
    \frac{
        \left|R_\mathrm{m}^{n}\right|
    }{
        \max\left(
            \left|
                \dfrac{M_\mathrm{p}^{n}-M_\mathrm{p}^{n-1}}{\Delta t}
            \right|,
            \left|\dot m_\mathrm{s}^{n}\right|,
            \left|\dot m_{\mathrm{p\rightarrow f}}^{n}\right|
        \right)
    }
    \, .
    \label{gl:porous_mass_balance_rel}
\end{align}
This choice avoids an artificially small normalization scale caused by
cancellation between individual terms of the mass balance.
Across all calculations of the numerical verification campaign, the maximum relative shell-mass residual during the final excitation period is $\varepsilon_{\mathrm{m,max}} = 2.96\times10^{-7}~\%$. Hence, no systematic conservation defect is observed in the transient Darcy discretization.

\subsubsection{Spatial discretization}
\label{kap:spatial_discretization}

The spatial discretization is assessed by refining each coordinate direction
separately, allowing the dominant discretization error to be identified
without the computational cost of simultaneous three-dimensional refinement.
The employed meshes are summarized in Tab.~\ref{tab:grid_refinement}. For
the radial refinement, $N_r=\{13,25,49\}$ corresponds to 12, 24, and 48
control-volume layers within the porous shell. All cases use 300 time steps
per excitation period.
\begin{table}[h]
    \centering
    \caption{Meshes used for the directional spatial refinement study.}
    \label{tab:grid_refinement}
    \setlength{\tabcolsep}{4pt}
    \begin{tabular}{lccc}
    \hline
    Direction & $N_\varphi$ & $N_z$ & $N_r$ \\
    \hline
    $\varphi$ & $75,150,300$ & $24$ & $25$ \\
    $z$       & $75$         & $24,48,96$ & $25$ \\
    $r$       & $75$         & $24$ & $13,25,49$ \\
    \hline
    \end{tabular}
\end{table}
For three successively refined solutions \(q_\mathrm{c}\), \(q_\mathrm{m}\), and \(q_\mathrm{f}\), denoting the coarse, medium, and fine grids, respectively, with refinement ratio \(r=2\), the observed convergence order is calculated from
\begin{align}
    p_q
    =
    \frac{
        \ln\left(
            \dfrac{|q_\mathrm{c}-q_\mathrm{m}|}
                  {|q_\mathrm{m}-q_\mathrm{f}|}
        \right)
    }{
        \ln(r)
    }
    \, .
    \label{gl:observed_order}
\end{align}
The fine-grid discretization uncertainty is estimated using the grid-convergence index (GCI), with the recommended safety factor \(F_\mathrm{s}=1.25\) for three systematically refined grids~\cite{Roache1998,Celik2008}
\begin{align}\label{gl:gci}
    \mathrm{GCI}_\mathrm{f}
    =
    F_\mathrm{s}
    \frac{
        |q_\mathrm{f}-q_\mathrm{m}|
    }{
        |q_\mathrm{f}|
        \left(r^{p_q}-1\right)
    }
    \, .
\end{align}
For the complex bearing response, the discretization error is evaluated
using Eq.~\eqref{eq:H_vector_error}, with the finest investigated
discretization taken as reference.
%

\begin{figure}[h]
    \centering
    \includegraphics[width=\linewidth]
    {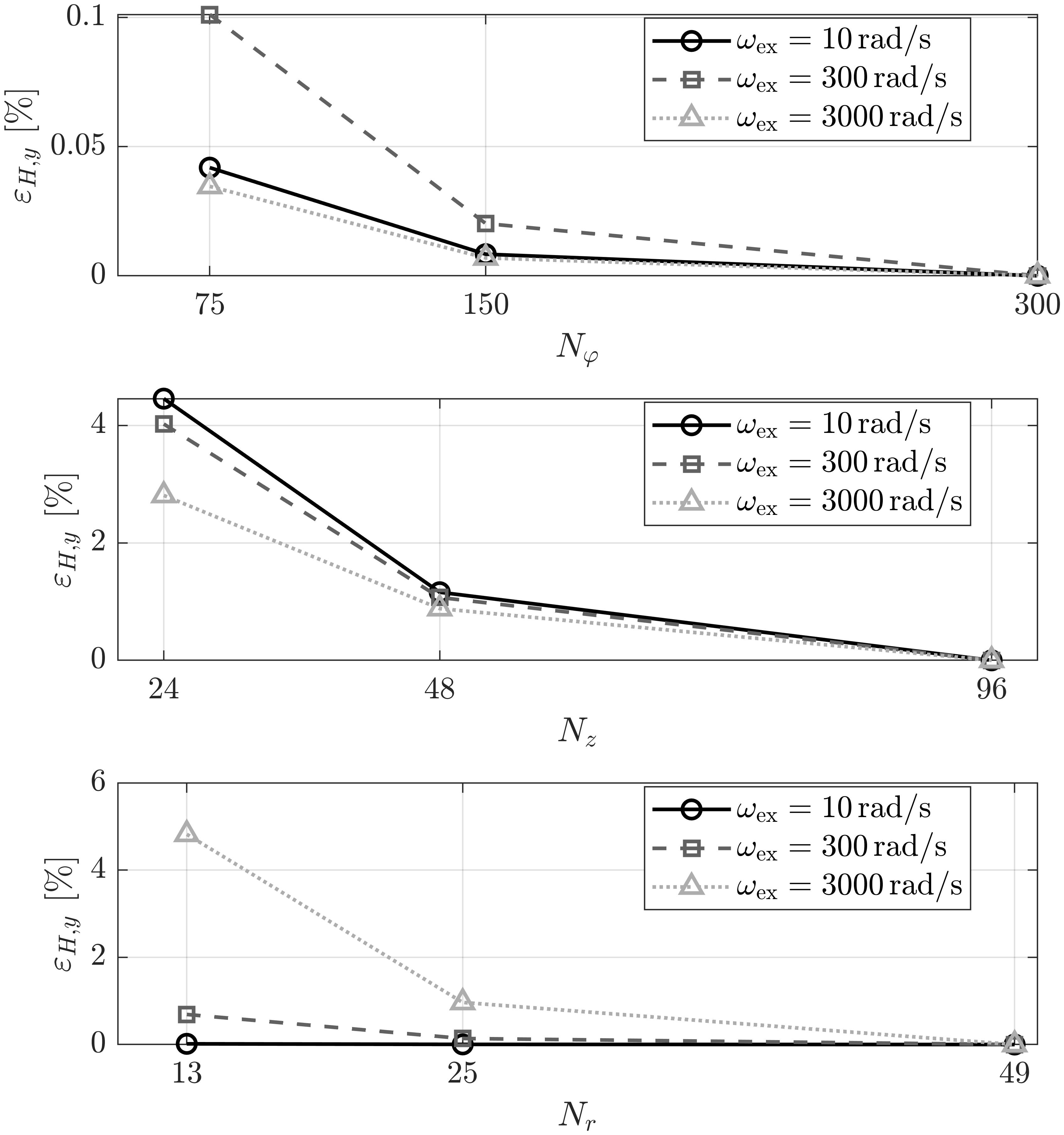}
    \caption{Relative change of the complex dynamic response vector for the
    selected spatial resolution with respect to the finest investigated
    resolution in each coordinate direction.}
    \label{fig:spatial_hvec_error}
\end{figure}

The spatial discretization errors are summarized in
Tab.~\ref{tab:spatial_convergence}, while
Fig.~\ref{fig:spatial_hvec_error} illustrates their dependence on excitation
frequency.

The circumferential refinement exhibits the expected second-order behavior
over the complete frequency range, with observed convergence orders close to
two and a maximum fine-grid GCI below $0.01\,\%$. Accordingly, the selected
resolution $N_\varphi=75$ is already essentially grid independent.

The radial refinement also shows approximately second-order convergence,
with observed orders between $1.99$ and $2.03$. However, the radial
sensitivity increases considerably with excitation frequency
(Fig.~\ref{fig:spatial_hvec_error}), consistent with the increasing relevance
of transient gas storage within the porous shell at high frequencies.

The axial direction exhibits the slowest asymptotic convergence, with
observed orders between approximately $1.10$ and $1.71$. This reduced order
is attributed to the steep axial pressure gradients near the bearing edges.
Nevertheless, the remaining deviations are limited to a few percent over
the investigated frequency range.

\begin{table}[h]
    \centering
    \caption{Relative response change of the coarse resolution in each directional refinement sequence with respect to the corresponding finest resolution.}
    \renewcommand*{\arraystretch}{1.2}
    \setlength{\tabcolsep}{4pt}
    \label{tab:spatial_convergence}
    \begin{tabular}{cccc}
        \hline
        & $\varphi$ & $z$ & $r$ \\
        Refinement
        & $75/150/300$
        & $24/48/96$
        & $13/25/49$ \\
        \hline
        $\omega_\mathrm{ex}$ [rad/s]
        & \multicolumn{3}{c}{$\varepsilon_{H,y}$ [\%]} \\
        \hline
        10   & 0.042 & 4.455 & 0.016 \\
        300  & 0.101 & 4.023 & 0.690 \\
        3000 & 0.035 & 2.807 & 4.820 \\
        \hline
    \end{tabular}
\end{table}

Overall, the circumferential discretization contributes only negligibly to
the remaining grid dependence, whereas the axial and radial directions
retain errors of a few percent. A fully refined grid such as
$75\times96\times49$ would, however, substantially increase the
computational cost of the transient simulations.

Consequently, the production grid is chosen as
$N_\varphi\times N_z\times N_r=75\times24\times13$.
The directional deviations of the complex response vector remain below
$5\,\%$ over the investigated frequency range. The robustness of the
EOS-induced differences with respect to this remaining discretization
dependence is assessed separately in Sec.~\ref{sec:eos_discretization_robustness}.

\subsubsection{Temporal discretization and perturbation amplitude}
\label{kap:temporal_discretization}

The temporal discretization is assessed at the three excitation frequencies
on a $75\times24\times25$ grid. Compared with the production grid
$75\times24\times13$, the increased radial resolution is used only for this
convergence study to reduce the influence of spatial discretization errors
on the temporal-error estimate.
Calculations are performed with $N_t={150,300,600,1200}$ time steps per period, while the convergence analysis uses $N_t={300,600,1200}$.

\begin{figure}[h]
    \centering
    \includegraphics[width=\columnwidth]
    {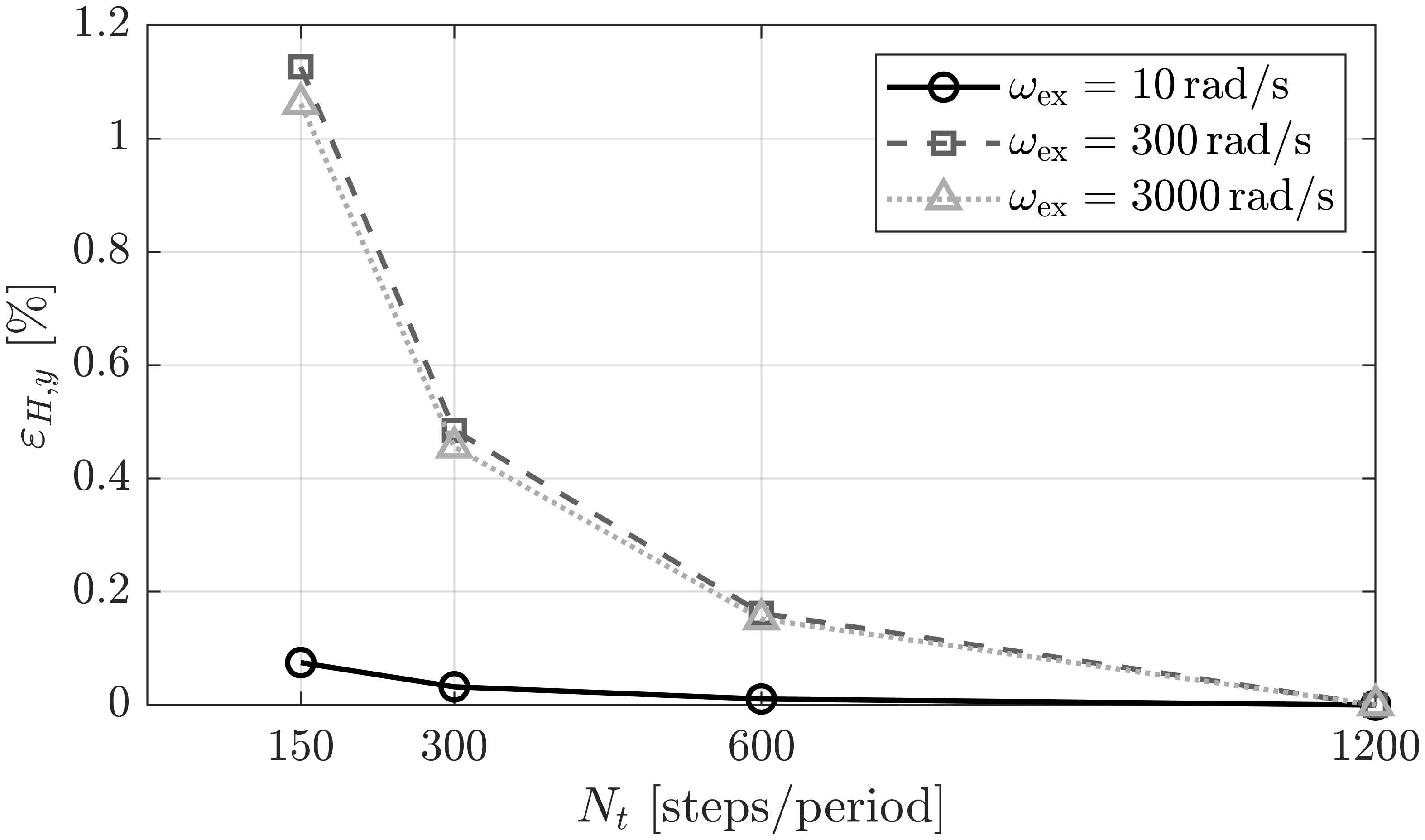}
    \caption{Temporal discretization error of the complex response vector
    relative to the solution obtained with 1200 time steps per excitation
    period.}
    \label{fig:temporal_hvec_error}
\end{figure}

As expected for the backward-Euler discretization, the temporal refinement
shows approximately first-order convergence. The corresponding convergence
orders and GCIs are summarized in Tab.~\ref{tab:temporal_convergence}.
For the selected resolution of $N_t=300$, the deviation from the
$N_t=1200$ solution remains below $0.49\,\%$ for the complex response
vector, as also illustrated in Fig.~\ref{fig:temporal_hvec_error}.

\begin{table}[h]
    \centering
    \caption{Temporal convergence of the complex dynamic response.
    The error $\varepsilon_{H,y}^{N_t=300}$ is evaluated relative to
    the solution obtained with $N_t=1200$ time steps per excitation period.}
    \renewcommand*{\arraystretch}{1.2}
    \setlength{\tabcolsep}{4pt}
    \label{tab:temporal_convergence}
    \begin{tabular}{lccc}
        \hline
        $\omega_\mathrm{ex}$ [rad/s]
        & 10 & 300 & 3000 \\
        \hline
        $p_{H_{yy}}$ [-]
        & 0.9996 & 0.9955 & 0.9993 \\

        $p_{H_{xy}}$ [-]
        & 0.9985 & 0.9918 & 1.0028 \\

        $\max(\mathrm{GCI}_{H_{yy}},
        \mathrm{GCI}_{H_{xy}})$ [\%]
        & 0.0141 & 0.2480 & 0.1943 \\

        $\varepsilon_{H,y}^{N_t=300}$ [\%]
        & 0.0321 & 0.4852 & 0.4553 \\
        \hline
    \end{tabular}
\end{table}

Thus, $N_t=300$ provides a temporal error well below the remaining spatial
discretization error at moderate computational cost. The periodic state is
well established, with a maximum period-to-period response change of
$1.99\times10^{-5}$.

\begin{figure}[h]
    \centering
    \includegraphics[width=\columnwidth]
    {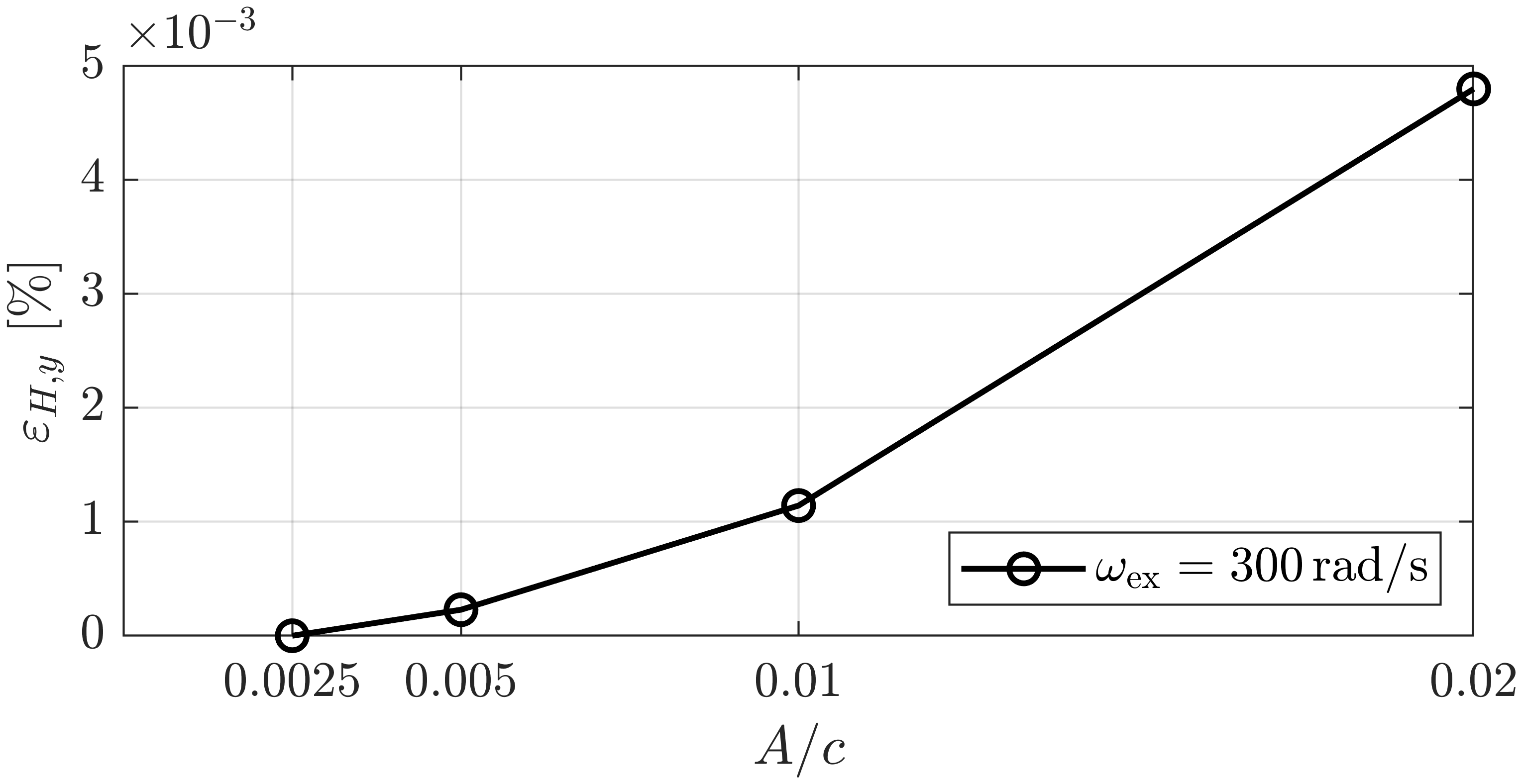}
    \caption{Influence of the perturbation amplitude on the identified
    complex bearing response at
    $\omega_\mathrm{ex}=300~\mathrm{rad/s}$, relative to the response at
    $A/c=0.0025$.}
    \label{fig:amplitude_linearity}
\end{figure}

The linearity of the perturbation response is examined at
$\omega_\mathrm{ex}=300~\mathrm{rad/s}$ using
$A/c=\{0.0025,0.005,0.01,0.02\}$. The calculations employ a
$75\times24\times25$ grid and 600 time steps per excitation period.
Using the smallest amplitude, $A/c=0.0025$, as an approximation of the
small-amplitude limit, the selected value $A/c=0.01$ differs by only
$0.00114\,\%$ in the complex response vector. Even at $A/c=0.02$, the
deviation remains below $0.005\,\%$, confirming linear response over the
investigated range (cf. Fig.~\ref{fig:amplitude_linearity}).

Consequently, $A/c=0.01$ is used for the subsequent dynamic
calculations.

\subsubsection{Robustness of the EOS-induced deviations}
\label{sec:eos_discretization_robustness}

Since the EOS-induced deviations of the more accurate cubic formulations
are smaller than the remaining discretization dependence of the absolute
dynamic response, the robustness of the relative EOS comparison is assessed
separately. The case $\omega_\mathrm{ex}=10~\mathrm{rad\,s^{-1}}$ is
considered, for which the preceding grid study exhibits the largest axial
discretization sensitivity and the PR formulation shows its maximum dynamic
deviation from the Helmholtz reference for both refrigerants.

The production discretization
$75\times24\times13$ with $N_t=300$ is compared with a spatially refined
$75\times48\times25$ grid and with $N_t=600$ on the production grid.
For each discretization, the approximate EOS and the corresponding
Helmholtz reference are evaluated on exactly the same grid and time step.

\begin{table}[h]
    \centering
    \caption{Robustness of the EOS-induced complex dynamic response error
    $\varepsilon_{H,y}^{\mathrm{EOS}}$ at
    $\omega_\mathrm{ex}=10~\mathrm{rad\,s^{-1}}$.}
    \label{tab:eos_discretization_robustness}
    \renewcommand*{\arraystretch}{1.05}
    \setlength{\tabcolsep}{4pt}
    \footnotesize

    \begin{tabular}{llrrr}
        \hline
        Fluid & Discretization
        & Ideal
        & SRK
        & PR \\
        & & [\%] & [\%] & [\%] \\
        \hline

        R1234yf
        & Production       & 5.5406 & 0.6359 & 0.3456 \\
        & Spatial refined  & 5.5810 & 0.6389 & 0.3464 \\
        & Temporal refined & 5.5415 & 0.6360 & 0.3457 \\

        \hline

        R1234ze(E)
        & Production       & 6.3287 & 0.8178 & 0.5293 \\
        & Spatial refined  & 6.3708 & 0.8213 & 0.5307 \\
        & Temporal refined & 6.3298 & 0.8179 & 0.5294 \\

        \hline
    \end{tabular}
\end{table}

The results in Tab.~\ref{tab:eos_discretization_robustness} show that the
EOS-induced relative deviations are substantially less sensitive to the
discretization than the absolute bearing response. Spatial refinement changes
$\varepsilon_{H,y}^{\mathrm{EOS}}$ by less than $0.043$ percentage points
for all investigated EOS. For PR, the changes amount to only $0.0007$ and
$0.0014$ percentage points for R1234yf and R1234ze(E), respectively.
Temporal refinement produces changes below $0.0011$ percentage points in
all cases.

Hence, for the critical low-frequency case considered here, the
sub-percent deviations of SRK and PR from the Helmholtz reference are
robust with respect to the investigated spatial and temporal refinements,
despite the remaining discretization dependence of the absolute dynamic
response.

\subsection{Thermodynamic implementation}
\label{kap:thermodynamic_verification}

The multiparameter Helmholtz-energy equations of state introduced in
Sec.~\ref{sec:eos_helmholtz} are verified independently of the bearing solver against
CoolProp 8.0.0~\cite{Bell2014}.

A total of 135 single-phase thermodynamic states are considered. For each
fluid, 15 pressure states are evaluated at
$T=\{293.15,313.15,333.15\}~\mathrm{K}$. The pressure is varied between
$1$ and $8~\mathrm{bar}$; for the refrigerants, the upper pressure is
restricted to $0.95p_\mathrm{sat}$ whenever this limit is lower than
$8~\mathrm{bar}$ in order to remain within the single-phase vapor region.

The verification is deliberately separated into a direct evaluation and an
inverse evaluation. First, CoolProp provides the reference density
$\varrho_\mathrm{CP}(p,T)$. This density is inserted directly into the present
Helmholtz implementation, and the resulting pressure
$p(\varrho_\mathrm{CP},T)$ and isothermal density derivative from Eq.~\eqref{eq:pressure_gradient_isothermal}
are compared with the corresponding CoolProp values. This direct test avoids
mixing an error of the Helmholtz formulation with an error of the numerical
$p\rightarrow\varrho$ inversion. In a second step, the density obtained from the
implemented inversion routine is compared directly with
$\varrho_\mathrm{CP}$, and the pressure residual after the
$p\rightarrow\varrho\rightarrow p$ cycle is evaluated.

The maximum deviations over all investigated states are summarized in
\tab{tab:thermodynamic_verification}. For R1234yf and R1234ze(E), both the
pressure and $p_\varrho$ agree with CoolProp to approximately machine
precision.
Across all three fluids, the maximum pressure deviation remains below
$4.4\times10^{-14}\,\%$, and the maximum deviation of $p_\varrho$ remains
below $8.4\times10^{-14}\,\%$.

The inverse density calculation exhibits maximum relative deviations below
$9\times10^{-11}\,\%$. The corresponding
$p\rightarrow\varrho\rightarrow p$ residual remains below
$7.5\times10^{-13}$. Hence, both the Helmholtz-energy implementation and
the analytical density derivative entering the Newton linearization and the
transient compressibility terms are reproduced to numerical precision.

\begin{table}[h]
    \centering
    \caption{Maximum deviations of the implemented multiparameter
    Helmholtz-energy equations of state from CoolProp 8.0.0~\cite{Bell2014}.}
    \label{tab:thermodynamic_verification}
    \renewcommand*{\arraystretch}{1.2}
    \setlength{\tabcolsep}{6pt}
    \small
    \begin{tabular}{lccc}
        \hline
        & Air & R1234yf & R1234ze(E) \\
        \hline
        $\varepsilon_{\varrho,\max}$ [\%]
        & $3.73\times10^{-11}$
        & $7.58\times10^{-11}$
        & $8.84\times10^{-11}$ \\

        $\varepsilon_{p,\max}$ [\%]
        & $2.33\times10^{-14}$
        & $4.37\times10^{-14}$
        & $3.42\times10^{-14}$ \\

        $\varepsilon_{p_\varrho,\max}$ [\%]
        & $3.05\times10^{-14}$
        & $8.31\times10^{-14}$
        & $2.14\times10^{-14}$ \\

        $R_{p\varrho p,\max}$ [-]
        & $3.73\times10^{-13}$
        & $7.44\times10^{-13}$
        & $7.43\times10^{-13}$ \\
        \hline
    \end{tabular}
\end{table}

The saturation-pressure correlations used to delimit the single-phase
operating range are checked separately. Their maximum deviations from
CoolProp are $0.0215\,\%$ for R1234yf and $0.0163\,\%$ for R1234ze(E).
These deviations are small compared with the thermodynamic effects examined
in the subsequent bearing calculations.

The viscosity is not included in the CoolProp verification because CoolProp 8.0.0 employs the viscosity model of \textsc{Bell} and \textsc{Laesecke}~\cite{Bell2016} for the refrigerants. The comparison is therefore restricted to thermodynamic quantities for which the same Helmholtz-energy formulations are implemented independently in both codes.

%
\subsection{Verification by the method of manufactured solutions}\label{sec:mms}
The coupled Reynolds--Darcy implementation is additionally verified using the method of manufactured solutions (MMS)~\cite{Salari2000,Roache2001}. Prescribed analytical fields are enforced as exact solutions through artificial source terms, allowing the consistency and formal convergence order of the discretization to be assessed.

For the MMS, an isothermal ideal gas is considered. In the following, the
superscript $\star$ denotes the prescribed manufactured solution. With the
nondimensionalization introduced in Sec.~\ref{sec:nondimensionalization}, the manufactured
dimensionless pressure and density are identical,
\begin{align}
    P^\star
    =
    \frac{p^\star}{p_{\mathrm{amb}}}
    =
    \frac{\varrho^\star}{\varrho_{\mathrm{amb}}}.
\end{align}
This choice separates the verification of the spatial and temporal
operators from the verification of the real-gas thermodynamic models,
which is performed independently in Sec.~\ref{kap:thermodynamic_verification}.

The axial and radial coordinates used for the manufactured fields are
\begin{align}
    \zeta
    &=
    \frac{z}{L},
    &
    \xi
    &=
    \frac{r-R_b}{t_p},
\end{align}
with $0 \leq \zeta \leq 1$ and $0 \leq \xi \leq 1$.

The manufactured pressure field in the lubrication film is prescribed as
\begin{align}\label{eq:mms_pressure_film}
\begin{split}
    P_f^\star(&\varphi,\zeta,\tau)
    = 1 \\
    &+
    A
    \left[
        1-a_\varphi\cos(m_\varphi\varphi)
    \right]
    g(\zeta)
    \left[
        1+a_t\sin(\omega_t\tau)
    \right],
\end{split}
\end{align}
where the axial shape function is defined by
\begin{align}
    g(\zeta)
    =
    256\,\zeta^4(1-\zeta)^4.
    \label{eq:mms_axial_function}
\end{align}
The fourth-order polynomial is selected such that
\begin{align*}
    g(0)
    =
    g(1)
    =
    0,
    \qquad
    \left.
    \frac{\partial g}{\partial\zeta}
    \right|_{\zeta=0}
    =
    \left.
    \frac{\partial g}{\partial\zeta}
    \right|_{\zeta=1}
    =
    0.
\end{align*}
Thus, the prescribed ambient-pressure condition at both axial boundaries
of the lubrication film and the zero axial pressure-gradient condition at
the corresponding boundaries of the porous domain are satisfied
simultaneously.

The manufactured pressure field in the porous shell is obtained by a
continuous interpolation between the film pressure at the Reynolds--Darcy
interface and the prescribed supply pressure at the outer shell boundary,
\begin{align}
    P_p^\star(\varphi,\zeta,\xi,\tau)
    =
    P_s
    +
    (1-\xi)
    \left[
        P_f^\star(\varphi,\zeta,\tau)-P_s
    \right].
    \label{eq:mms_pressure_shell}
\end{align}
Consequently, $P_p^\star(\xi=0) = P_f^\star$ and $P_p^\star(\xi=1) = P_s$
such that pressure continuity at the Reynolds--Darcy interface and the
Dirichlet condition at the outer porous-shell surface are fulfilled
exactly.

In addition to the pressure fields, a nonuniform manufactured film
thickness is imposed,
\begin{align}
    H^\star(\varphi,\tau)
    =
    1
    +
    a_h\cos(m_h\varphi)
    \left[
        1+a_{ht}\sin(\omega_h\tau)
    \right].
    \label{eq:mms_gap}
\end{align}
For $m_h=1$, this function has the same circumferential form as the film
thickness of an eccentrically positioned journal, while its time-dependent
amplitude introduces a nonzero squeeze contribution.
The selected amplitude is sufficiently small to ensure
$H^\star>0$ throughout the complete excitation cycle.

\begin{figure*}[h]
    \centering
    \includegraphics[width=\linewidth]
        {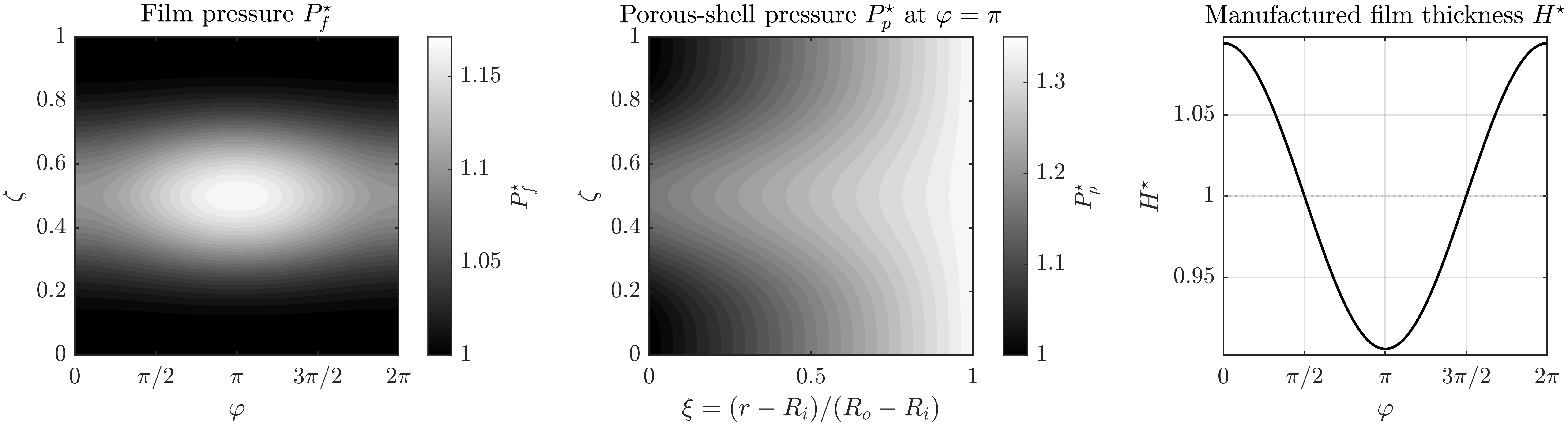}
    \caption{Manufactured fields used for the verification of the coupled
    Reynolds--Darcy implementation at $\tau=1/8$:
    (a) nondimensional film pressure $P_f^\star$,
    (b) nondimensional porous-shell pressure $P_p^\star$ at
    $\varphi=0$, and
    (c) nondimensional manufactured film thickness $H^\star$.}
    \label{fig:mms_fields}
\end{figure*}

The parameters used for the manufactured solution are
\begin{align}\label{eq:mms_parameters}
    \begin{split}
        A &= 0.12, \qquad a_\varphi = 0.25, \qquad m_\varphi = 1,\\
        a_t &= 0.20, \qquad  \omega_t = 2\pi, \qquad \quad  a_h = 0.08,\\
        m_h &= 1, \qquad \quad  a_{ht} = 0.25, \qquad \ \omega_h = 2\pi,\\
        P_s &= 1.35.
    \end{split}
\end{align}
Furthermore, the dimensionless Couette parameter is set to $\Lambda_\Omega=1$, corresponding to an equivalent shaft speed of approximately $n_s=600~\mathrm{min}^{-1}$ for the present parameters, thereby activating the circumferential transport term.


Fig.~\ref{fig:mms_fields} illustrates the manufactured fields at
$\tau=1/8$. The film pressure satisfies the prescribed axial boundary
conditions and varies continuously into the porous domain, where it
approaches the uniform supply pressure at the outer shell surface.
The nonuniform film thickness activates both circumferential transport and,
in the transient case, the squeeze term. Thus, the manufactured fields
exercise the principal spatial and transient operators of the coupled
Reynolds--Darcy formulation.

The corresponding source terms are obtained by inserting
Eqs.~\eqref{eq:mms_pressure_film}--\eqref{eq:mms_gap} into the continuous
Reynolds and Darcy operators.
Denoting the two differential operators by
$\mathcal{L}_R$ and $\mathcal{L}_D$, respectively, the artificial source
terms follow as
\begin{align}
    S_R^\star
    &=
    \mathcal{L}_R
    \left(
        P_f^\star,
        P_p^\star,
        H^\star
    \right),
    \\
    S_D^\star
    &=
    \mathcal{L}_D
    \left(
        P_p^\star
    \right).
\end{align}
The source terms are evaluated analytically and added only to the
right-hand side of the discrete governing equations.
The spatial operators themselves are therefore identical to those used
for the bearing simulations.

For the stationary MMS, the transient terms are deactivated while the
nonuniform film thickness and the Couette term remain active.
The spatial grid is successively refined from
$24\times12\times5$ to $48\times24\times9$ and
$96\times48\times17$, where $N_r$ includes the lubrication-film layer.
The error is evaluated directly with respect to the analytical
manufactured solution using the volume-weighted relative $L_2$ norm
\begin{align}
    E_2
    =
    \sqrt{
        \frac{
            \sum_P V_P
            \left(
                P_P-P_P^\star
            \right)^2
        }{
            \sum_P V_P
            \left(
                P_P^\star
            \right)^2
        }
    }.
    \label{eq:mms_l2}
\end{align}

\begin{figure}[h]
    \centering
    \includegraphics[width=\linewidth]
        {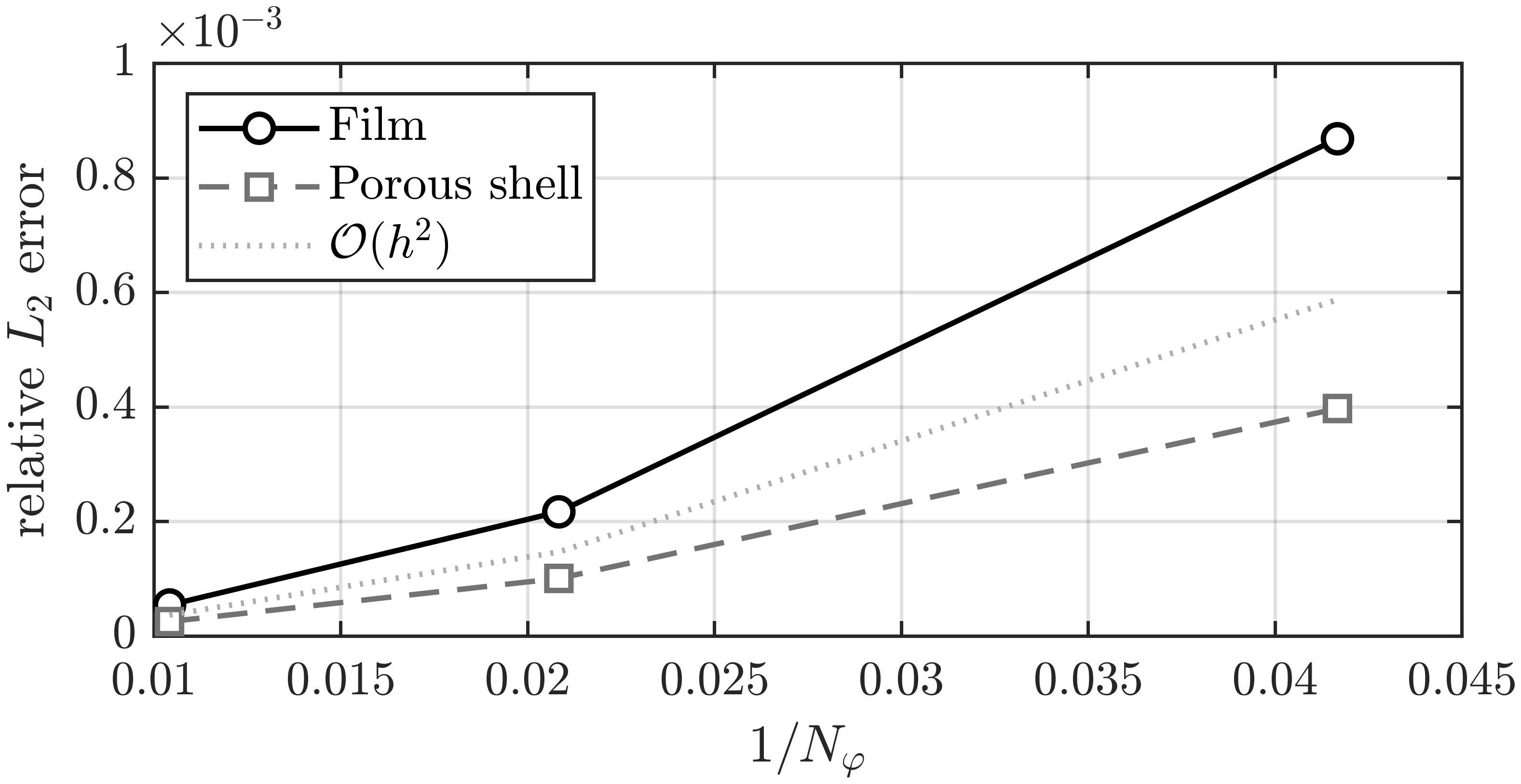}
    \caption{Spatial convergence of the stationary MMS.
    The errors are evaluated directly with respect to the prescribed
    analytical manufactured solution.}
    \label{fig:mms_spatial}
\end{figure}

As shown in Fig.~\ref{fig:mms_spatial}, the numerical solutions in both
domains approach the prescribed manufactured solution systematically with
grid refinement.
For the lubrication film, the observed spatial convergence orders are
$2.002$ and $2.000$, whereas values of $1.983$ and $1.995$ are obtained
for the porous shell.
The spatial discretization therefore recovers the expected second-order
convergence in both parts of the coupled system.

For the transient MMS, the storage terms in both domains are activated on
the intermediate $48\times24\times9$ grid of the preceding spatial
refinement study, while the temporal resolution is increased from 16 to 512
time steps per excitation period.

Two error measures are used. The direct error with respect to the analytical
manufactured solution is evaluated using Eq.~\eqref{eq:mms_l2}, while the
difference between successive temporal resolutions is used to isolate the
temporal discretization error once the fixed spatial error becomes dominant.

To isolate the temporal discretization error, the difference between two
successive temporal resolutions is therefore additionally evaluated,
\begin{align}
    D_{\Delta\tau}
    =
    \sqrt{
        \frac{
            \sum_P V_P
            \left(
                P_{\Delta\tau,P}
                -
                P_{\Delta\tau/2,P}
            \right)^2
        }{
            \sum_P V_P
            P_{\Delta\tau/2,P}^{\,2}
        }
    }.
    \label{eq:mms_temporal_difference}
\end{align}
The observed temporal convergence order $p_t$ is then obtained from
\begin{align}
    p_t
    =
    \frac{
        \ln
        \left(
            D_{\Delta\tau}/D_{\Delta\tau/2}
        \right)
    }{
        \ln 2
    }.
    \label{eq:mms_temporal_order}
\end{align}

\begin{figure}[h]
    \centering
    \includegraphics[width=\linewidth]
        {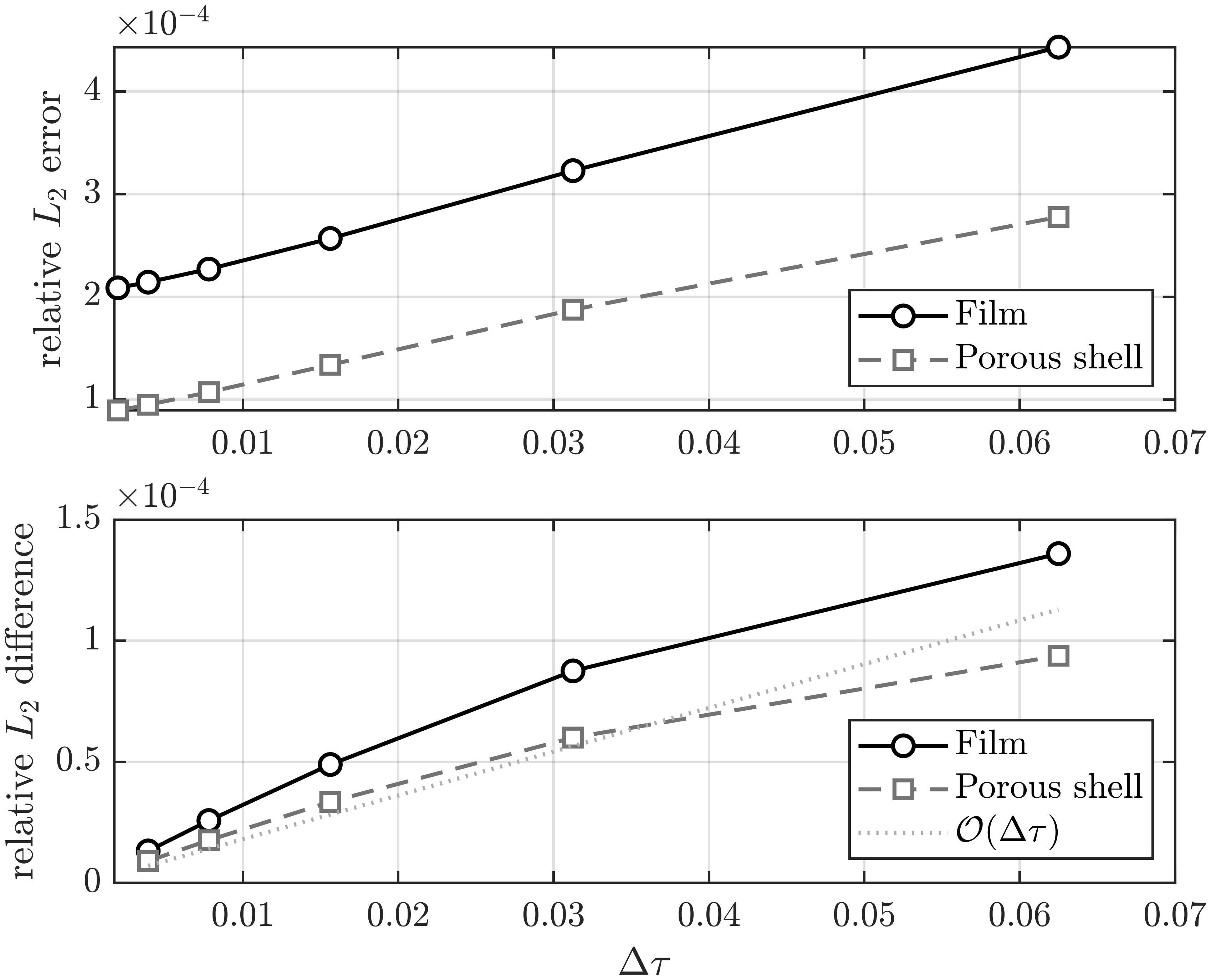}
    \caption{Temporal convergence of the transient MMS:
    (a) relative $L_2$ error with respect to the prescribed manufactured
    solution and (b) differences between successive temporal refinements.
    The $\mathcal{O}(\Delta\tau)$ reference slope indicates the expected
first-order convergence of the backward-Euler discretization.}
    \label{fig:mms_temporal}
\end{figure}

Fig.~\ref{fig:mms_temporal} shows the temporal refinement study. The direct error decreases monotonically with decreasing time-step size and approaches the remaining spatial-discretization error. Meanwhile, the successive-solution differences exhibit the expected first-order behavior, with observed convergence orders approaching unity for both the lubrication film and porous shell. This confirms the expected temporal accuracy of the backward-Euler scheme with observed convergence orders approaching $0.963$ for the lubrication film and $0.964$ for the porous shell.

Overall, the MMS confirms the expected convergence behavior of the coupled Reynolds--Darcy implementation, with approximately second-order spatial and first-order temporal accuracy. The verification covers the Reynolds and Darcy operators, interface coupling, Couette transport, time-dependent film thickness, and transient storage in both domains.

\subsection{Bearing-level reference comparison}
\label{sec:bearing_reference}

The porous gas-bearing model has previously been verified against the
analytical and numerical reference solutions of San Andrés et
al.~\cite{SanAndres2021}. Both stationary bearing forces and
frequency-dependent direct and cross-coupled dynamic coefficients showed
good agreement with the reference data. Since the underlying
Reynolds--Darcy formulation is unchanged in the present work, the
corresponding model adaptations and quantitative comparisons reported in
Stellmach et al.~\cite{Stellmach2026a} are not repeated here.

The present study complements this bearing-level verification by the MMS,
which directly assesses the numerical implementation and its formal spatial
and temporal convergence behavior.

\section{Bearing response for different equations of state}
\label{sec:results}

The influence of the thermodynamic closure on the bearing response is evaluated
using the Helmholtz-energy formulation as reference. The ideal-gas, van der
Waals (VDW), Soave--Redlich--Kwong (SRK), and Peng--Robinson (PR) equations
of state are assessed according to Eq.~\eqref{eq:relative_deviation}.
Stationary load capacity and supplied mass flow are considered first, followed
by the frequency-dependent direct and cross-coupled dynamic response and the
associated supplied mass flow. The thermodynamic origin of the observed
differences is discussed separately in Sec.~\ref{sec:thermodynamic_interpretation}.

\subsection{Stationary bearing response}
\label{sec:results_stationary}

To isolate the influence of the thermodynamic closure on pressure-driven
aerostatic transport, the stationary analysis is performed at an eccentricity
ratio of $\varepsilon=0.5$ with a non-rotating journal.
Temperatures of $T=\{293.15,313.15,333.15\}\,\mathrm{K}$ and absolute supply pressures
of $p_\mathrm{s}=\{2,4,6,8\}\,\mathrm{bar}$ are considered wherever the single-phase
restriction is satisfied.
For each admissible thermodynamic state, all five equations of state are
evaluated using the same bearing configuration and numerical discretization.

\begin{figure*}[h]
    \centering
    \includegraphics[width=\textwidth]
        {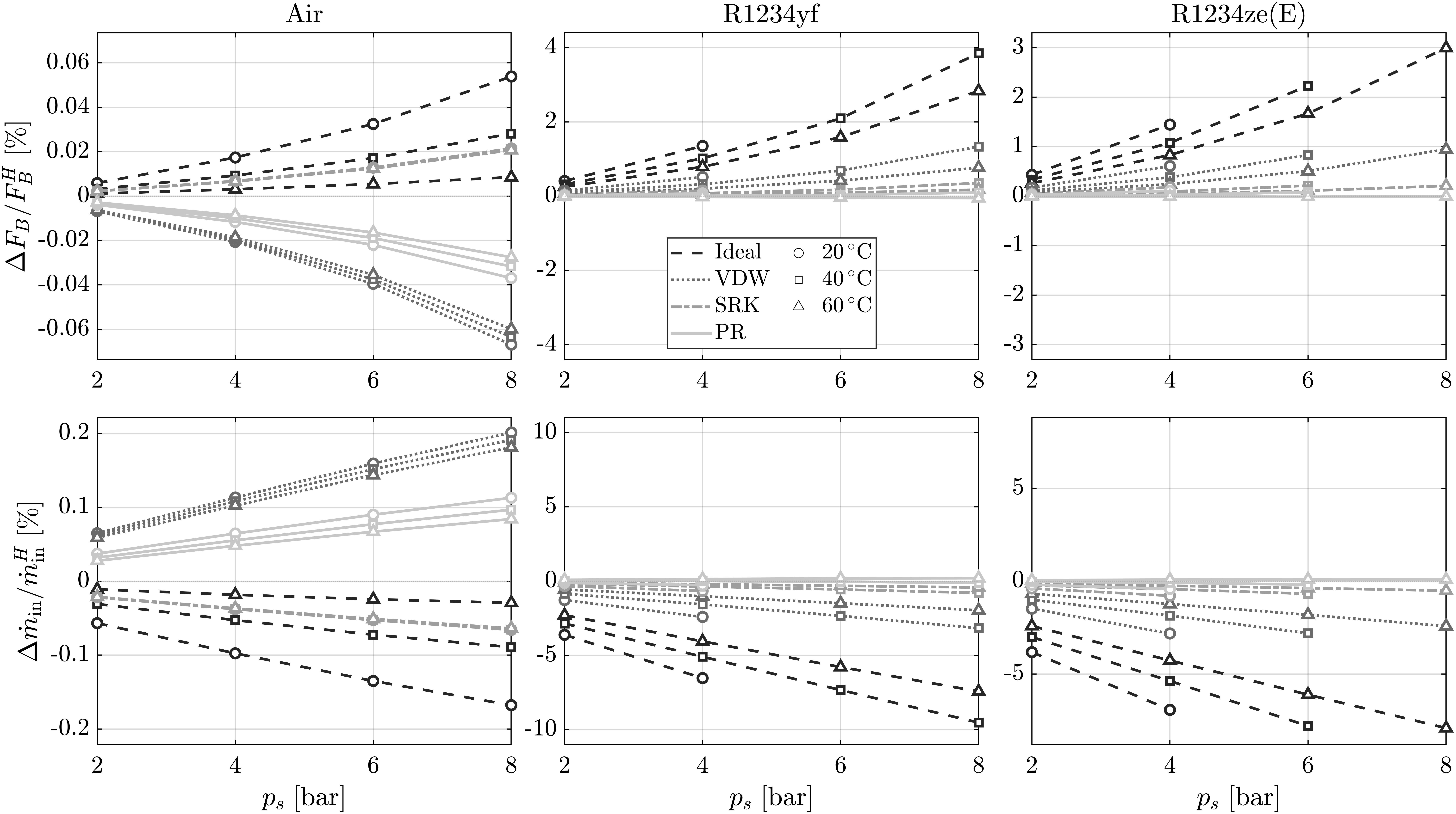}
    \caption{Relative deviations of the stationary bearing load (upper row)
    and supplied mass flow (lower row) from the Helmholtz reference as a
    function of supply pressure for air, R1234yf, and R1234ze(E).
    Only operating states satisfying the single-phase restriction are
    included.}
    \label{fig:static_eos_deviations}
\end{figure*}

Fig.~\ref{fig:static_eos_deviations} shows the resulting deviations of the
bearing load and supplied mass flow from the corresponding Helmholtz
solution. For air, the curves remain close to the reference throughout the
investigated pressure and temperature range. In contrast, both refrigerants
show a systematic separation between the thermodynamic models. The
differences generally increase with increasing supply pressure and become
more pronounced at lower temperature.


A clear difference is observed between the sensitivity of the two bearing
quantities. For both refrigerants, the supplied mass flow reacts considerably
more strongly to the thermodynamic closure than the integrated bearing load.
The ideal-gas approximation exhibits the largest deviations, followed by
VDW. SRK substantially improves the agreement with the Helmholtz reference,
while PR provides the closest approximation among the cubic equations of
state. The same qualitative ordering is obtained for both refrigerants.

The maximum deviations over the complete admissible stationary operating
range are summarized in Tab.~\ref{tab:static_eos_max_error}. In addition to
the bearing load and supplied mass flow, the maximum film pressure is included
to distinguish changes of the overall pressure distribution from changes of
its local maximum.

\begin{table*}[h]
    \centering
    \caption{Maximum absolute deviations from the Helmholtz reference over
    the investigated stationary operating range.}
    \label{tab:static_eos_max_error}
    \renewcommand*{\arraystretch}{1.2}
    \setlength{\tabcolsep}{7pt}
    \begin{tabular}{llrrr}
        \hline
        Fluid & EOS
        & $\max|\varepsilon_{F}|$ [\%]
        & $\max|\varepsilon_{\dot m}|$ [\%]
        & $\max|\varepsilon_{p_{\max}}|$ [\%] \\
        \hline

        Air
        & Ideal & 0.054 & 0.168 & 0.00016 \\
        & VDW   & 0.067 & 0.201 & 0.00020 \\
        & SRK   & 0.021 & 0.066 & 0.00006 \\
        & PR    & 0.037 & 0.113 & 0.00011 \\
        \hline

        R1234yf
        & Ideal & 3.848 & 9.522 & 0.01067 \\
        & VDW   & 1.332 & 3.165 & 0.00359 \\
        & SRK   & 0.352 & 0.794 & 0.00095 \\
        & PR    & 0.096 & 0.272 & 0.00030 \\
        \hline

        R1234ze(E)
        & Ideal & 2.993 & 7.914 & 0.00882 \\
        & VDW   & 0.945 & 2.819 & 0.00298 \\
        & SRK   & 0.212 & 0.787 & 0.00076 \\
        & PR    & 0.107 & 0.420 & 0.00045 \\
        \hline
    \end{tabular}
\end{table*}

Whereas the bearing load and particularly the mass flow exhibit measurable
EOS dependence for the refrigerants, the maximum film pressure remains
virtually unchanged. The nearly unchanged peak pressure does not imply an EOS-independent pressure field, since changes in the pressure distribution may still affect the integrated bearing force and supplied mass flow.

For reference, Tab.~\ref{tab:static_reference_state} gives the absolute
Helmholtz results at the common operating state
$T=293.15\,\mathrm{K}$ and $p_\mathrm{s}=4\,\mathrm{bar}$.
The comparable bearing loads demonstrate that the substantially different
mass flows of the three working fluids do not simply result from a
corresponding change in the pressure level within the lubrication film.

\begin{table}[h]
    \centering
    \caption{Stationary Helmholtz-reference results at
    $T=293.15\,\mathrm{K}$ and $p_\mathrm{s}=4\,\mathrm{bar}$ for
    $\varepsilon=0.5$.}
    \label{tab:static_reference_state}
    \renewcommand*{\arraystretch}{1.2}
    \begin{tabular}{lrrr}
        \hline
        Fluid
        & $F_B$ [N]
        & $\dot m_\mathrm{s}$ [g/s]
        & $p_{\max}$ [bar] \\
        \hline
        Air        & 586.55 & 0.11505 & 3.9949 \\
        R1234yf    & 578.41 & 0.79036 & 3.9951 \\
        R1234ze(E) & 579.80 & 0.75283 & 3.9950 \\
        \hline
    \end{tabular}
\end{table}

Overall, the stationary calculations show that the relevance of the
thermodynamic closure depends not only on the working fluid but also on the
quantity of interest. While the ideal-gas approximation is effectively
sufficient for air within the investigated range, its deviations become
clearly visible for both refrigerants.

\subsection{Dynamic bearing response}
\label{sec:results_dynamic}

The dynamic bearing response is evaluated at
$T=293.15~\mathrm{K}$, $p_\mathrm{s}=4~\mathrm{bar}$, and
$n_s=25,000~\mathrm{min}^{-1}$ using a centered harmonic perturbation with
$A/c=0.01$. Transient gas storage is retained in both the film and porous shell.

The excitation angular frequency is varied over
$\omega_\mathrm{ex} = \{10,30,100,300,1000,3000\}~\mathrm{rad/s}.$
For each fluid, all five equations of state are evaluated at identical
operating conditions.

\subsubsection{Dynamic force response}

\begin{figure*}[h]
    \centering
    \includegraphics[width=\textwidth]
        {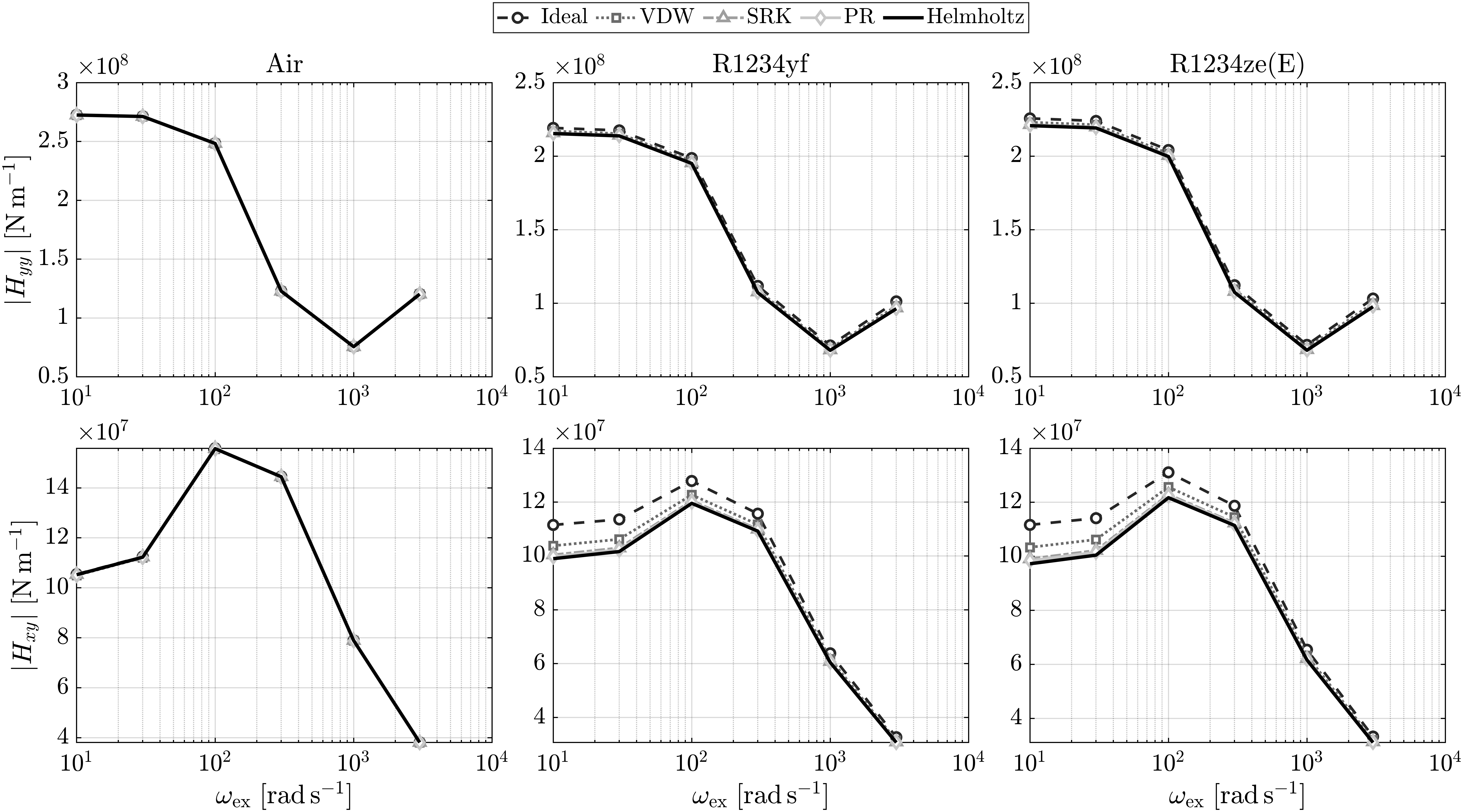}
    \caption{Magnitude of the direct and cross-coupled complex dynamic
    force responses, $|H_{yy}|$ and $|H_{xy}|$, as functions of excitation
    frequency for air, R1234yf, and R1234ze(E).}
    \label{fig:dynamic_complex_response}
\end{figure*}

Fig.~\ref{fig:dynamic_complex_response} shows the magnitudes of the direct
and cross-coupled complex force responses,
$|H_{yy}|$ and $|H_{xy}|$, respectively.
For air, the curves obtained with the different equations of state are
nearly indistinguishable over the complete frequency range.
The thermodynamic closure therefore has only a minor influence on the
dynamic bearing response for this fluid.

For both refrigerants, a systematic separation between the equations of
state becomes visible.
The ideal-gas approximation shows the largest deviation from the Helmholtz
reference, followed by VDW.
SRK and particularly PR remain considerably closer to the reference
solution.
This ordering is observed for both the direct and cross-coupled response
components.

\begin{figure*}[h]
    \centering
    \includegraphics[width=\textwidth]
        {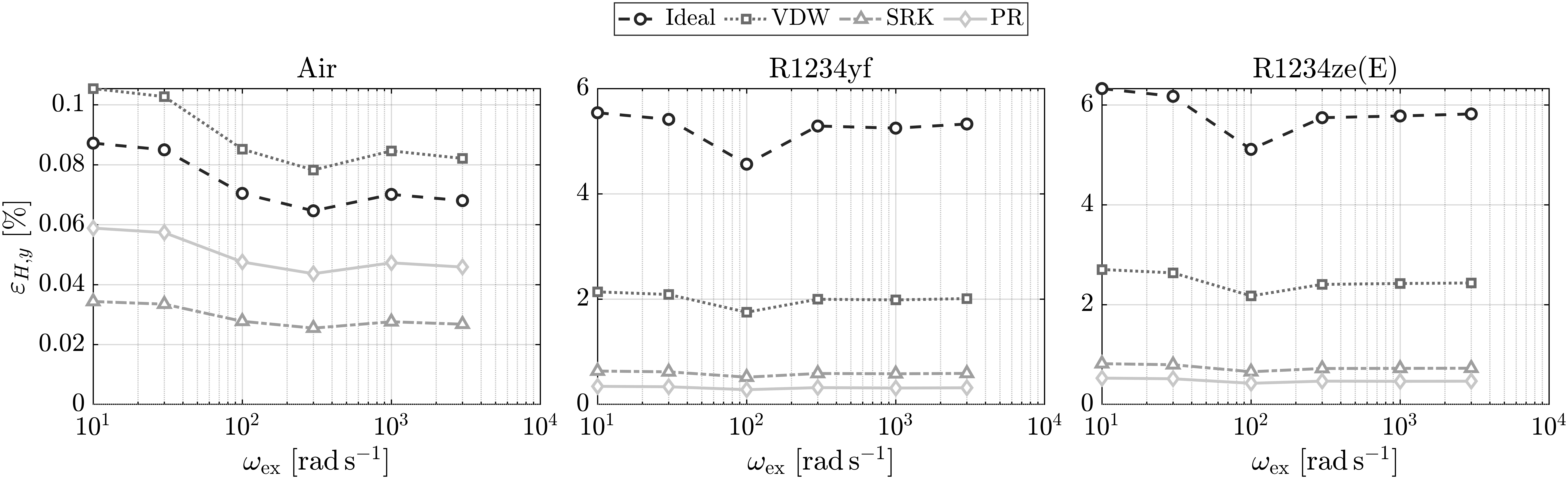}
    \caption{Relative error of the complete complex dynamic force response
    $\varepsilon_{H,y}$ with respect to the Helmholtz reference as a
    function of excitation frequency.}
    \label{fig:eos_error_hvec}
\end{figure*}

To quantify the combined influence on the force response, the vector error
$\varepsilon_{H,y}$ defined in
Eq.~\eqref{eq:H_vector_error} is evaluated relative to the Helmholtz
reference.
Fig.~\ref{fig:eos_error_hvec} shows its frequency dependence, while the
maximum deviations over the investigated range are summarized in
Tab.~\ref{tab:max_dynamic_eos_error}.

\begin{table}[h]
    \centering
    \caption{Maximum complex dynamic response error
    $\varepsilon_{H,y}$ relative to the Helmholtz reference over the
    investigated excitation-frequency range.}
    \label{tab:max_dynamic_eos_error}
    \renewcommand*{\arraystretch}{1.2}
    \begin{tabular}{lrrr}
        \hline
        EOS
        & Air [\%]
        & R1234yf [\%]
        & R1234ze(E) [\%] \\
        \hline
        Ideal & 0.087 & 5.541 & 6.329 \\
        VDW   & 0.105 & 2.139 & 2.704 \\
        SRK   & 0.034 & 0.636 & 0.818 \\
        PR    & 0.059 & 0.346 & 0.529 \\
        \hline
    \end{tabular}
\end{table}

The results confirm that the relevance of the thermodynamic model depends
strongly on the working fluid.
For air, all investigated closures yield essentially the same dynamic
response.
For the refrigerants, however, the simplified equations of state produce
measurable differences over the complete frequency range.
For both refrigerants, PR provides the closest agreement with the Helmholtz
reference among the cubic equations of state, followed by SRK. For air,
SRK yields the smallest deviation, although all EOS-induced differences
remain negligible.

The EOS influence is not distributed uniformly between the direct and
cross-coupled components.
This is illustrated in Tab.~\ref{tab:dynamic_component_error}, where the
ideal-gas approximation is considered at the lowest and highest excitation
frequencies.
At low frequency, the cross-coupled response is substantially more
sensitive to the thermodynamic closure than the direct response.
At higher excitation frequency, the difference between the sensitivities of
the two components becomes smaller.

\begin{table}[h]
    \centering
    \caption{Relative deviations of the magnitudes of the direct and
    cross-coupled complex responses obtained with the ideal-gas approximation
    at selected excitation frequencies.}
    \label{tab:dynamic_component_error}
    \renewcommand*{\arraystretch}{1.2}
    \setlength{\tabcolsep}{2.5pt}
    \begin{tabular}{lrrrr}
        \hline
        & \multicolumn{2}{c}{$10~\mathrm{rad/s}$}
        & \multicolumn{2}{c}{$3000~\mathrm{rad/s}$} \\
        \cline{2-5}
        Fluid
        & $\Delta_{|H_{yy}|}$ [\%]
        & $\Delta_{|H_{xy}|}$ [\%]
        & $\Delta_{|H_{yy}|}$ [\%]
        & $\Delta_{|H_{xy}|}$ [\%] \\
        \hline
        Air
        & 0.051 & 0.202
        & 0.066 & 0.085 \\
        R1234yf
        & 1.777 & 12.691
        & 5.184 & 6.571 \\
        R1234ze(E)
        & 2.229 & 14.862
        & 5.659 & 7.228 \\
        \hline
    \end{tabular}
\end{table}

Consequently, evaluating only the direct response can underestimate the
influence of the thermodynamic closure.
In particular, the cross-coupled component reveals substantially larger
EOS-induced deviations at low excitation frequencies for both refrigerants.
This is particularly relevant for rotordynamic applications, since
cross-coupled bearing forces can significantly affect rotor stability and
whirl behavior.
The combined response measure $\varepsilon_{H,y}$ therefore provides a
more complete assessment of the thermodynamic-model error than any
individual force coefficient alone.

\subsubsection{Dynamic supplied mass flow}

In addition to the force response, the supplied mass flow is evaluated over
the periodic simulations. Tab.~\ref{tab:dynamic_massflow} summarizes its
mean value, frequency dependence, and first-harmonic content.
For compactness, only the ideal-gas and PR deviations are listed,
representing the largest and smallest deviations among the approximate
equations of state for the refrigerants.

\begin{table*}[h]
    \centering
    \caption{Dynamic supplied mass-flow characteristics. The frequency
    variation refers to the Helmholtz solution, while the maximum
    first-harmonic ratio is evaluated over all investigated equations of
    state. The EOS errors of the mean mass flow are given relative to the
    Helmholtz reference.}
    \label{tab:dynamic_massflow}
    \renewcommand*{\arraystretch}{1.2}
    \setlength{\tabcolsep}{6pt}
    \begin{tabular}{lrrrrr}
        \hline
        Fluid
        & $\overline{\dot m}_{\mathrm{s},H}$ [g/s]
        & $\Delta_\omega\overline{\dot m}_{\mathrm{s},H}$ [\%]
        & $\max |\hat{\dot m}_{\mathrm{s},1}|/
          \overline{\dot m}_{\mathrm{s}}$ [\%]
        & $\varepsilon_{\dot m,\mathrm{Ideal}}$ [\%]
        & $\varepsilon_{\dot m,\mathrm{PR}}$ [\%] \\
        \hline
        Air
        & 0.11271
        & 0.00556
        & $8.88\times10^{-6}$
        & -0.097
        & 0.064 \\
        R1234yf
        & 0.77418
        & 0.00388
        & $3.72\times10^{-6}$
        & -6.514
        & -0.271 \\
        R1234ze(E)
        & 0.73750
        & 0.00399
        & $4.32\times10^{-6}$
        & -6.919
        & -0.419 \\
        \hline
    \end{tabular}
\end{table*}

The supplied mass flow is essentially independent of excitation frequency,
and its first harmonic is negligible for all investigated fluids and
equations of state. The thermodynamic closure nevertheless affects the mean
mass-flow level, with deviations comparable to the stationary results.

Overall, the EOS influence is therefore primarily reflected in the dynamic
force response. It remains negligible for air but becomes pronounced for
both refrigerants, particularly for the cross-coupled response.
For both refrigerants, PR provides the closest cubic approximation to the
Helmholtz reference.
%
%
%
\section{Thermodynamic interpretation and EOS assessment}
\label{sec:thermodynamic_interpretation}

The bearing-level results of Sec.~\ref{sec:results} are related to the
underlying thermodynamic properties and their propagation through the
coupled Reynolds--Darcy formulation.

Particular attention is given to the density $\varrho$, the isothermal
pressure--density derivative $p_\varrho=(\partial p/\partial\varrho)_T$,
and the nondimensional isothermal compressibility $\chi_T$.
Their influence on the bearing response is subsequently assessed together
with the computational cost to derive a practical EOS selection strategy.

\subsection{Thermodynamic differences between the equations of state}
\label{sec:thermodynamic_differences}

The thermodynamic comparison covers the same temperature range as the
stationary bearing study,
$T=\{293.15,313.15,333.15\}\,\mathrm{K}$, with pressures between
$1$ and $8~\mathrm{bar}$ wherever the single-phase condition is satisfied.
For the refrigerants, the upper pressure is therefore restricted by
$0.95p_\mathrm{sat}$.

The compressibility factor $Z$ introduced in Eq.~\eqref{eq:compressibility_factor}
quantifies deviations from the ideal-gas pressure--density relation, whereas
$\chi_T$ from Eq.~\eqref{eq:chi_T} additionally reflects its local slope and
is therefore particularly relevant to transient gas storage.

\begin{figure*}[h]
    \centering
    \includegraphics[width=\textwidth]
        {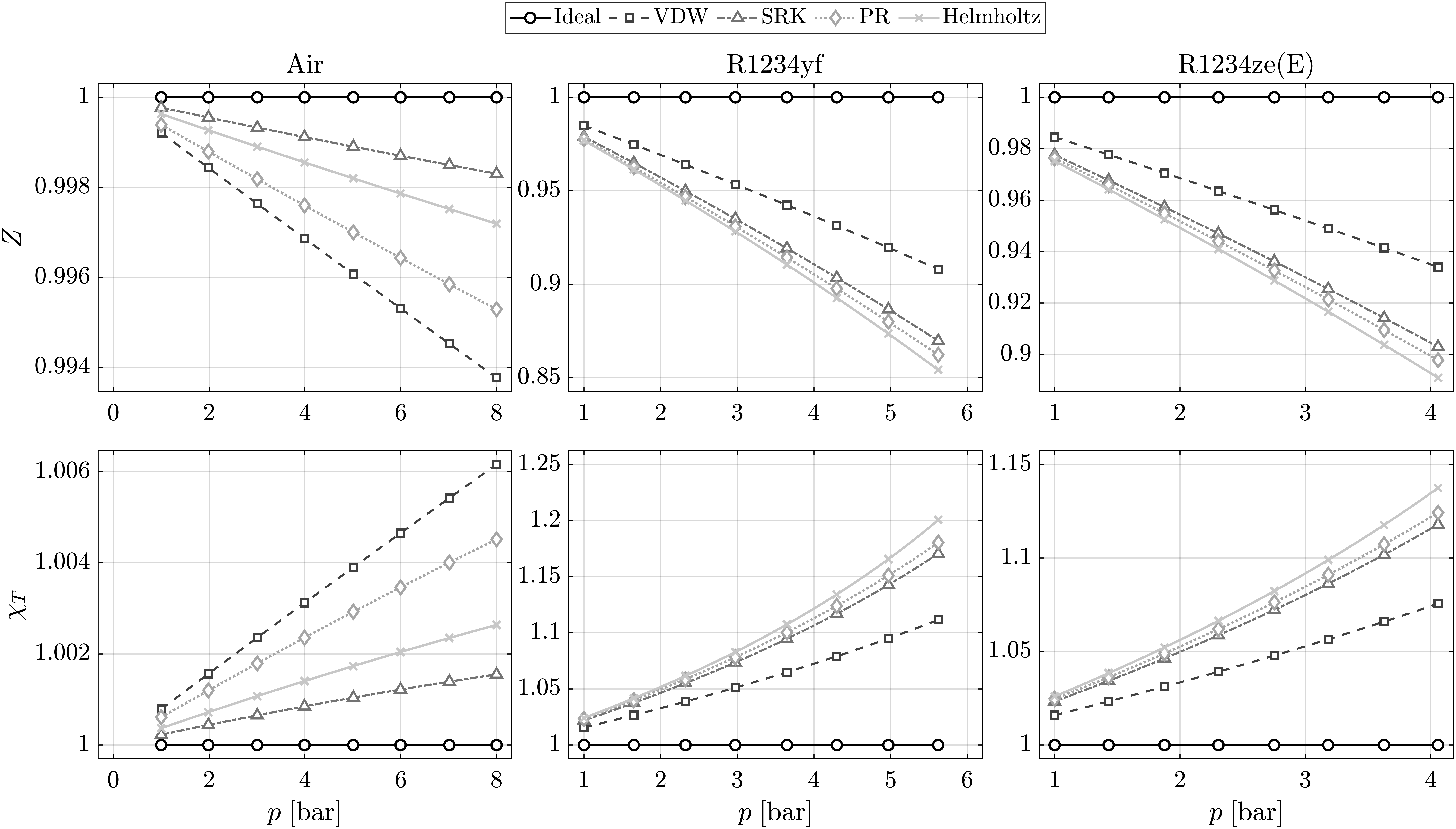}
    \caption{Thermodynamic behavior of the investigated equations of state
    at $T=293.15~\mathrm{K}$: compressibility factor $Z$ and
    nondimensional isothermal compressibility $\chi_T$ as functions of
    pressure for air, R1234yf, and R1234ze(E).}
    \label{fig:thermodynamic_eos_characteristics}
\end{figure*}

Fig.~\ref{fig:thermodynamic_eos_characteristics} illustrates both
quantities at $T=293.15~\mathrm{K}$. Air remains close to the ideal-gas
limit throughout the investigated pressure range, whereas both
refrigerants exhibit increasingly non-ideal behavior with increasing
pressure. The ideal-gas and VDW formulations deviate most strongly from
the Helmholtz reference, while SRK and particularly PR reproduce the
reference behavior considerably more closely.

The differences are quantified at the operating state used for the dynamic
bearing calculations in Tab.~\ref{tab:thermo_reference_state}.
For air, the Helmholtz values are
$Z=0.99855$ and $\chi_T=1.00141$, confirming that the fluid remains
very close to ideal behavior.
For R1234yf and R1234ze(E), the corresponding compressibility factors are
$0.90094$ and $0.89274$, respectively, while
$\chi_T=1.12156$ and $1.13463$.
The real-gas character is therefore substantially more pronounced for the
two refrigerants.

\begin{table*}[h]
    \centering
    \caption{Thermodynamic deviations from the Helmholtz reference at
    $T=293.15~\mathrm{K}$ and $p=4~\mathrm{bar}$.}
    \label{tab:thermo_reference_state}
    \renewcommand*{\arraystretch}{1.2}
    \setlength{\tabcolsep}{6pt}
    \begin{tabular}{llrrr}
        \hline
        Fluid & EOS
        & $\Delta\varrho/\varrho_H$ [\%]
        & $\Delta p_\varrho/p_{\varrho,H}$ [\%]
        & $\Delta\chi_T/\chi_{T,H}$ [\%] \\
        \hline
        Air
        & Ideal & -0.145 &  0.286 & -0.140 \\
        & VDW   &  0.170 & -0.341 &  0.172 \\
        & SRK   & -0.056 &  0.112 & -0.056 \\
        & PR    &  0.096 & -0.191 &  0.095 \\
        \hline
        R1234yf
        & Ideal & -9.906 & 24.488 & -10.839 \\
        & VDW   & -3.782 &  8.692 &  -4.381 \\
        & SRK   & -1.073 &  2.463 &  -1.345 \\
        & PR    & -0.483 &  1.262 &  -0.767 \\
        \hline
        R1234ze(E)
        & Ideal & -10.726 & 27.094 & -11.865 \\
        & VDW   &  -4.523 & 10.629 &  -5.326 \\
        & SRK   &  -1.307 &  3.049 &  -1.674 \\
        & PR    &  -0.743 &  1.893 &  -1.124 \\
        \hline
    \end{tabular}
\end{table*}

The same hierarchy persists over the wider pressure and temperature range.
For air, deviations remain below the percent level, whereas they increase
for the refrigerants with rising pressure and decreasing temperature. The
ideal-gas model shows the largest deviations; VDW captures part of the
real-gas behavior, while SRK and PR reproduce both $\varrho$ and
$p_\varrho$ more accurately.

The two quantities affect the bearing model differently: $\varrho$ governs
the mass associated with a local state and enters the mass transport terms,
whereas $p_\varrho$ determines the density response to pressure variations
and is therefore particularly relevant to transient gas storage.

For R1234yf and R1234ze(E), EOS-induced density differences also affect the
density-dependent viscosity, despite using the same transport-property
correlation for all EOS. This additional coupling is absent for air, for
which the isothermal Sutherland relation depends only on temperature.

\subsection{Propagation into the stationary bearing response}
\label{sec:thermodynamic_stationary}

The different sensitivities of stationary load capacity and supplied mass
flow observed in Sec.~\ref{sec:results_stationary} follow directly from
the manner in which the thermodynamic state enters the governing mass
fluxes.

For pressure-driven flow, the local Reynolds and Darcy mass fluxes contain
a factor proportional to $\varrho/\eta\,\nabla p$.
The quantity $\varrho/\eta$ can therefore be regarded as a local mass
mobility. Since the stationary calculations are performed with a
non-rotating journal, the Couette contribution vanishes, thereby isolating
the pressure-driven transport particularly clearly.

An EOS-induced density deviation acts directly on the transported mass.
For the two refrigerants, the ideal-gas approximation predicts a
substantially lower density than the Helmholtz formulation over much of the
investigated range. This is consistent with the systematic
underprediction of the supplied mass flow reported in
Sec.~\ref{sec:results_stationary}.
The cubic equations progressively recover the real-gas density and,
accordingly, the Helmholtz mass-flow level.

The load capacity responds differently. According to Eq.~\eqref{eq:force_xy},
it is determined by the integrated pressure distribution and is therefore affected by the equation of state only indirectly through the solution of the coupled pressure field.
A considerable error in density does not translate one-to-one into an
equally large pressure or force error because the physical supply and
ambient pressures are prescribed identically for every EOS.

This explains the hierarchy observed in
Tab.~\ref{tab:static_eos_max_error}: the mass flow is substantially more
EOS-sensitive than the load capacity, whereas the maximum pressure is
almost unchanged.
The peak pressure is strongly constrained by the imposed supply pressure
and therefore provides little information about the integrated influence
of the thermodynamic model.
Small EOS-induced modifications of the pressure distribution can affect
the integrated force even when its maximum value remains practically
constant.

The stationary results therefore reveal an important distinction for model
assessment. Agreement in pressure maxima or even in load capacity does not
necessarily imply comparable accuracy in the predicted gas consumption.
For porous aerostatic bearings, the supplied mass flow constitutes a
particularly sensitive indicator of the thermodynamic closure because
density enters the transported mass directly.

\subsection{Propagation into the dynamic bearing response}
\label{sec:thermodynamic_dynamic}

Under transient conditions, the thermodynamic closure additionally affects
gas storage. For an isothermal perturbation about an operating state, the density
perturbation is given by
\begin{align}
    \varrho'
    &=
    \left(
        \frac{\partial\varrho}{\partial p}
    \right)_T p'
    =
    \frac{p'}{p_\varrho},
    \label{eq:density_pressure_perturbation}
\end{align}
Using Eq.~\eqref{eq:chi_T}, the reciprocal pressure--density derivative can
be expressed as
\begin{align}
    \frac{1}{p_\varrho}
    =
    \frac{\varrho\chi_T}{p}.
    \label{eq:storage_compliance_chi}
\end{align}
Thus, $1/p_\varrho$ represents the local isothermal storage compliance.

A first-order linearization of the film and porous-shell storage terms gives
\begin{align}\label{eq:film_shell_storage_linearized}
\begin{split}
    \frac{\partial(\varrho h)}{\partial t}
    &\approx
    \varrho_0\frac{\partial h'}{\partial t}
    +
    \frac{h_0}{p_{\varrho,0}}
    \frac{\partial p'}{\partial t},\\
    \varepsilon_\mathrm{p}
    \frac{\partial\varrho}{\partial t}
    &\approx
    \frac{\varepsilon_\mathrm{p}}{p_{\varrho,0}}
    \frac{\partial p'}{\partial t}.
    \end{split}
\end{align}
The dynamic response therefore depends not only on stationary mass transport
but also explicitly on $p_\varrho$, or equivalently $\chi_T$.

For both refrigerants, the ideal-gas model overpredicts $p_\varrho$ at the
dynamic reference state (cf. Tab.~\ref{tab:thermo_reference_state}) and
therefore underpredicts the corresponding storage compliance. SRK and,
particularly, PR reproduce the Helmholtz value more closely, consistent
with their smaller dynamic-response errors in
Tab.~\ref{tab:max_dynamic_eos_error}.

However, gas storage alone cannot explain the observed frequency dependence.
In harmonic motion, the storage terms scale with excitation frequency and
become less important toward the quasi-static limit. The pronounced
low-frequency deviation of $|H_{xy}|$ in
Tab.~\ref{tab:dynamic_component_error} is therefore unlikely to originate
primarily from transient storage. Instead, it is consistent with a stronger
sensitivity of the cross-coupled force to EOS-dependent pressure
redistribution and circumferential transport. With increasing excitation
frequency, storage becomes more important and the sensitivities of the
direct and cross-coupled responses become more similar.

Accordingly, the non-monotonic behavior of $\varepsilon_{H,y}$ in
Fig.~\ref{fig:eos_error_hvec} is consistent with the coexistence of two
mechanisms. At lower frequencies, EOS-dependent quasi-steady pressure
redistribution and circumferential transport can contribute strongly to the
response, whereas differential compressibility becomes increasingly
important for gas storage in the film and porous shell as the excitation
frequency increases. The observed dynamic EOS error therefore reflects the
combined influence of transport and storage.

The supplied mass flow exhibits a different behavior. Although its mean
level remains EOS dependent, no relevant first harmonic is observed at the
centered operating point. Owing to rotational symmetry, the first-order circumferential variations of
pressure and local supply flux occur with opposite signs on opposing sides of
the bearing and therefore cancel when integrated over the complete
circumference. Local supply fluxes may therefore
oscillate while the total supplied mass flow remains unchanged to first
order, consistent with Tab.~\ref{tab:dynamic_massflow}.

Hence, the thermodynamic closure determines the mean amount of transported
gas, whereas the harmonic force response additionally depends on its
frequency-dependent redistribution and storage.

\subsection{EOS agreement--cost trade-off and practical EOS selection}
\label{sec:eos_assessment}

\begin{figure*}[h]
    \centering
    \includegraphics[width=\textwidth]
        {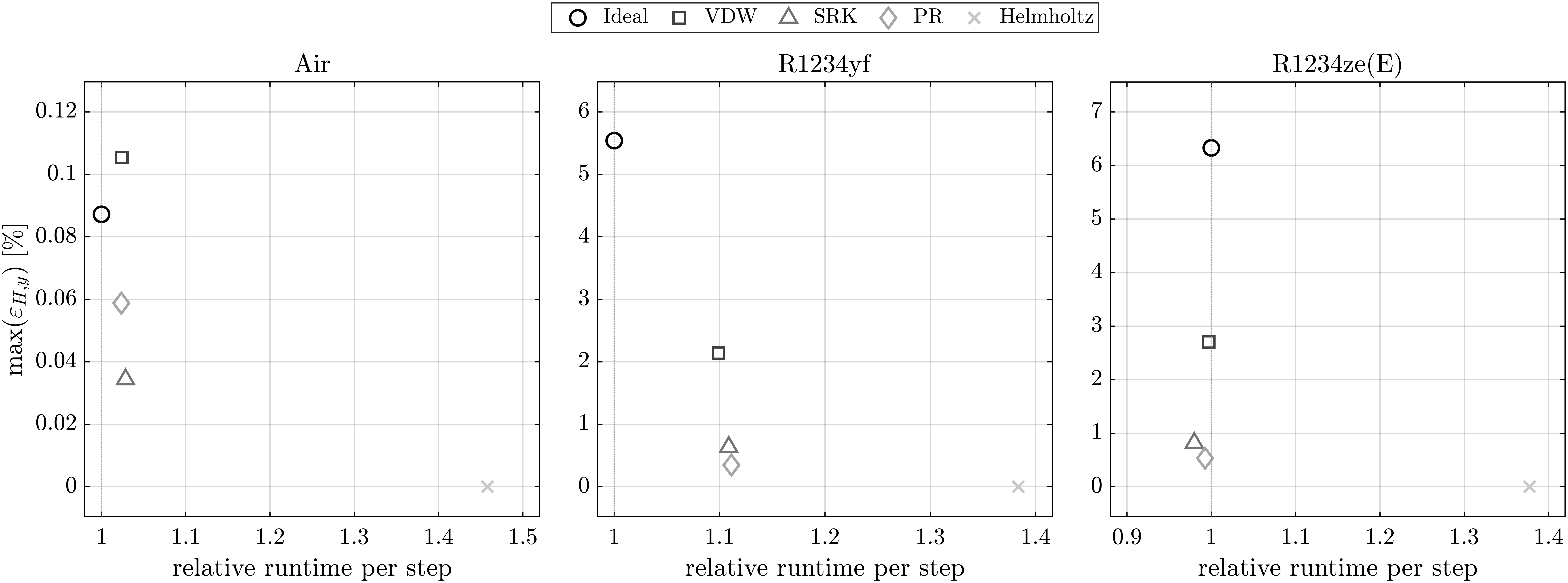}
    \caption{EOS-deviation--cost comparison of the investigated equations of state.
    The ordinate shows the maximum complex dynamic-response deviation from the
    Helmholtz reference, and the abscissa the mean runtime per time step
    normalized by the corresponding ideal-gas calculation.}
    \label{fig:eos_accuracy_cost}
\end{figure*}
The improvement in thermodynamic fidelity must ultimately be considered
together with its computational cost.
Fig.~\ref{fig:eos_accuracy_cost} compares the maximum dynamic response
error with the mean computational effort per time step normalized by the
ideal-gas calculation of the same fluid.

The runtime measurements were obtained using MATLAB R2023b on a workstation
equipped with an Intel Core i9-13900K processor and 64 GB of RAM. All
thermodynamic models were evaluated using the same solver implementation,
discretization, and hardware configuration. The comparison is therefore
intended to quantify the relative computational cost within the present
implementation rather than absolute implementation-independent performance.

The cubic equations exhibit computational costs close to those of the ideal-gas formulation and, more importantly, very similar runtimes among one another.
Consequently, model selection between VDW, SRK, and PR is governed
primarily by accuracy rather than computational cost.
VDW provides no relevant runtime advantage over the more accurate cubic
formulations.
For both refrigerants, PR gives the smallest stationary and dynamic
deviations among the cubic equations without a meaningful increase in
computational effort relative to SRK. For air, SRK yields slightly smaller
deviations than PR, although the differences between all investigated EOS
are negligible.

The multiparameter Helmholtz formulation is computationally more expensive.
Depending on the working fluid, the additional cost of the dynamic
calculations is approximately several tens of percent relative to the
simpler formulations.
This overhead is primarily associated with the more elaborate evaluation
of the residual Helmholtz-energy terms and their derivatives.
It does not, however, alter the Reynolds--Darcy discretization itself.

The resulting model hierarchy is summarized in
Tab.~\ref{tab:eos_recommendation}. While the quantitative deviations depend
on the working fluid, operating state, and bearing configuration, the
underlying trends are expected to remain applicable more generally: the
required EOS fidelity increases with increasing non-ideal thermodynamic
behavior and with the sensitivity of the bearing quantity to density and
differential compressibility.

\begin{table*}[h]
    \centering
    \caption{Practical assessment of the investigated equations of state
    within the single-phase operating range and bearing configuration
    considered in this study.}
    \label{tab:eos_recommendation}
    \renewcommand*{\arraystretch}{1.10}
    \setlength{\tabcolsep}{5pt}
    \footnotesize

    \begin{tabular}{lccccc}
        \hline
        Application / requirement
        & Ideal
        & VDW
        & SRK
        & PR
        & Helmholtz \\
        \hline

        Air in investigated range
        & Suitable
        & Suitable
        & Suitable
        & Suitable
        & Reference \\


        HFO stationary deviation from Helmholtz
        $\lesssim 1\,\%$
        & No
        & No
        & Yes
        & Yes
        & Yes \\

        HFO dynamic response deviation from Helmholtz
        $\lesssim 1\,\%$
        & No
        & No
        & Yes
        & Yes
        & Yes \\

        Smallest EOS-induced deviation
        among cubic models for HFOs
        & --
        & No
        & No
        & Yes
        & -- \\

        Thermodynamic reference calculations
        & No
        & No
        & No
        & No
        & Reference \\

        \hline
    \end{tabular}
\end{table*}

For air, the ideal-gas formulation is therefore sufficient within the
investigated pressure and temperature range.
For air, all EOS-induced bearing-level differences remain below approximately
$0.2\,\%$ within the investigated range, so that the additional real-gas
complexity provides no meaningful benefit for the present application.

For R1234yf and R1234ze(E), the ideal-gas approximation remains useful for
qualitative studies but is not appropriate when quantitatively accurate
predictions are required, for example for gas-supply sizing or for
rotordynamic analyses based on the predicted bearing force coefficients.
VDW improves the prediction but offers an unfavorable accuracy--cost
compromise.

SRK provides a useful reduced-order real-gas model when EOS-induced
deviations of approximately one percent relative to the Helmholtz reference
are acceptable.
PR reduces these deviations further at essentially the same computational
cost and is therefore the preferred cubic formulation for the two
refrigerants considered here.

The multiparameter Helmholtz formulation remains the appropriate reference
when the highest thermodynamic fidelity is required, particularly for
single-phase states approaching the saturation boundary, for model
verification, or when accurate thermodynamic derivatives are of primary
importance.
Its additional computational cost is therefore most justified for
reference calculations and operating regimes in which the residual deviations of the more accurate cubic formulations are relevant.

\section{Summary and conclusions}
\label{sec:conclusions}

The influence of the thermodynamic equation of state on porous aerostatic
journal bearings was investigated using a common Reynolds--Darcy model.
Ideal-gas, VDW, SRK, and PR (using the Stryjek--Vera base correlation) formulations were compared with multiparameter
Helmholtz-energy equations of state for air, R1234yf, and R1234ze(E).

For air, all thermodynamic closures yield practically identical stationary
and dynamic bearing characteristics over the investigated operating range,
confirming the adequacy of the ideal-gas approximation. In contrast, the HFO
refrigerants exhibit appreciable real-gas effects. Under stationary
conditions, the supplied mass flow is considerably more EOS-sensitive than
the load capacity. The maximum ideal-gas deviations from the Helmholtz
reference reach approximately $9.5\,\%$ and $7.9\,\%$ in mass flow for
R1234yf and R1234ze(E), respectively, while the corresponding load deviations
remain below approximately $3.9\,\%$. The nearly unchanged peak pressure
further shows that agreement in maximum pressure does not imply agreement in
integrated bearing quantities.

The dynamic response shows a similarly pronounced EOS dependence for the
refrigerants. The maximum error of the ideal-gas model in the combined
complex response reaches approximately $5.5\,\%$ for R1234yf and
$6.3\,\%$ for R1234ze(E), with particularly strong sensitivity of the
cross-coupled response at low frequencies. 
The observed frequency dependence is consistent with the combined influence
of EOS-dependent mass transport and transient gas storage, the latter being
governed by the pressure--density derivative
$p_\varrho=(\partial p/\partial\varrho)_T$.

Among the simplified real-gas models, VDW improves upon the ideal-gas law,
whereas SRK and PR provide substantially closer agreement with the Helmholtz reference.
PR gives the closest agreement with the Helmholtz reference when using refrigerants, with maximum dynamic-response
deviations of approximately $0.35\,\%$ for R1234yf and $0.53\,\%$ for
R1234ze(E). Since the investigated cubic equations of state have comparable
computational cost, PR offers the most favorable accuracy--cost compromise,
while the Helmholtz formulation remains preferable when the highest
thermodynamic fidelity is required.

Overall, the appropriate EOS depends on the working fluid, operating state,
and quantity of interest. The ideal-gas law is sufficient for the
investigated air-bearing conditions, whereas for R1234yf and R1234ze(E), a real-gas thermodynamic closure is required to limit EOS-induced deviations to the percent level.
Future work should extend the present isothermal, single-phase model toward
non-isothermal conditions, states closer to the phase boundary, and dynamic
response about eccentric operating points.

\newpage
%
\section*{Nomenclature}

\begin{table}[!h]
\caption{Nomenclature -- Latin symbols ($A$--$T$)}
\label{tab:nom_lat_AT}
\centering
\renewcommand*{\arraystretch}{1.0}
\setlength{\tabcolsep}{3pt}
\footnotesize
\begin{tabularx}{\columnwidth}{@{}lXl@{}}
\hline
\multicolumn{3}{@{}l}{\textit{Latin symbols}} \\
\hline
$A$                 & excitation or manufactured-solution amplitude
                    & m, -- \\
$a$                 & attractive parameter of cubic EOS
                    & \si{Pa\,m^6\,mol^{-2}} \\
$b$                 & co-volume parameter of cubic EOS
                    & \si{m^3\,mol^{-1}} \\
$c$                 & radial bearing clearance
                    & m \\
$D_{ij}$            & bearing damping coefficient
                    & \si{N\,s\,m^{-1}} \\
$E$                 & dimensionless dynamic viscosity
                    & -- \\
$F_x,F_y$           & Cartesian bearing-force components
                    & N \\
$F_B$               & bearing load capacity
                    & N \\
$G$                 & dimensionless EOS derivative,
                      $\partial\Pi/\partial P$
                    & -- \\
$H$                 & dimensionless film thickness
                    & -- \\
$H_{ij}$            & complex dynamic stiffness
                    & \si{N\,m^{-1}} \\
$h$                 & local lubrication-film thickness
                    & m \\
$K_{ij}$            & bearing stiffness coefficient
                    & \si{N\,m^{-1}} \\
$\bbm{K}$           & permeability tensor
                    & \si{m^2} \\
$K$                 & homogeneous, isotropic permeability
                    & \si{m^2}\\
$K_0$               & reference permeability
                    & \si{m^2} \\
$L$                 & axial bearing length
                    & m \\
$M$                 & molar mass
                    & \si{kg\,mol^{-1}} \\
$\dot m_{s}$        & supplied mass flow
                    & \si{kg\,s^{-1}} \\
$n_s$               & shaft rotational speed
                    & \si{min^{-1}} \\
$n_i$               & coefficient of Helmholtz-energy formulation
                    & -- \\
$\bbm{n}$           & outward unit normal vector
                    & -- \\
$N_{\mathrm{film}}$ & number of lubrication-film control volumes
                    & -- \\
$N_{\mathrm{por}}$  & number of porous-domain control volumes
                    & -- \\
$N_r$               & number of radial control volumes
                    & --\\
$N_t$               & number of time steps per excitation period
                    & -- \\
$N_z$               & number of axial control volumes
                    & --\\
$N_\varphi$         & number of circumferential control volumes
                    & --\\
$p$                 & pressure
                    & Pa \\
$p_c$               & critical pressure
                    & Pa \\
$p_q$               & mesh convergence order
                    & --\\
$p_{\mathrm{sat}}$  & saturation pressure
                    & Pa \\
$p_t$               & temporal convergence order
                    & --\\
$p_\varrho$         & isothermal pressure--density derivative,
                      $(\partial p/\partial\varrho)_T$
                    & \si{m^2\,s^{-2}} \\
$p$                 & reference pressure scale
                    & Pa \\
$P$                 & dimensionless density
                    & -- \\
$q$                 & generic quantity
                    & -- \\
$R$                 & universal gas constant
                    & \si{J\,mol^{-1}\,K^{-1}} \\
$R_b$               & inner radius of porous bearing shell
                    & m \\
$R_s$               & shaft radius
                    & m \\
$R_{\mathrm{spc}}$   & specific gas constant, $R/M$
                    & \si{J\,kg^{-1}\,K^{-1}} \\
$Re_K$              & permeability-based Reynolds number
                    & -- \\
$r$                 & radial coordinate
                    & m \\
$\bar r$            & dimensionless radial coordinate
                    & -- \\
$T$                 & temperature
                    & K \\
$T_c$               & critical temperature
                    & K \\
$T_{\mathrm{eval}}$ & evaluation interval
                    & s \\
$T_r$               & reducing temperature
                    & K \\
$Ta$                & Taylor number
                    & -- \\
$t$                 & time
                    & s \\
$t_0$               & characteristic time scale
                    & s \\
$t_p$               & porous-layer thickness
                    & m \\
$U$                 & dimensionless mean circumferential velocity
                    & -- \\
$u_m$               & mean circumferential surface velocity
                    & \si{m\,s^{-1}} \\
\hline
\end{tabularx}
\end{table}

\begin{table}[!h]
\caption{Nomenclature -- Latin symbols ($U$--$Z$)}
\label{tab:nom_lat_AT}
\centering
\renewcommand*{\arraystretch}{1.0}
\setlength{\tabcolsep}{3pt}
\footnotesize
\begin{tabularx}{\columnwidth}{@{}lXl@{}}
\hline
\multicolumn{3}{@{}l}{\textit{Latin symbols}} \\
\hline
$u_D,v_D,w_D$       & circumferential, axial, and radial Darcy velocities
                    & \si{m\,s^{-1}} \\
$V_\mathrm{pore}$   & volume of the gas in the porous shell
                    & $\mathrm{m}^3$ \\
$V_\mathrm{total}$  & total volume of the porous shell
                    & $\mathrm{m}^3$ \\
$\bbm{v}_D$        & superficial Darcy-velocity vector
                    & \si{m\,s^{-1}} \\
$x,y$               & Cartesian journal coordinates
                    & m \\
$\ov{z}$                 & dimensionless axial coordinate
                    & -- \\
$Z$                 & compressibility factor
                    & -- \\
$z$                 & axial coordinate
                    & m \\
\hline
\end{tabularx}
\end{table}

\begin{table}[!h]
\caption{Nomenclature -- Greek and other symbols}
\label{tab:nom_greek}
\centering
\renewcommand*{\arraystretch}{1.0}
\setlength{\tabcolsep}{3pt}
\footnotesize
\begin{tabularx}{\columnwidth}{@{}lXl@{}}
\hline
\multicolumn{3}{@{}l}{\textit{Greek symbols}} \\
\hline
$\alpha$            & dimensionless molar Helmholtz energy
                    & -- \\
$\alpha^0$          & ideal-gas contribution to Helmholtz energy
                    & -- \\
$\alpha^r$          & residual contribution to Helmholtz energy
                    & -- \\
$\Gamma_f$          & geometric face-flux coefficient
                    & -- \\
$\delta$            & reduced density
                    & -- \\
$\Delta$            & difference or deviation
                    & -- \\
$\Delta\tau$        & dimensionless time-step size
                    & -- \\
$\varepsilon$       & journal eccentricity ratio
                    & -- \\
$\varepsilon_p$     & porosity of porous material
                    & -- \\
$\varepsilon_{H,y}$ & relative complex-response error
                    & \% \\
$\eta$               & dynamic viscosity
                    & \si{Pa\,s} \\
$\kappa_T$          & isothermal compressibility
                    & \si{Pa^{-1}} \\
$\kappa_{r,\varphi,z}$
                    & dimensionless permeability components
                    & -- \\
$\Lambda_z$         & axial Reynolds parameter
                    & -- \\
$\Lambda_\Omega$    & Couette parameter
                    & -- \\
$\Lambda_\kappa$    & porous-supply parameter
                    & -- \\
$\mathcal{M}_f$     & face density--viscosity mobility
                    & -- \\
$\mathcal{P}_{r,\varphi,z}$ & porous-domain transport parameter 
                    & -- \\
$\Pi$               & dimensionless pressure
                    & -- \\
$\rho_{\mathrm{mol}}$
                    & molar density
                    & \si{mol\,m^{-3}} \\
$\varrho$           & mass density
                    & \si{kg\,m^{-3}} \\
$\varrho_{\mathrm{r,mol}}$
                    & reducing molar density
                    & \si{mol\,m^{-3}} \\
$\tau$              & dimensionless time
                    & --\\
$\tau_H$            & inverse reduced temperature, $T_r/T$
                    & -- \\
$\Theta$            & dimensionless temperature
                    & -- \\
$\varphi$           & circumferential coordinate
                    & rad \\
$\chi_T$            & dimensionless isothermal compressibility
                    & -- \\
$\Omega$            & shaft angular velocity
                    & \si{rad\,s^{-1}} \\
$\omega_{\mathrm{ac}}$
                    & acentric factor
                    & -- \\
$\omega_\mathrm{ex}$
                    & excitation angular frequency
                    & \si{rad\,s^{-1}} \\
\hline
\end{tabularx}
\end{table}

\begin{table}[!h]
\caption{Nomenclature -- Abbreviations}
\label{tab:nom_abbreviations}
\centering
\renewcommand*{\arraystretch}{1.0}
\setlength{\tabcolsep}{3pt}
\footnotesize
\begin{tabularx}{\columnwidth}{@{}lX@{}}
\hline
\multicolumn{2}{@{}l}{\textit{Abbreviations}} \\
\hline
EOS        & equation of state \\
FVM        & finite-volume method \\
GCI        & grid-convergence index\\
HFO        & hydrofluoroolefin \\
MMS        & method of manufactured solutions \\
PR         & Peng--Robinson with Stryjek--Vera base \\
PRSV       & full Peng--Robinson--Stryjek--Vera \\
SRK        & Soave--Redlich--Kwong \\
VDW        & van der Waals \\
\hline
\end{tabularx}
\end{table}


\appendix

\section{Parameters of the multiparameter Helmholtz equations of state}\label{app:helmholtz_parameters}
\setcounter{table}{0}
\renewcommand{\thetable}{A.\arabic{table}}

\setcounter{equation}{0}
\renewcommand{\theequation}{A.\arabic{equation}}
For reproducibility, the parameters of the multiparameter Helmholtz-energy
equations of state employed in the present calculations are summarized
below. The parameter sets correspond to the formulations of
\textsc{Lemmon} et al.~\cite{Lemmon2000} for dry air,
\textsc{Lemmon} and \textsc{Akasaka}~\cite{Lemmon2022} for R1234yf,
and \textsc{Thol} and \textsc{Lemmon}~\cite{Thol2016} for R1234ze(E).

Only the residual Helmholtz contribution is required for the pressure
$p(\varrho,T)$ and the isothermal derivative $p_\varrho$ employed in the
present Reynolds--Darcy formulation. Parameters of the ideal Helmholtz
contribution required for caloric properties are therefore not listed.

The individual residual terms correspond to
\begin{align}
    f_i^{\mathcal P}
    &=
    n_i \delta^{d_i}\tau_\mathrm{H}^{t_i},
    \\
    f_i^{\mathcal E}
    &=
    n_i \delta^{d_i}\tau_\mathrm{H}^{t_i}
    \exp\left(-\delta^{l_i}\right),
    \\
    f_i^{\mathcal G}
    &=
    n_i \delta^{d_i}\tau_\mathrm{H}^{t_i}
    \exp\left[
        -\eta_i(\delta-\epsilon_i)^2
        -\beta_i(\tau_\mathrm{H}-\gamma_i)^2
    \right],
\end{align}
where $\mathcal P$, $\mathcal E$, and $\mathcal G$ denote polynomial,
exponentially damped, and Gaussian terms, respectively.

\begin{table}[!h]
    \centering
    \caption{Reducing parameters and fluid constants used for the
    multiparameter Helmholtz equations of state with $N$ number of terms.}
    \label{tab:helmholtz_general_parameters}
    \renewcommand*{\arraystretch}{1.15}
    \setlength{\tabcolsep}{3pt}
    \footnotesize
    \begin{tabular}{lccccc}
        \hline
        Fluid
        & $R$
        & $M$
        & $T_r$
        & $\varrho_{\mathrm{r,mol}}$
        & $N$
        \\
        & $\mathrm{J/(mol\,K)}$
        & $\mathrm{kg/mol}$
        & $\mathrm{K}$
        & $\mathrm{mol/m^3}$
        & --
        \\
        \hline
        Air
        & 8.314462618
        & 0.02896546
        & 132.6312
        & 10447.7
        & 19
        \\
        R1234yf
        & 8.314462618
        & 0.1140416
        & 367.85
        & 4180
        & 17
        \\
        R1234ze(E)
        & 8.314472
        & 0.1140415928
        & 382.513
        & 4290
        & 16
        \\
        \hline
    \end{tabular}
\end{table}

\begin{table}[!h]
    \centering
    \caption{Residual Helmholtz-energy coefficients for R1234yf according
    to \textsc{Lemmon} and \textsc{Akasaka}~\cite{Lemmon2022}.}
    \label{tab:helmholtz_r1234yf}
    \renewcommand*{\arraystretch}{1.2}
    \setlength{\tabcolsep}{3pt}
    \scriptsize
    \begin{tabular}{rcrrrrrrrr}
        \hline
        $i$ & Type & $n_i$ & $d_i$ & $t_i$ & $l_i$
            & $\eta_i$ & $\beta_i$ & $\gamma_i$ & $\epsilon_i$ \\
        \hline
         1 & $\mathcal P$ &  0.0340387         & 4 & 1.000 & -- & --   & --     & --    & --    \\
         2 & $\mathcal P$ &  1.912859          & 1 & 0.222 & -- & --   & --     & --    & --    \\
         3 & $\mathcal P$ & -2.1984719881617   & 1 & 0.610 & -- & --   & --     & --    & --    \\
         4 & $\mathcal P$ & -0.8147645305033   & 2 & 1.122 & -- & --   & --     & --    & --    \\
         5 & $\mathcal P$ &  0.2288282         & 3 & 0.500 & -- & --   & --     & --    & --    \\

         6 & $\mathcal E$ & -1.65436           & 1 & 2.140 & 2  & --   & --     & --    & --    \\
         7 & $\mathcal E$ & -1.446628          & 3 & 2.284 & 2  & --   & --     & --    & --    \\
         8 & $\mathcal E$ &  0.5961722         & 2 & 1.050 & 1  & --   & --     & --    & --    \\
         9 & $\mathcal E$ & -0.6148209         & 2 & 2.330 & 2  & --   & --     & --    & --    \\
        10 & $\mathcal E$ & -0.0180069         & 7 & 0.778 & 1  & --   & --     & --    & --    \\

        11 & $\mathcal G$ &  0.2845275         & 1 & 1.000 & -- & 28.1 & 1016.0 & 1.061 & 0.960 \\
        12 & $\mathcal G$ & -0.3050809         & 1 & 2.000 & -- & 28.0 & 1000.0 & 1.062 & 0.960 \\
        13 & $\mathcal G$ &  2.198935          & 1 & 1.236 & -- & 1.307& 1.16   & 1.300 & 0.853 \\
        14 & $\mathcal G$ & -0.4270329         & 1 & 1.600 & -- & 1.96 & 1.20   & 1.040 & 1.150 \\
        15 & $\mathcal G$ & -0.4015581         & 1 & 1.850 & -- & 1.25 & 1.26   & 1.000 & 1.300 \\
        16 & $\mathcal G$ &  0.1179587         & 1 & 0.700 & -- & 1.00 & 1.60   & 1.140 & 1.500 \\
        17 & $\mathcal G$ & -0.3141002         & 1 & 1.750 & -- & 2.20 & 0.87   & 1.100 & 0.762 \\
        \hline
    \end{tabular}
\end{table}

\begin{table}[!h]
    \centering
    \caption{Residual Helmholtz-energy coefficients for R1234ze(E)
    according to \textsc{Thol} and \textsc{Lemmon}~\cite{Thol2016}.}
    \label{tab:helmholtz_r1234zee}
    \renewcommand*{\arraystretch}{1.2}
    \setlength{\tabcolsep}{3pt}
    \scriptsize
    \begin{tabular}{rcrrrrrrrr}
        \hline
        $i$ & Type & $n_i$ & $d_i$ & $t_i$ & $l_i$
            & $\eta_i$ & $\beta_i$ & $\gamma_i$ & $\epsilon_i$ \\
        \hline
         1 & $\mathcal P$ &  0.03982797   & 4 & 1.000 & -- & --   & --    & --    & --    \\
         2 & $\mathcal P$ &  1.812227     & 1 & 0.223 & -- & --   & --    & --    & --    \\
         3 & $\mathcal P$ & -2.537512     & 1 & 0.755 & -- & --   & --    & --    & --    \\
         4 & $\mathcal P$ & -0.5333254    & 2 & 1.240 & -- & --   & --    & --    & --    \\
         5 & $\mathcal P$ &  0.1677031    & 3 & 0.440 & -- & --   & --    & --    & --    \\

         6 & $\mathcal E$ & -1.323801     & 1 & 2.000 & 2  & --   & --    & --    & --    \\
         7 & $\mathcal E$ & -0.6694654    & 3 & 2.200 & 2  & --   & --    & --    & --    \\
         8 & $\mathcal E$ &  0.8072718    & 2 & 1.200 & 1  & --   & --    & --    & --    \\
         9 & $\mathcal E$ & -0.7740229    & 2 & 1.500 & 2  & --   & --    & --    & --    \\
        10 & $\mathcal E$ & -0.01843846   & 7 & 0.900 & 1  & --   & --    & --    & --    \\

        11 & $\mathcal G$ &  1.407916     & 1 & 1.330 & -- & 1.00 & 1.21  & 0.943 & 0.728 \\
        12 & $\mathcal G$ & -0.4237082    & 1 & 1.750 & -- & 1.61 & 1.37  & 0.642 & 0.870 \\
        13 & $\mathcal G$ & -0.2270068    & 3 & 2.110 & -- & 1.24 & 0.98  & 0.590 & 0.855 \\
        14 & $\mathcal G$ & -0.805213     & 3 & 1.000 & -- & 9.34 & 171.0 & 1.200 & 0.790 \\
        15 & $\mathcal G$ &  0.00994318   & 2 & 1.500 & -- & 5.78 & 47.4  & 1.330 & 1.300 \\
        16 & $\mathcal G$ & -0.008798793  & 1 & 1.000 & -- & 3.08 & 15.4  & 0.640 & 0.710 \\
        \hline
    \end{tabular}
\end{table}

\begin{table}[!h]
    \centering
    \caption{Residual Helmholtz-energy coefficients for dry air according
    to \textsc{Lemmon} et al.~\cite{Lemmon2000}.}
    \label{tab:helmholtz_air}
    \renewcommand*{\arraystretch}{1.2}
    \setlength{\tabcolsep}{5pt}
    \footnotesize
    \begin{tabular}{rccrrrr}
        \hline
        $i$ & Type & $n_i$ & $d_i$ & $t_i$ & $l_i$ \\
        \hline
         1 & $\mathcal P$ &  $ 0.118160747229$    &  1 &  0.00 & -- \\
         2 & $\mathcal P$ &  $ 0.713116392079$    &  1 &  0.33 & -- \\
         3 & $\mathcal P$ &  $-1.61824192067$     &  1 &  1.01 & -- \\
         4 & $\mathcal P$ &  $ 0.0714140178971$   &  2 &  0.00 & -- \\
         5 & $\mathcal P$ &  $-0.0865421396646$   &  3 &  0.00 & -- \\
         6 & $\mathcal P$ &  $ 0.134211176704$    &  3 &  0.15 & -- \\
         7 & $\mathcal P$ &  $ 0.0112626704218$   &  4 &  0.00 & -- \\
         8 & $\mathcal P$ &  $-0.0420533228842$   &  4 &  0.20 & -- \\
         9 & $\mathcal P$ &  $ 0.0349008431982$   &  4 &  0.35 & -- \\
        10 & $\mathcal P$ &  $ 0.000164957183186$ &  6 &  1.35 & -- \\
        11 & $\mathcal E$ &  $-0.101365037912$    &  1 &  1.60 & 1 \\
        12 & $\mathcal E$ &  $-0.17381369097$     &  3 &  0.80 & 1 \\
        13 & $\mathcal E$ &  $-0.0472103183731$   &  5 &  0.95 & 1 \\
        14 & $\mathcal E$ &  $-0.0122523554253$   &  6 &  1.25 & 1 \\
        15 & $\mathcal E$ &  $-0.146629609713$    &  1 &  3.60 & 2 \\
        16 & $\mathcal E$ &  $-0.0316055879821$   &  3 &  6.00 & 2 \\
        17 & $\mathcal E$ &  $ 0.000233594806142$ & 11 &  3.25 & 2 \\
        18 & $\mathcal E$ &  $ 0.0148287891978$   &  1 &  3.50 & 3 \\
        19 & $\mathcal E$ &  $-0.00938782884667$  &  3 & 15.00 & 3 \\
        \hline
    \end{tabular}
\end{table}

\section{Parameters of the viscosity correlations}
\label{app:viscosity_parameters}
\setcounter{table}{0}
\renewcommand{\thetable}{B.\arabic{table}}

\setcounter{equation}{0}
\renewcommand{\theequation}{B.\arabic{equation}}
The parameters of the viscosity correlations employed in
Sec.~\ref{sec:transport_properties} are summarized in the following.
For dry air, Sutherland's law is used, whereas the viscosity of R1234yf
and R1234ze(E) is evaluated using the correlations of
\textsc{Huber} and \textsc{Assael}~\cite{Huber2016}.
The coefficient sets reported below correspond to the values used in the
numerical implementation.

\subsection{Dry air}

The dynamic viscosity of dry air is evaluated from
Eq.~\eqref{eq:sutherland}. The parameters used in the present
implementation are listed in Tab.~\ref{tab:viscosity_air_parameters}.

\begin{table}[!h]
    \centering
    \caption{Parameters of Sutherland's viscosity correlation for dry air.}
    \label{tab:viscosity_air_parameters}
    \renewcommand*{\arraystretch}{1.1}
    \setlength{\tabcolsep}{5pt}
    \footnotesize
    \begin{tabular}{lcc}
        \hline
        Parameter & Value & Unit \\
        \hline
        $T_{\mathrm{ref}}$    & 293.15          & $\mathrm{K}$ \\
        $\eta_{\mathrm{ref}}$ & $18.3\times10^{-6}$ & $\mathrm{Pa\,s}$ \\
        $S$                   & 110.4           & $\mathrm{K}$ \\
        \hline
    \end{tabular}
\end{table}

\subsection{HFO refrigerants}
For R1234yf and R1234ze(E), the density-dependent viscosity correlations
of \textsc{Huber} and \textsc{Assael}~\cite{Huber2016} are employed.
Consistent with Eq.~\eqref{eq:hfo_viscosity}, the viscosity is decomposed as
\begin{align}
    \eta(\varrho_{\mathrm{mol}},T)
    =
    \eta_0(T)
    +
    \eta_1(T)\varrho_{\mathrm{mol}}
    +
    \Delta\eta(\varrho_{\mathrm{mol}},T),
    \label{eq:app_hfo_viscosity}
\end{align}
where the critical enhancement is neglected.
The correlation is evaluated internally in $\mu\mathrm{Pa\,s}$ and
subsequently converted to $\mathrm{Pa\,s}$.

The dilute-gas contribution is represented by
\begin{align}
    \eta_0(T)
    =
    \frac{
        a_0+a_1T+a_2T^2+a_3T^3
    }{
        a_4+a_5T+T^2
    },
    \label{eq:app_hfo_eta0}
\end{align}
where $T$ is given in kelvin and $\eta_0$ is obtained in
$\mu\mathrm{Pa\,s}$. The coefficients used for both refrigerants are
listed in Tab.~\ref{tab:viscosity_hfo_dilute}.

\begin{table}[!h]
    \centering
    \caption{Coefficients of the dilute-gas viscosity correlation,
    Eq.~\eqref{eq:app_hfo_eta0}.}
    \label{tab:viscosity_hfo_dilute}
    \renewcommand*{\arraystretch}{1.1}
    \setlength{\tabcolsep}{7pt}
    \footnotesize
    \begin{tabular}{lrr}
        \hline
        Coefficient
        & R1234yf
        & R1234ze(E) \\
        \hline
        $a_0$ & $-836950$   & $-963382$   \\
        $a_1$ & $6336.28$   & $9614.09$   \\
        $a_2$ & $-2.3547$   & $-13.233$   \\
        $a_3$ & $0.0395563$ & $0.0360562$ \\
        $a_4$ & $39509.1$   & $122059$    \\
        $a_5$ & $121.018$   & $-224.741$  \\
        \hline
    \end{tabular}
\end{table}

The initial-density contribution is evaluated using the reduced
temperature
\begin{align}
    T^\star
    =
    \frac{T}{\epsilon_\mathrm{LJ}/k_{\mathrm B}},
    \label{eq:app_hfo_Tstar}
\end{align}
and the reduced second viscosity virial coefficient
\begin{align}
    B_\eta^\star(T^\star)
    =
    \sum_{i=0}^{8}
    b_i
    (T^\star)^{s_i}.
    \label{eq:app_hfo_Bstar}
\end{align}
The corresponding contribution to the viscosity is written in the
implementation as
\begin{align}
    \eta_1(T)\varrho_{\mathrm{mol}}
    =
    \eta_0(T)
    B_\eta^\star(T^\star)
    C_B
    \frac{\varrho_{\mathrm{mol}}}
         {1000},
    \label{eq:app_hfo_eta1}
\end{align}
where $\varrho_{\mathrm{mol}}$ is given in $\mathrm{mol/m^3}$ and
$C_B$ denotes the second-viscosity-virial scaling
factor expressed in $\mathrm{L\,mol^{-1}}$. All fluid-specific scaling parameters
are summarized in Tab.~\ref{tab:viscosity_hfo_scaling}.
The reducing quantities of the viscosity correlation are independent of
the reducing parameters of the Helmholtz EOS and may therefore differ
slightly. The gas-constant values are retained exactly as specified by the respective
reference formulations.

\begin{table}[!h]
    \centering
    \caption{Fluid-specific scaling parameters of the HFO viscosity
    correlations.}
    \label{tab:viscosity_hfo_scaling}
    \renewcommand*{\arraystretch}{1.1}
    \setlength{\tabcolsep}{4pt}
    \footnotesize
    \begin{tabular}{lcc}
        \hline
        Parameter & R1234yf & R1234ze(E) \\
        \hline
        $\epsilon_\mathrm{LJ}/k_{\mathrm B}$ [K]
        & 275.0
        & 340.0 \\

        $C_B$ [$\mathrm{L/mol}$]
        & 0.090164213
        & 0.075276713 \\

        $T_{\eta,r}$ [K]
        & 367.85
        & 382.513 \\

        $\varrho_{\eta,r}$ [$\mathrm{mol/m^3}$]
        & 4170
        & 4290 \\

        $T_{\min}$ [K]
        & 220
        & 169 \\

        $T_{\max}$ [K]
        & 410
        & 420 \\

        $p_{\max}$ [MPa]
        & 30
        & 100 \\
        \hline
    \end{tabular}
\end{table}

The coefficients $b_i$ and exponents $s_i$ are identical for both
refrigerants and are given in
Tab.~\ref{tab:viscosity_hfo_Bstar}.

\begin{table}[!h]
    \centering
    \caption{Coefficients of the reduced second viscosity virial
    correlation, Eq.~\eqref{eq:app_hfo_Bstar}.}
    \label{tab:viscosity_hfo_Bstar}
    \renewcommand*{\arraystretch}{1.05}
    \setlength{\tabcolsep}{7pt}
    \footnotesize
    \begin{tabular}{rrr}
        \hline
        $i$ & $b_i$ & $s_i$ \\
        \hline
        0 & $-19.572881$  & $0$     \\
        1 & $219.73999$   & $-0.25$ \\
        2 & $-1015.3226$  & $-0.50$ \\
        3 & $2471.01251$  & $-0.75$ \\
        4 & $-3375.1717$  & $-1.00$ \\
        5 & $2491.6597$   & $-1.25$ \\
        6 & $-787.26086$  & $-1.50$ \\
        7 & $14.085455$   & $-2.50$ \\
        8 & $-0.34664158$ & $-5.50$ \\
        \hline
    \end{tabular}
\end{table}

For the residual higher-density contribution, the reduced variables
\begin{align}
    T_\eta
    &=
    \frac{T}{T_{\eta,r}},
    &
    \delta_\eta
    &=
    \frac{\varrho_{\mathrm{mol}}}
         {\varrho_{\eta,r}}
    \label{eq:app_hfo_reduced}
\end{align}
are introduced. To employ the same numerical routine for both HFOs, the
residual correlation is implemented in the generic rational form
\begin{align}
    \Delta\eta
    =
    \frac{
        \displaystyle
        \sum_{i=1}^{N_c}
        c_i
        T_\eta^{\,t_i}
        \delta_\eta^{\,d_i/3}
    }{
        \displaystyle
        \sum_{j=1}^{N_q}
        q_j
        T_\eta^{\,u_j}
        \delta_\eta^{\,v_j}
    }.
    \label{eq:app_hfo_residual}
\end{align}
This expression is algebraically equivalent to the fluid-specific
residual correlations of \textsc{Huber} and
\textsc{Assael}~\cite{Huber2016}, but allows both fluids to be evaluated
using the same implementation. The numerical coefficients in
Tabs.~\ref{tab:viscosity_hfo_res_yf} and
\ref{tab:viscosity_hfo_res_ze} are the coefficients stored directly in
the numerical model.

\begin{table}[!h]
    \centering
    \caption{Residual-viscosity coefficients for R1234yf in
    Eq.~\eqref{eq:app_hfo_residual}.}
    \label{tab:viscosity_hfo_res_yf}
    \renewcommand*{\arraystretch}{1.05}
    \setlength{\tabcolsep}{5pt}
    \footnotesize
    \begin{tabular}{rrrr}
        \hline
        $i$ & $c_i$ & $t_i$ & $d_i$ \\
        \hline
        1 & $ 0.68706100665$ & 1.5 & 2 \\
        2 & $ 7.1608477411$  & 1.5 & 5 \\
        3 & $-2.079577245$   & 1.5 & 8 \\
        4 & $-43.47027288$   & 0.5 & 5 \\
        \hline
    \end{tabular}

    \vspace{2mm}

    \begin{tabular}{rrrr}
        \hline
        $j$ & $q_j$ & $u_j$ & $v_j$ \\
        \hline
        1 & $-3.53682791$ & 1.0 & 0 \\
        2 & $ 1.0$        & 1.0 & 1 \\
        \hline
    \end{tabular}
\end{table}

\begin{table}[!h]
    \centering
    \caption{Residual-viscosity coefficients for R1234ze(E) in
    Eq.~\eqref{eq:app_hfo_residual}.}
    \label{tab:viscosity_hfo_res_ze}
    \renewcommand*{\arraystretch}{1.05}
    \setlength{\tabcolsep}{5pt}
    \footnotesize
    \begin{tabular}{rrrr}
        \hline
        $i$ & $c_i$ & $t_i$ & $d_i$ \\
        \hline
        1 & $ 8.61691913$    &  0.5 &  2 \\
        2 & $20.83024738$    &  0.5 &  8 \\
        3 & $ 0.5424369$     & -0.5 & 20 \\
        4 & $-10.49684841$   &  1.5 &  5 \\
        5 & $-1.38137689$    &  0.5 & 17 \\
        \hline
    \end{tabular}

    \vspace{2mm}

    \begin{tabular}{rrrr}
        \hline
        $j$ & $q_j$ & $u_j$ & $v_j$ \\
        \hline
        1 & $1.0$ & 0.0 & 0 \\
        \hline
    \end{tabular}
\end{table}

For R1234ze(E), the denominator of
Eq.~\eqref{eq:app_hfo_residual} is therefore identically unity.
The rational representation is retained nevertheless so that both
refrigerants can be evaluated by the same transport-property routine.
Finally, the viscosity returned by
Eqs.~\eqref{eq:app_hfo_eta0}--\eqref{eq:app_hfo_residual} is converted
according to $\eta~[\mathrm{Pa\,s}] = 10^{-6} \eta~[\mu\mathrm{Pa\,s}]$.

\newpage









\bibliographystyle{elsarticle-num} 
\bibliography{literatur}

@Article{Sun1975,
  author    = {Sun, Dah-chen},
  journal   = {Journal of Lubrication Technology},
  title     = {Analysis of the Steady State Characteristics of Gas-Lubricated, Porous Journal Bearings},
  year      = {1975},
  month     = {Jan},
  number    = {1},
  pages     = {44--51},
  volume    = {97},
  doi       = {10.1115/1.3452533},
  publisher = {{ASME} International},
}

@PhdThesis{Buuren2014,
  author    = {van Buuren, Sietze},
  school    = {Karlsruher Instituts f\"ur Technologie},
  title     = {Modeling and simulation of porous journal bearings in multibody systems},
  year      = {2014},
  date      = {2014-07-30},
  ean       = {9783731500841},
  isbn      = {3731500841},
  pagetotal = {180},
  publisher = {Karlsruher Institut f\"ur Technologie},
}

@Article{Guenat2019,
  author    = {Eliott Guenat and J\"urg Schiffmann},
  journal   = {Journal of Tribology},
  title     = {Thin Gas Film Isothermal Condensation in Aerodynamic Bearings},
  year      = {2019},
  month     = {Sep},
  number    = {11},
  volume    = {141},
  doi       = {10.1115/1.4044447},
  publisher = {{ASME} International},
}

@Article{Kwan1998,
  author    = {Y.B.P Kwan and J Corbett},
  month     = {Nov},
  title     = {Porous aerostatic bearings{\textendash}an updated review},
  year      = {1998},
  doi       = {10.1016/s0043-1648(98)00285-3},
  journal   = {Wear},
  number    = {2},
  pages     = {69--73},
  publisher = {Elsevier {BV}},
  volume    = {222},
}

@Book{Ferziger2020b,
  author    = {Joel H. Ferziger and Milovan Peri{\'{c}} and Robert L. Street},
  publisher = {Springer International Publishing},
  title     = {Computational Methods for Fluid Dynamics},
  year      = {2020},
  doi       = {10.1007/978-3-319-99693-6},
}

@Article{Gu2022,
  author    = {Yandong Gu and Jinwu Cheng and Chaojie Xie and Longyu Li and Changgeng Zheng},
  journal   = {Machines},
  title     = {Theoretical and Numerical Investigations on Static Characteristics of Aerostatic Porous Journal Bearings},
  year      = {2022},
  month     = {Feb},
  number    = {3},
  pages     = {171},
  volume    = {10},
  doi       = {10.3390/machines10030171},
  publisher = {{MDPI} {AG}},
}

@Article{Whitaker1986,
  author    = {Stephen Whitaker},
  journal   = {Transport in Porous Media},
  title     = {Flow in porous media I: A theoretical derivation of Darcy{\textquotesingle}s law},
  year      = {1986},
  number    = {1},
  pages     = {3--25},
  volume    = {1},
  doi       = {10.1007/bf01036523},
  publisher = {Springer Science and Business Media {LLC}},
}

@Book{Darcy1856,
  author    = {Henry Darcy},
  publisher = {Les fontaines publiques de la ville de Dijon},
  title     = {The flow of homogeneous fluids through porous media},
  year      = {1856},
}

@Article{Liu2022,
  author    = {Hao Liu and Ying Xu},
  journal   = {Materials Research Express},
  title     = {Gas permeability measurement in porous graphite under steady-state flow},
  year      = {2022},
  month     = {Feb},
  number    = {2},
  pages     = {025603},
  volume    = {9},
  doi       = {10.1088/2053-1591/ac50d7},
  publisher = {{IOP} Publishing},
}

@Article{Guenat2018,
  author    = {Guenat, Eliott and Schiffmann, Jürg},
  journal   = {Tribology International},
  title     = {Real-gas effects on aerodynamic bearings},
  year      = {2018},
  issn      = {0301-679X},
  month     = apr,
  pages     = {358--368},
  volume    = {120},
  doi       = {10.1016/j.triboint.2018.01.008},
  publisher = {Elsevier BV},
}

@Article{Redlich1949,
  author    = {Redlich, Otto. and Kwong, J. N. S.},
  journal   = {Chemical Reviews},
  title     = {On the Thermodynamics of Solutions. V. An Equation of State. Fugacities of Gaseous Solutions.},
  year      = {1949},
  issn      = {1520-6890},
  month     = feb,
  number    = {1},
  pages     = {233--244},
  volume    = {44},
  doi       = {10.1021/cr60137a013},
  publisher = {American Chemical Society (ACS)},
}

@Article{Rao1978,
  author    = {Rao, N. S. and Majumdar, B. C.},
  journal   = {Journal of Mechanical Engineering Science},
  title     = {Dynamic Stiffness and Damping Coefficients of Aerostatic, Porous, Journal Bearings},
  year      = {1978},
  issn      = {2058-3389},
  month     = oct,
  number    = {5},
  pages     = {291--296},
  volume    = {20},
  doi       = {10.1243/jmes_jour_1978_020_049_02},
  publisher = {SAGE Publications},
}

@Book{Moukalled2016,
  author    = {Moukalled, F. and Mangani, L. and Darwish, M.},
  publisher = {Springer International Publishing},
  title     = {The Finite Volume Method in Computational Fluid Dynamics: An Advanced Introduction with OpenFOAM® and Matlab},
  year      = {2016},
  isbn      = {9783319168746},
  doi       = {10.1007/978-3-319-16874-6},
  issn      = {2215-0056},
  journal   = {Fluid Mechanics and Its Applications},
}

@Article{SanAndres2021,
  author    = {Andrés, Luis San and Yang, Jing and Devitt, Andrew},
  journal   = {Applied Sciences},
  title     = {Porous Gas Journal Bearings: An Exact Solution Revisited and Force Coefficients for Stable Rotordynamic Performance},
  year      = {2021},
  issn      = {2076-3417},
  month     = Aug,
  number    = {17},
  pages     = {7949},
  volume    = {11},
  doi       = {10.3390/app11177949},
  publisher = {MDPI AG},
}

@PhdThesis{Waals1873,
  author  = {Johannes Diderik van der Waals},
  school  = {University of Leiden},
  title   = {Over de Continuiteit van den Gas- en Vloeistoftoestand},
  year    = {1873},
  address = {Leiden},
}

@Article{Soave1972,
  author    = {Soave, Giorgio},
  journal   = {Chemical Engineering Science},
  title     = {Equilibrium constants from a modified Redlich-Kwong equation of state},
  year      = {1972},
  issn      = {0009-2509},
  month     = June,
  number    = {6},
  pages     = {1197--1203},
  volume    = {27},
  doi       = {10.1016/0009-2509(72)80096-4},
  publisher = {Elsevier BV},
}

@Article{Lund1968,
  author    = {Lund, J. W.},
  journal   = {Journal of Lubrication Technology},
  title     = {Calculation of Stiffness and Damping Properties of Gas Bearings},
  year      = {1968},
  issn      = {0022-2305},
  month     = Oct,
  number    = {4},
  pages     = {793--803},
  volume    = {90},
  doi       = {10.1115/1.3601723},
  publisher = {ASME International},
}

@Article{Lund1987,
  author    = {Lund, J. W.},
  journal   = {Journal of Tribology},
  title     = {Review of the Concept of Dynamic Coefficients for Fluid Film Journal Bearings},
  year      = {1987},
  issn      = {1528-8897},
  month     = Jan,
  number    = {1},
  pages     = {37--41},
  volume    = {109},
  doi       = {10.1115/1.3261324},
  publisher = {ASME International},
}

@Article{Snyder2019,
  author    = {Snyder, Troy and Braun, Minel},
  journal   = {Lubricants},
  title     = {A CFD-Based Frequency Response Method Applied in the Determination of Dynamic Coefficients of Hydrodynamic Bearings. Part 1: Theory},
  year      = {2019},
  issn      = {2075-4442},
  month     = Mar,
  number    = {3},
  pages     = {23},
  volume    = {7},
  doi       = {10.3390/lubricants7030023},
  publisher = {MDPI AG},
}

@Article{Rao1977,
  author    = {Rao, N. S.},
  journal   = {Journal of Lubrication Technology},
  title     = {Analysis of the Stiffness and Damping Characteristics of an Externally Pressurized Porous Gas Journal Bearing},
  year      = {1977},
  issn      = {0022-2305},
  month     = Apr,
  number    = {2},
  pages     = {295--301},
  volume    = {99},
  doi       = {10.1115/1.3453085},
  publisher = {ASME International},
}

@Article{Stellmach2026a,
  author    = {Stellmach, Laurenz and Nitzschke, Steffen and Niemeyer, Mascha and Woschke, Elmar},
  title     = {Transient gas storage in porous aerostatic journal bearings and the limits of the quasi-steady approximation},
  year      = {2026},
  doi       = {10.2139/ssrn.7550203},
  publisher = {Elsevier BV},
}

@Article{Garcia2013,
  author    = {Garcia, Mathieu and Bou-Saïd, Benyebka and Rocchi, Jérôme and Grau, Grégory},
  journal   = {Tribology International},
  title     = {Refrigerant foil bearing behavior—A Thermo-HydroDynamic study},
  year      = {2013},
  issn      = {0301-679X},
  month     = {Sept},
  pages     = {363--369},
  volume    = {65},
  doi       = {10.1016/j.triboint.2012.12.006},
  publisher = {Elsevier BV},
}

@Article{Bouchehit2016,
  author    = {Bouchehit, Bachir and Bou-Saïd, Benyebka and Garcia, Mathieu},
  journal   = {Tribology International},
  title     = {Static and dynamic performances of refrigerant-lubricated bearings},
  year      = {2016},
  issn      = {0301-679X},
  month     = Apr,
  pages     = {326--348},
  volume    = {96},
  doi       = {10.1016/j.triboint.2015.12.035},
  publisher = {Elsevier BV},
}

@Article{Conboy2013,
  author    = {Conboy, T. M.},
  journal   = {Journal of Tribology},
  title     = {Real-Gas Effects in Foil Thrust Bearings Operating in the Turbulent Regime},
  year      = {2013},
  issn      = {1528-8897},
  month     = Mar,
  number    = {3},
  volume    = {135},
  doi       = {10.1115/1.4024048},
  publisher = {ASME International},
}

@Article{Kim2015,
  author    = {Kim, Daejong},
  journal   = {Journal of Engineering for Gas Turbines and Power},
  title     = {Design Space of Foil Bearings for Closed-Loop Supercritical CO2 Power Cycles Based on Three-Dimensional Thermohydrodynamic Analyses},
  year      = {2015},
  issn      = {1528-8919},
  month     = Oct,
  number    = {3},
  volume    = {138},
  doi       = {10.1115/1.4031433},
  publisher = {ASME International},
}

@Article{deWaart2024,
  author    = {de Waart, Wessel and Pini, Matteo},
  journal   = {Journal of Engineering for Gas Turbines and Power},
  title     = {The Role of the Working Fluid and Nonideal Thermodynamic Effects on Performance of Gas-Lubricated Bearings},
  year      = {2025},
  issn      = {1528-8919},
  month     = Dec,
  number    = {6},
  volume    = {147},
  doi       = {10.1115/1.4066822},
  publisher = {ASME International},
}

@Article{Peng1976,
  author    = {Peng, Ding-Yu and Robinson, Donald B.},
  journal   = {Industrial \& Engineering Chemistry Fundamentals},
  title     = {A New Two-Constant Equation of State},
  year      = {1976},
  issn      = {1541-4833},
  month     = Feb,
  number    = {1},
  pages     = {59--64},
  volume    = {15},
  doi       = {10.1021/i160057a011},
  publisher = {American Chemical Society (ACS)},
}

@Article{Lemmon2022,
  author    = {Lemmon, Eric W. and Akasaka, Ryo},
  journal   = {International Journal of Thermophysics},
  title     = {An International Standard Formulation for 2,3,3,3-Tetrafluoroprop-1-ene (R1234yf) Covering Temperatures from the Triple Point Temperature to 410 K and Pressures Up to 100 MPa},
  year      = {2022},
  issn      = {1572-9567},
  month     = June,
  number    = {8},
  volume    = {43},
  doi       = {10.1007/s10765-022-03015-y},
  publisher = {Springer Science and Business Media LLC},
}

@Article{Thol2016,
  author    = {Thol, Monika and Lemmon, Eric W.},
  journal   = {International Journal of Thermophysics},
  title     = {Equation of State for the Thermodynamic Properties of trans-1,3,3,3-Tetrafluoropropene [R-1234ze(E)]},
  year      = {2016},
  issn      = {1572-9567},
  month     = Jan,
  number    = {3},
  volume    = {37},
  doi       = {10.1007/s10765-016-2040-6},
  publisher = {Springer Science and Business Media LLC},
}

@Article{Huber2016,
  author    = {Huber, Marcia L. and Assael, Marc J.},
  journal   = {International Journal of Refrigeration},
  title     = {Correlations for the viscosity of 2,3,3,3-tetrafluoroprop-1-ene (R1234yf) and trans-1,3,3,3-tetrafluoropropene (R1234ze(E))},
  year      = {2016},
  issn      = {0140-7007},
  month     = Nov,
  pages     = {39--45},
  volume    = {71},
  doi       = {10.1016/j.ijrefrig.2016.08.007},
  publisher = {Elsevier BV},
}

@Article{Bell2014,
  author    = {Bell, Ian H. and Wronski, Jorrit and Quoilin, Sylvain and Lemort, Vincent},
  journal   = {Industrial \& Engineering Chemistry Research},
  title     = {Pure and Pseudo-pure Fluid Thermophysical Property Evaluation and the Open-Source Thermophysical Property Library CoolProp},
  year      = {2014},
  issn      = {1520-5045},
  month     = Jan,
  number    = {6},
  pages     = {2498--2508},
  volume    = {53},
  doi       = {10.1021/ie4033999},
  publisher = {American Chemical Society (ACS)},
}

@Article{Bechiri2026,
  author    = {Bechiri, S. and Bouchehit, B. and Bou‐Saïd, B.},
  journal   = {Lubrication Science},
  title     = {Static and Dynamic Analysis of Porous Gas Bearings Lubricated With Air, Refrigerant R‐134a, Helium and Hydrogen},
  year      = {2026},
  issn      = {1557-6833},
  month     = Mar,
  number    = {5},
  pages     = {301--326},
  volume    = {38},
  doi       = {10.1002/ls.70032},
  publisher = {Wiley},
}

@Article{Gao2019,
  author    = {Gao, Qiang and Chen, Wanqun and Lu, Lihua and Huo, Dehong and Cheng, Kai},
  journal   = {Tribology International},
  title     = {Aerostatic bearings design and analysis with the application to precision engineering: State-of-the-art and future perspectives},
  year      = {2019},
  issn      = {0301-679X},
  month     = July,
  pages     = {1--17},
  volume    = {135},
  doi       = {10.1016/j.triboint.2019.02.020},
  publisher = {Elsevier BV},
}

@Article{Lemmon2000,
  author    = {Lemmon, Eric W. and Jacobsen, Richard T and Penoncello, Steven G. and Friend, Daniel G.},
  journal   = {Journal of Physical and Chemical Reference Data},
  title     = {Thermodynamic Properties of Air and Mixtures of Nitrogen, Argon, and Oxygen From 60 to 2000 K at Pressures to 2000 MPa},
  year      = {2000},
  issn      = {1529-7845},
  month     = May,
  number    = {3},
  pages     = {331--385},
  volume    = {29},
  doi       = {10.1063/1.1285884},
  publisher = {AIP Publishing},
}

@Article{Stryjek1986,
  author    = {Stryjek, R. and Vera, J. H.},
  journal   = {The Canadian Journal of Chemical Engineering},
  title     = {PRSV: An improved peng—Robinson equation of state for pure compounds and mixtures},
  year      = {1986},
  issn      = {1939-019X},
  month     = Apr,
  number    = {2},
  pages     = {323--333},
  volume    = {64},
  doi       = {10.1002/cjce.5450640224},
  publisher = {Wiley},
}

@Article{Sutherland1893,
  author    = {Sutherland, William},
  journal   = {The London, Edinburgh, and Dublin Philosophical Magazine and Journal of Science},
  title     = {LII. The viscosity of gases and molecular force},
  year      = {1893},
  issn      = {1941-5990},
  month     = Dec,
  number    = {223},
  pages     = {507--531},
  volume    = {36},
  doi       = {10.1080/14786449308620508},
  publisher = {Informa UK Limited},
}

@Article{Celik2008,
  author  = {Celik, Ismail B. and Ghia, Urmila and Roache, Patrick J. and Freitas, Christopher J. and Coleman, Hugh and Raad, Peter E.},
  title   = {Procedure for Estimation and Reporting of Uncertainty Due to Discretization in CFD Applications},
  journal   = {Journal of Fluids Engineering},
  year      = {2008},
  issn      = {0098-2202},
  number    = {7},
  pages     = {078001},
  volume    = {130},
  doi       = {10.1115/1.2960953},
  publisher = {ASME International},
}

@Book{White1991,
  author    = {White, Frank M.},
  publisher = {McGraw-Hill},
  title     = {Viscous fluid flow},
  year      = {1991},
  address   = {New York [u.a.]},
  edition   = {2nd ed.},
  isbn      = {0070697124},
  pagetotal = {614},
  ppn_gvk   = {02714965X},
}

@Book{Roache1998,
  author    = {Roache, Patrick J.},
  publisher = {Hermosa Publ.},
  title     = {Verification and validation in computational science and engineering},
  year      = {1998},
  address   = {Albuquerque, NM},
  isbn      = {0913478083},
  note      = {Includes bibliographical references and index},
  pagetotal = {446},
  ppn_gvk   = {250414368},
}

@Conference{Bell2016,
  author    = {Ian Bell and Arno Laesecke},
  title     = {Viscosity of refrigerants and other working fluids from residual entropy scaling},
  year      = {2016},
  month     = {2016-07-15},
  publisher = {16th International Refrigeration and Air Conditioning Conference at Purdue, July 11-14, 2016, West Lafayette, IN},
  language  = {en},
  url       = {https://tsapps.nist.gov/publication/get_pdf.cfm?pub_id=920655},
}

@Book{Salari2000,
  author = {Salari, Kambiz and Knupp, Patrick},
  title  = {Code Verification by the Method of Manufactured Solutions},
  year   = {2000},
  month  = June,
  doi    = {10.2172/759450},
  school = {Office of Scientific and Technical Information (OSTI)},
}

@Article{Roache2001,
  author    = {Roache, Patrick J.},
  journal   = {Journal of Fluids Engineering},
  title     = {Code Verification by the Method of Manufactured Solutions},
  year      = {2001},
  issn      = {1528-901X},
  month     = Nov,
  number    = {1},
  pages     = {4--10},
  volume    = {124},
  doi       = {10.1115/1.1436090},
  publisher = {ASME International},
}

@Article{Bouchehit2021,
  author    = {Bouchehit, Bachir and Bou-Saïd, Benyebka and Tichy, John},
  journal   = {Lubricants},
  title     = {Towards Ecological Alternatives in Bearing Lubrication},
  year      = {2021},
  issn      = {2075-4442},
  month     = June,
  number    = {6},
  pages     = {62},
  volume    = {9},
  doi       = {10.3390/lubricants9060062},
  publisher = {MDPI AG},
}

@Article{Wu2026,
  author    = {Wu, Guifang and Tang, Mingsheng and Zou, Huiming and Liu, Shanquan and Ren, Zhijie and Zhang, Shuo},
  journal   = {International Journal of Refrigeration},
  title     = {Transient aerodynamic characteristics and parameter optimization of porous aerostatic bearings for JT linear compressors},
  year      = {2026},
  issn      = {0140-7007},
  month     = Oct,
  pages     = {107025},
  volume    = {190},
  doi       = {10.1016/j.ijrefrig.2026.107025},
  publisher = {Elsevier BV},
}

@Article{DePaoli2023,
  author    = {De Paoli, Marco},
  journal   = {The European Physical Journal E},
  title     = {Convective mixing in porous media: a review of Darcy, pore-scale and Hele-Shaw studies},
  year      = {2023},
  issn      = {1292-895X},
  month     = Dec,
  number    = {12},
  volume    = {46},
  doi       = {10.1140/epje/s10189-023-00390-8},
  publisher = {Springer Science and Business Media LLC},
}

\end{document}